\documentclass[aps,prx,twocolumn,superscriptaddress,showpacs,footnotebib]{revtex4-2}
\usepackage{amsmath}
\usepackage{bm}
\usepackage{color}
\usepackage{graphicx}
\usepackage{epstopdf}
\usepackage{epsfig}
\usepackage{amsfonts}
\usepackage[naturalnames]{hyperref}
\usepackage{hypcap}
\usepackage{verbatim}
\usepackage{tabularx}
\usepackage{bbm}
\usepackage{esvect}
\usepackage{subfigure}
\usepackage{slashed}
\usepackage{comment}

\usepackage{leftindex}
\usepackage{bbm}

\usepackage{comment}
\usepackage{booktabs}
\usepackage{multirow}
\usepackage{makecell} 
\usepackage{tikz}
\usetikzlibrary{arrows.meta,positioning,fit}

\def\bea{\begin{eqnarray}}
\def\eea{\end{eqnarray}}

\def\ba{\begin{array}}
\def\ea{\end{array}}

\hypersetup{colorlinks=true, citecolor=blue, urlcolor=blue, linkcolor=blue}
\begin{document}
\title{Analytic Bootstrap for $O(N)$ Boundary Conformal Field Theories with Interacting Boundaries}
\author{Xinyu Sun}
\affiliation{Institute for Advanced Study, Tsinghua University, Beijing 100084, China}
\author{Shao-Kai Jian}
\email{sjian@tulane.edu}
\affiliation{Department of Physics and Engineering Physics, Tulane University, New Orleans, Louisiana, 70118, USA}
\author{Hong Yao}
\email{yaohong@tsinghua.edu.cn}
\affiliation{Institute for Advanced Study, Tsinghua University, Beijing 100084, China}

\date{\today}

\begin{abstract}
We investigate $O(N)$ boundary conformal field theories (BCFTs) with boundary interactions in $d=4-\epsilon$ and $d=3-\epsilon$ employing the analytic bootstrap. 
By deriving universal constraints on conformal data, we show that infinitely many operator expansions can be expressed in terms of a finite set of inputs. 
Complementing the analytic bootstrap with a perturbative renormalization-group analysis, we identify totally new boundary fixed points in $d=4-\epsilon$, including non-unitary ones, generated by a boundary cubic coupling, and compute their conformal data to leading order. 
Moreover, we leverage our solution in $d=3-\epsilon$ to extract, for the first time, the boundary conformal data for the tricritical $O(N)$ model.  
Altogether, our approach provides a unified prescription for BCFTs with interacting boundaries and streamlines the determination of bulk and boundary operator expansions.
\end{abstract}

\maketitle

\section{Introduction}
\label{sec:intro}

Conformal field theory (CFT) plays a central role across many areas of physics and exhibits universal behavior governed by conformal data, most notably scaling dimensions and operator product expansion (OPE) coefficients~\cite{francesco2012conformal,amit2005field}.
The presence of a boundary in CFTs introduces substantially richer structure~\cite{cardy1984conformal,Domb1986PHASE,affleck1991universal,bariev1992surface,diehl1997theory,cardy2004boundary,herzog2016universal,hogervorst2017crossing,andrei2020boundary}, adding boundary-specific conformal data, such as boundary primary fields and the boundary operator expansion (BOE)~\cite{lubensky1975critical,ohno1984The,diehi1987walks,lauria2018radial,dey2020operator,krishnan2023plane}, while preserving a subgroup of the bulk conformal symmetry~\cite{ishibashi1989boundary,cardy1989boundary}.
In close analogy with the role of crossing symmetry for four-point correlators in CFT, which constrains scaling dimensions and OPE coefficients~\cite{polyakov1974nonhamiltonian,belavin1984infinite,dolan2004conformal,poland2019conformal}, the presence of a boundary allows two-point correlators in boundary CFTs (BCFTs) to put strong constraints on the conformal data, including not only the bulk ones but also the boundary primary fields and BOE coefficients~\cite{cardy1991bulk,rastelli2017mellin,mazavc2019analytic,pagani2020operator,kaviraj2020functional,nishioka2023comments}. 
Building on this observation, Ref.~\cite{liendo2013bootstrap} introduced an analytic bootstrap method for BCFT, with subsequent work focusing on the $\phi^4$ theory and generalizations to interfaces~\cite{gliozzi2015boundary,behan2020bootstrapping,bissi2019analytic,dey2021analytic,behan2022bootstrapping}.

Conventional boundary phase transitions for the $O(N)$ model are commonly classified into three types: ordinary, special, and extraordinary~\cite{diehl1981field,diehl1983multicritical,burkhardt1987surface,batchelor1995surface,diehl1998massive,parisen2022boundary}. 
Recently, more exotic boundary phases have been identified, including the extraordinary-log phase~\cite{metlitski2022boundary,hu2021extraordinary,padayasi2022extraordinary,sun2023extraordinary,parisen2024universal}, continuous symmetry breaking at the boundary~\cite{cuomo2024spontaneous,sun2025boundary}, and boundary phase transitions involving fermions~\cite{van2018analytic,giombi2022fermions,herzog2023fermions,giombi2023line,shen2024new,jiang2025boundary,barrat2025line}. 
Moreover, when additional interactions localized on the boundary are introduced, it can lead to new boundary fixed points and critical behavior~\cite{diehl1991surface,eisenriegler1988surface,herzog2019marginal,harribey2023boundaries}.
While the analytic bootstrap has been applied to explore $O(N)$ BCFTs, its use in studying BCFTs with nontrivial boundary interactions remains relatively underdeveloped. In view of the substantial progress in understanding BCFTs beyond conventional boundary transitions, it is therefore important to generalize the analytic bootstrap framework to systematically treat such boundary interactions.

In this paper, we clarify certain ambiguities in earlier implementations of the analytic bootstrap, and discuss the analytic bootstrap for %general composite operators $O^{(k)}\sim\phi^k$ without model-specific assumptions.
$O(N)$ BCFTs with boundary interactions. 
As a starting point, we obtain operator expansions for general composite operators $O^{(k)} \sim \phi^k$ for a free $O(N)$ theory, which already contains infinitely many conformal blocks in general. 
These expansions serve as a zeroth order solution to the crossing symmetry equation in $\epsilon=d_0-d$ expansion with an integer $d$.  
In particular, the expansion for $k=1$ consists of finitely many conformal blocks and enables the calculation of the next order. 
Hence, we focus on bootstrapping the crossing symmetry equation for $\langle\phi(x)\phi(y)\rangle$ correlation in $O(N)$ BCFTs in $d=4-\epsilon$ as well as in $d=3-\epsilon$~\cite{speth1983tricritical,eisenriegler1988surface} with interacting boundaries. 
We use a finite set of conformal data as input to the solutions, and express all remaining conformal data in terms of these inputs.
Notably, our solution in $d=3-\epsilon$ applies to the tricritical point, where a boundary quartic coupling is necessary at the special transition~\cite{eisenriegler1988surface}. 

In addition, to construct a concrete example for the $O(N)$ BCFT in $d=4-\epsilon$ with interacting boundaries, we consider a boundary cubic coupling that respects $S_{N+1}$ symmetry. 
Note that the BCFT of cubic theories has been studied in Ref.~\cite{dey2021analytic,guo2026boundary}.
This boundary interaction can be realized by adding a Potts anisotropy on the boundary to the $O(N)$ model. 
By performing renormalization group (RG) analysis, we identify two fixed points arising from boundary interactions, including a \textit{long-range Potts} fixed point where the bulk is free, and a nontrivial \textit{boundary Potts} fixed point with nonvanishing bulk and boundary interaction strength. 
We further compute the OPE and BOE coefficients for them. 
The RG calculation further provides the input for the bootstrap solution, which yields the full set of conformal data to leading order.

The organization of the paper is as follows.
We summarize our results in the remainder of the introduction, including the operator expansion for $O^{(k)} \sim \phi^k$, the conformal data for several BCFTs obtained from the bootstrap, and the fixed points found by RG calculations. 
In Section~\ref{sec:Framework of analytic bootstrap for boundary CFT}, we review the analytic bootstrap procedure for general BCFTs and clarify some subtleties along the way. 
In Section~\ref{sec:General operator expansion}, we solve the operator expansion for $\langle O^{(k)}O^{(k)}\rangle$ at zeroth order, which already involves infinitely many conformal blocks when $k>1$. 
Then, starting from the finitely many expansions for $k=1$, we focus on bootstrapping $\langle\phi(x)\phi(y)\rangle$ correlation in Section~\ref{sec:d 4-epsilon expansion}. 
We analytically bootstrap the $O(N)$ theory with interacting boundaries in $d=4-\epsilon$, and obtain the solution for infinitely many BOE and OPE up to order $\mathcal{O}(\epsilon^2)$, with only a handful of inputs. 
In addition, we perform the RG analysis for a boundary cubic coupling that respects $S_{N+1}$ symmetry in $d=4-\epsilon$, revealing several new boundary fixed points. 
Section~\ref{sec:d 3-epsilon expansion} presents the analytic bootstrap for $d=3-\epsilon$, with an application to the tricritical $O(N)$ model, where the boundary interaction is unavoidably generated along the RG~\cite{eisenriegler1988surface}.  
Finally, we conclude in Section~\ref{sec:Conclusion} with a discussion of possible future directions for our theory. 
The appendices provide technical details of the calculations.

\subsection{Summary of the results}
\label{sec:Summary of the results}

\begin{figure}
    \centering
    \includegraphics[width=1\linewidth]{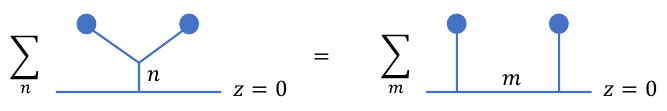}
    \caption{The crossing symmetry equation in Eq.~\eqref{eq:equivalence_bulk_boundary_channel_summary}.
    The left- and right-hand sides are the operator expansion in the bulk and boundary channels, and blue circles label bulk operators.}
    \label{fig:bootstrap_eq}
\end{figure}

Consider a BCFT defined on $\mathbb{R}^d_+=\{x=(r,z)\,|\,r\in\mathbb{R}^{d-1},\,z>0\}$.
The two-point function, $\langle O(x)O(y)\rangle=\frac{F(\xi)}{|x-y|^{2\Delta_O}}$, of a primary operator $O$ with scaling dimension $\Delta_O$ admits equivalent decompositions in the bulk and boundary channels,
\begin{eqnarray}
\label{eq:equivalence_bulk_boundary_channel_summary}
    F(\xi)&=&1+\sum_n \tilde \lambda_n f_b(\Delta_n,\xi) \nonumber\\
    &=&\xi^{\Delta_O}\left(a_O^2+\sum_m \tilde{\mu}_m f_i(\hat{\Delta}_m,\xi)\right)\,,
\end{eqnarray}
where $f_b(\Delta_n,\xi)$ and $f_i(\hat{\Delta}_m,\xi)$ are the bulk and boundary conformal blocks, where  $\xi$ is the cross-ratio and $\Delta_n$ and $\hat{\Delta}_m$ are the scaling dimensions of the operators appearing in the expansions. 
The expansion coefficients, $\tilde{\lambda}_n$, $\tilde{\mu}_m$ are defined through the OPE and BOE coefficients:
$\tilde{\lambda}_n = \lambda_n a_n$, and $\tilde \mu_m = \mu_m^2$, where $\lambda_n$ and $\mu_m$ are the OPE and BOE coefficients, respectively, and $a_n$ denotes the nonvanishing one-point function in the presence of boundaries. 
We will refer to $\tilde{\lambda}_n$ and $\tilde{\mu}_m$ as the OPE and BOE coefficients, respectively, without further ambiguity. 
The crossing symmetry equation~\eqref{eq:equivalence_bulk_boundary_channel_summary} as illustrated in Fig.~\ref{fig:bootstrap_eq} puts stringent constraint among the conformal data, and thus, serves as the cornerstone of the analytic bootstrap.  

We solve the crossing symmetry equation for the $O(N)$ theory supplemented with boundary interactions via the $\epsilon$ expansion.  
Our contributions include the operator expansion for the general composite operator $O^{(k)} \sim \phi^k$ in the free $O(N)$ theory, which is also the zeroth order solution to the crossing symmetry equation, and provides the starting point for the $\epsilon$ expansion. 
We analytically bootstrap the $O(N)$ BCFT with interacting boundaries in $\epsilon = 4-d$ expansion up to $\mathcal O(\epsilon^2)$. 
We are able to express the infinitely many OPE and BOE coefficients in terms of finitely many inputs. 
Moreover, as a concrete example of BCFT with interacting boundaries,  we present a RG calculation for the $O(N)$ model with a boundary coupling that preserves $S_{N+1}$ symmetry and identify several new boundary fixed points along with their corresponding conformal data at leading order. 
In addition, we bootstrap the $O(N)$ BCFT in $d=3-\epsilon$ and obtain the conformal data for several boundary fixed points, including the ordinary and special fixed point in the tricritical $O(N)$ model.

\subsubsection{General operator expansion at zeroth order}
\label{General operator expansion at zeroth order}

We compute the operator expansions for $O^{(k)}\sim\phi^k$ in both the bulk and boundary channels at zeroth order, or equivalently, in a free theory, under either the Neumann or the Dirichlet boundary condition.  
In the $O(N)$ field theory, the normalized operators $O^{(k)}$ are defined as 
\begin{eqnarray}
\label{eq:O_k_definition_summary}
    O^{(2m)}&=&\sqrt{\frac{\Gamma(N/2)2^{-2m}}{\Gamma(m+N/2)\Gamma(m+1)}}:\left(\sum_{i=1}^N \phi_i^2\right)^m:\,,\\ O^{(2m+1)}_i&=&\sqrt{\frac{\Gamma(N/2+1)2^{-2m}}{\Gamma(m+N/2+1)\Gamma(m+1)}}:\left(\sum_{i=1}^N \phi_i^2\right)^m\phi_i:\,.  \nonumber
\end{eqnarray}
The corresponding two-point function at zeroth order reads $\langle O^{(k)}(x)O^{(k)}(y)\rangle_s=\frac{F_s^{(k)}(\xi)}{|x-y|^{2\Delta_{O^{(k)}}}}$, with
\begin{eqnarray}
\label{eq:F_phik_summary}
    F_s^{(k)}(\xi)=\left(1+s\left(\frac{\xi}{\xi+1}\right)^{\Delta_\phi}\right)^k+\alpha_{k,N}\xi^{k\Delta_\phi}\,, \\
    \alpha_{2m+1,N} = 0 \,,  \qquad \alpha_{2m,N}=\frac{\Gamma(m+N/2)}{\Gamma(N/2)\Gamma(m+1)}\,.
\end{eqnarray}
Here, $\langle\cdot\rangle_s$ with $s=\pm$ denotes the expectation value under the Neumann and Dirichlet boundary conditions, respectively.

As in Eq.~\eqref{eq:equivalence_bulk_boundary_channel_summary}, the correlator admits equivalent decompositions in the bulk and boundary channels.
In the bulk channel, there are $k$ families of conformal blocks with scaling dimensions $\Delta=2l\Delta_\phi+2n$, where $l=1,\cdots,k$ and $n=0,1,\cdots$, corresponding to the scalar primaries $:\phi^l \partial^{2n}\phi^l:$.  
The OPE coefficients are
\begin{equation}
\label{eq:OPE_phi_k_summary}
\begin{split}
    \tilde{\lambda}_{2l\Delta_\phi+2n}=\frac{\Gamma(k+1)}{\Gamma(l+1)\Gamma(k-l+1)}s^l c_{l\Delta_\phi,n}'+\delta_{l,k} \alpha_{k,N}c_{k\Delta_\phi,n}\,,
\end{split}
\end{equation}
where $c_{a,n}$ and $c_{a,n}'$ are given in Eqs.~\eqref{eq:bulk_second_c_solution} and \eqref{eq:bulk_first_c_solution}.
For $\Delta_\phi=\frac{d-2}{2}$ with an integer $d$, the expansion will collapse:
if $d$ is even, then all families collapse to a single series $\Delta=d-2+2n$; while if $d$ is odd, then two families remain, $\Delta=d-2+2n$ and $\Delta=2d-4+2n$.

In the boundary channel, there is a single family with scaling dimensions $\hat{\Delta}=k\Delta_\phi+n$ for $n=0,1,\cdots$, corresponding to the boundary primaries $:\partial^n \hat \phi^k:$. 
The BOE coefficients are 
\begin{equation}
\label{eq:BOE_general_free_summary}
    \tilde{\mu}_{k\Delta_\phi+n}=\sum_{l=0}^{k}\frac{\Gamma(k+1)}{\Gamma(l+1)\Gamma(k-l+1)}s^l d_{l\Delta_\phi,k,n}.
\end{equation}
where $d_{a,k,n}$ is given in Eq.~\eqref{eq:BOE_mu_m_general}.

When $k=1$, the bulk and boundary primary operators that appear in the operator expansions are $O_n\sim\phi\partial^{2n}\phi$ and $\hat O_m\sim\partial^m \hat\phi$, with scaling dimensions $\Delta_n=d-2+2n$ for $n=0,1,\cdots$ and $\hat{\Delta}_m=\frac{d-2}{2}+m$ for $m=0,1,\cdots$, respectively.
More importantly, the number of nonvanishing OPE and BOE coefficients in the zeroth-order solution is finite. 
To distinguish the correction at different orders, we use $\Delta_n^{(l)}$ and $\hat \Delta_m^{(l)}$ to denote the $\mathcal O(\epsilon^l)$ corrections to the bulk and boundary scaling dimensions, respectively. 
Similarly, we also use $\tilde \lambda_n^{(l)}$ and $\tilde \mu_m^{(l)}$ to denote the $\mathcal O(\epsilon^l)$ corrections to the OPE and BOE coefficients, respectively. 
The zeroth-order solution for $k=1$ is given in Eq.~\eqref{eq:0th_solution_general_bootstrap}.
This serves as the starting point for the subsequent analytic bootstrap at higher orders.

\subsubsection{Analytic bootstrap for $O(N)$ BCFTs in $d=4-\epsilon$}
\label{sec:Application of analytic bootstrap}

\begin{table}[t]
  \centering
  {\renewcommand{\arraystretch}{1.3}
  \begin{tabular}{lccccr} % l=left, c=center, r=right
    \toprule
    Fixed point & $\Delta_\phi^{(1)}$ & $\Delta_{0}^{(1)}$ & $\hat{\Delta}_{0}^{(1)}$ & $\tilde{\lambda}_{0}^{(1)}$ & other data \\
    \midrule
    special  & $0$ & $2\alpha$ & $-\alpha$ & $\alpha$ & Eq.~\eqref{eq:bulk_phi4_boundary_free_output} \\
    lrP/lrY-L   & 0 & 0 & 0 & $\alpha'$ &  Eq.~\eqref{eq:bulk_free_boundary_phi3_output}  \\
    bP/bY-L   & $0$ & $2\alpha$ & $-\alpha$ & $\beta$ & Eq.~\eqref{eq:bulk_phi4_boundary_phi3_output}   \\
    image (N)   & $\Delta_\phi^{(1)}$    & $\Delta_{0}^{(1)}$ & $-\frac{1}{2}\Delta_0^{(1)}+\Delta_{\phi}^{(1)}$ & $\frac{1}{2}\Delta_0^{(1)}$ &      \\
    \bottomrule
  \end{tabular}}
  \caption{Conformal data for BCFTs with $d=4-\epsilon$ expansion at $\mathcal{O}(\epsilon)$ order, with $\alpha=\frac{1}{2}\frac{N+2}{N+8}$, and $\alpha'=-\frac{1}{64\pi^2}$ ($\alpha'=-\frac{N-1}{64\pi^2(2-N)}$) for the long-range Yang-Lee (Potts) fixed point, and $ \beta = \alpha+\tilde{\alpha}'$, $\alpha' = - \frac1{32\pi^2}$ ($\tilde{\alpha}'=-\frac{(N-1)(5-2N)}{32\pi^2(N+8)(N-2)}$) for the boundary Yang-Lee (Potts) fixed point. 
  lrP/lrY-L (bP/bY-L) denote the long-range Potts/long-range Yang-Lee (boundary Potts/boundary Yang-Lee) fixed points, and image (N) corresponds to the fixed point with the image symmetry and the Neumann boundary condition.}
  \label{tab:4_epsilon_epsilon1}
\end{table}

We present the results for the $O(N)$ BCFT in $d=4-\epsilon$.
Solving the analytic bootstrap equations for $O^{(k=1)}_i = \phi_i$ yields universal constraints on the conformal data, which in turn provide the OPE and BOE coefficients at $\mathcal O(\epsilon)$,

\begin{equation}
\label{eq:OPE_d_4_epsilon_summary}
\begin{split}
    &\tilde{\lambda}_n^{(1)}=\begin{cases}
        \frac{1}{4}(A_1-2A_2),&\qquad n=1,\\[0.7ex]
        \frac{\Gamma(n)\Gamma(n+1)}{2n(-1)^{n+1} \Gamma(2n)}(A_1-A_2),&\qquad n>1,
    \end{cases}\\[1.ex]
    &\tilde{\mu}_m^{(1)}=\begin{cases}
        \hat{\Delta}_{0}^{(1)}-\Delta_\phi^{(1)}+\tilde{\lambda}_{0}^{(1)},& m=0,\\[0.7ex]
        \frac{5}{2}\Delta_\phi^{(1)}-\frac{3}{2}\hat{\Delta}_{0}^{(1)}-\frac{\tilde{\lambda}_{0}^{(1)}}{2}-\frac{\Delta_{0}^{(1)}}{2},& m=1,\\[0.7ex]
        \frac{4^{1-m}\sqrt{\pi}\Gamma(m-1)}{\Gamma(m-\frac{1}{2})}\left[(-1)^{1-m}B_3'\right.\\
        \left.\qquad\qquad\ +(1+(-1)^m) B_4'\right],& m>1,
    \end{cases}
\end{split}
\end{equation}
where $A_i$ and $B_j'$ are given in Eqs.~\eqref{eq:parameters_A_d_4_epsilon} and \eqref{eq:parameters_Bp_d_4_epsilon}.
Note that the solution is general rather than model specific, and, crucially, only four inputs $\Delta_\phi^{(1)}$, $ \Delta_{0}^{(1)}$, $\hat{\Delta}_{0}^{(1)}$, and $\tilde{\lambda}_{0}^{(1)}$, determine infinitely many OPE and BOE coefficients.

Now, we consider a concrete BCFT model in $d=4-\epsilon$ with interacting boundaries. 
The theory,  governed by the action Eq.~\eqref{eq:bulk_phi4_boundary_phi3}, features a cubic boundary interaction that respects $S_{N+1}$ symmetry.  %in Sec.~\ref{sec:Boundary phase transition with bulk-phi4/boundary-phi3 interaction} and Sec.~\ref{sec:RG of the bulk-phi4/boundary-phi3 theory}. 
It hosts two interesting fixed points, termed as the long-range Potts fixed point where the bulk is free and the boundary has nonvanishing cubic interaction and the boundary Potts fixed point where both bulk and boundary have nonzero couplings. 
Their conformal data to order $\mathcal{O}(\epsilon)$ are summarized in Table~\ref{tab:4_epsilon_epsilon1}.
%Note that the special and long-range Potts/Yang-Lee theories can be viewed as special cases of the boundary Potts/Yang-Lee theory, with $\tilde{\lambda}_{0}^{(1)}=\frac{1}{2}\Delta_{0}^{(1)}$ or $\Delta_{0}^{(1)}=\hat{\Delta}_{0}^{(1)}=0$.
Note that we also include the special transition of the Wilson-Fisher fixed point for completeness. 
In addition, a scalar field with boundary Yang-Lee type interactions has similar fixed points~\cite{diehl1991surface}, which are included as well.  

From the results above, we find that all fixed points involve only finitely many conformal blocks at order $\mathcal{O}(\epsilon)$.
Therefore, building on the boundary Potts fixed point, we solve the bootstrap equations to the next order $\mathcal{O}(\epsilon^2)$, which yields the OPE and BOE coefficients
\begin{equation}
\label{eq:OPE_d_4_epsilon2_summary}
\begin{split}
    &\tilde{\lambda}_n^{(2)}=\begin{cases}
        \frac{1}{8}(-2A_1+2A_2+3A_3+4A_4+4A_5),& n=1\\[0.7ex]
        \frac{\Gamma(n)\Gamma(n+1)}{8(-1)^n\Gamma(2n)}\left(\frac{4}{n}A_1+\frac{4(-1)^{n}((-1)^n+n)}{n(n^2-1)}A_2\right.\\
        \left.\qquad\ +\frac{4(-1)^{n}}{n^2-1}A_3-\frac{4}{n}A_4+\frac{4(-1)^{n}}{n^2}A_5\right),& n>1.
    \end{cases}\\[1.ex]
    &\tilde{\mu}_m^{(2)}=\begin{cases}
        B_1',& m=0,\\[0.7ex]
        B_2'-B_4'+B_5',& m=1,\\[0.7ex]
        \frac{4^{1-m}\sqrt{\pi}\Gamma(m-1)}{(-1)^{m-1}\Gamma(m-\frac{1}{2})}\left(B_2'+\frac{2(1+(-1)^m)}{m(m-1)}B_3'\right.\\
        \left.\qquad\ +\frac{(-1)^{m-1}}{m(m-1)}B_4'-(1+(-1)^m)B_6'\right),& m>1,
    \end{cases}
\end{split}
\end{equation}
where $A_i$ and $B_j'$ are given in Eqs.~\eqref{eq:parameters_A_d_4_epsilon2} and \eqref{eq:parameters_Bp_d_4_epsilon2}.
Here, $\Delta_0^{(1)},\, \tilde{\lambda}_0^{(1)}$ are the inputs at order $\mathcal{O}(\epsilon)$, while $\Delta_\phi^{(2)},\Delta_0^{(2)},\Delta_1^{(1)},\tilde{\lambda}_0^{(2)},\hat{\Delta}_0^{(2)},\hat{\Delta}_1^{(1)}$ are independent inputs at order $\mathcal{O}(\epsilon^2)$. 
The results for concrete fixed points are summarized in Table~\ref{tab:4_epsilon_epsilon2}. 

Finally, let's discuss the image symmetry introduced in Ref.~\cite{bissi2019analytic} and reviewed in Sec.~\ref{sec:Framework of analytic bootstrap for boundary CFT}. 
We find that it is not preserved for the BCFTs with boundary interactions.
Nevertheless, we explore the image symmetric fixed point by treating the image symmetry as a constraint. 
It imposes an additional constraint among the independent inputs in the general solutions. 
The results are shown in Table~\ref{tab:4_epsilon_epsilon1},~\ref{tab:4_epsilon_epsilon2} for completeness. 

% \begin{table}[t]
%   \centering
%   {\renewcommand{\arraystretch}{1.3}
%   \begin{tabular}{lccr} % l=left, c=center, r=right
%     \toprule
%     Fixed point & free inputs & fixed inputs & other data \\
%     \midrule
%     special  & $\Delta_\phi^{(2)},\Delta_0^{(2)}$ & Eq.~\eqref{eq:relation_conformal_data_two_image_symmetry_epsilon2} &  Eq.~\eqref{eq:bulk_phi4_bpundary_free_epsilon2_OPE_BOE} \\
%     lrP/lrY-L   & $\tilde{\lambda}_0^{(2)},\hat{\Delta}_1^{(1)}$ & trivial &  Eq.~\eqref{eq:bulk_free_boundary_phi3_epsilon2_confromal_data}  \\
%     bP/bY-L   &  \makecell[l]{$\Delta_\phi^{(2)},\Delta_0^{(2)},\Delta_1^{(1)},$\\
%     $\tilde{\lambda}_0^{(2)},\hat{\Delta}_0^{(2)},\hat{\Delta}_1^{(1)}$} & None &  Eqs.~\eqref{eq:OPE_d_4_epsilon2},\eqref{eq:BOE_d_4_epsilon2} \\
%     image (N)   & $\Delta_\phi^{(2)},\Delta_0^{(2)},\hat{\Delta}_1^{(1)}$  & Eq.~\eqref{eq:relation_conformal_data_image_symmetry_epsilon2} &    \\
%     \bottomrule
%   \end{tabular}}
%   \caption{Conformal data for BCFTs in the $d=4-\epsilon$ expansion at order $\mathcal{O}(\epsilon^2)$.
%   Only the new inputs at this order are shown. 
%   ``Free inputs'' are fixed by RG; ``fixed inputs'' are determined by the analytic bootstrap.
%   lrP/lrY-L (bP/bY-L) denote the long-range Potts/long-range Yang-Lee (boundary Potts/boundary Yang-Lee) fixed points, and image (N) corresponds to the fixed point with the image symmetry and Neumann boundary condition}
%   \label{tab:4_epsilon_epsilon2}
% \end{table}

\begin{table}[t]
  \centering
  {\renewcommand{\arraystretch}{1.3}
  \begin{tabular}{lccr} % l=left, c=center, r=right
    \toprule
    Fixed point & independent inputs &  other data \\
    \midrule
    special  & $\Delta_\phi^{(2)},\Delta_0^{(2)}$ & Eqs.~\eqref{eq:relation_conformal_data_two_image_symmetry_epsilon2},\ \eqref{eq:bulk_phi4_bpundary_free_epsilon2_OPE_BOE} \\
    lrP/lrY-L   & $\tilde{\lambda}_0^{(2)},\hat{\Delta}_1^{(1)}$ &  Eq.~\eqref{eq:bulk_free_boundary_phi3_epsilon2_confromal_data}  \\
    bP/bY-L   & $\Delta_\phi^{(2)},\Delta_0^{(2)},\Delta_1^{(1)},\tilde{\lambda}_0^{(2)},\hat{\Delta}_0^{(2)},\hat{\Delta}_1^{(1)}$ & Eqs.~\eqref{eq:OPE_d_4_epsilon2},\eqref{eq:BOE_d_4_epsilon2} \\
    image (N)   & $\Delta_\phi^{(2)},\Delta_0^{(2)},\hat{\Delta}_1^{(1)}$  & Eq.~\eqref{eq:relation_conformal_data_image_symmetry_epsilon2}  \\
    \bottomrule
  \end{tabular}}
  \caption{Conformal data for BCFTs in the $d=4-\epsilon$ expansion at order $\mathcal{O}(\epsilon^2)$.
  Only the independent inputs at this order are shown. 
  ``other data'' are determined by the analytic bootstrap.}
  %lrP/lrY-L (bP/bY-L) denote the long-range Potts/long-range Yang-Lee (boundary Potts/boundary Yang-Lee) fixed points, and image (N) corresponds to the fixed point with the image symmetry and Neumann boundary condition}
  \label{tab:4_epsilon_epsilon2}
\end{table}

\subsubsection{Analytic bootstrap for $O(N)$ BCFT in $d=3-\epsilon$}
\label{sec:summary_3d}

We summarize the analytic bootstrap results for the $O(N)$ BCFT in $d=3-\epsilon$. 
At order $\mathcal{O}(\epsilon)$, the general constraints from the bootstrap equations with the Neumann boundary condition give the OPE and BOE coefficients
\begin{eqnarray}
\label{eq:OPE_d_3_epsilon_Neumann_summary}
    && \tilde{\lambda}_n^{(1)}=\begin{cases}
        -\frac{1}{3}A_1-\frac{1}{2}A_2,&\qquad n=1,\\[0.7ex]
        \frac{\Gamma(n)\Gamma(n+\frac{1}{2})\Gamma(n-\frac{1}{2})}{2\sqrt{\pi}(-1)^n \Gamma(2n-\frac{1}{2})\Gamma(n+1)}A_2,&\qquad n>1.
    \end{cases}\nonumber\\
    \\
    &&\tilde{\mu}_m^{(1)}=\begin{cases}
        B_2',& m=0\,,\\[0.7ex]
        -\frac{1}{2}B_1'+B_3'+\frac{1-\log{4}}{4}B_4'+\frac{1}{4}B_5'\,,& m=1,\\[2.ex]
        -\frac{2^{-2m}}{m(m-1)}\left[2(1+(-1)^m(2m-1))B_3'\right.\\
        \left.\qquad\qquad\qquad\quad+B_4'-(2m-1)B_5'\right],& m>1, \nonumber
    \end{cases}
\end{eqnarray}
where $A_i$ and $B_j'$ are given in Eqs.~\eqref{eq:parameters_A_d_3_epsilon} and \eqref{eq:parameters_Bp_d_3_epsilon}. 
The independent inputs are $\Delta_\phi^{(1)},\, \Delta_{0}^{(1)},\, \hat{\Delta}_{0}^{(1)}$ and $\tilde{\lambda}_{0}^{(1)}$.

In addition, we have considered the expansion around the zeroth order Dirichlet solution.
The general constraints from the bootstrap equations determine the corresponding OPE and BOE coefficients.
They have the same functional form as Eq.~\eqref{eq:OPE_d_3_epsilon_Neumann_summary}, but with different $A_i$ and $B_j'$ as specified in Eqs.~\eqref{eq:parameters_A_d_3_epsilon_Dirichlet} and \eqref{eq:parameters_Bp_d_3_epsilon_Dirichlet}.

Above general results can be applied to the tricritical $O(N)$ model at $d=3-\epsilon$. 
At the special fixed point, the boundary $\phi^4$ coupling is necessarily generated under the RG flow, even if it is initially set to zero~\cite{eisenriegler1988surface}. 
We are able to obtain the infinitely many OPE and BOE exponents at order $\mathcal O(\epsilon)$. 
Additionally, we have considered the ordinary fixed point. 
The results are summarized in Table~\ref{tab:3_epsilon_epsilon}.

Another interesting case is given by the free $O(N)$ model supplemented by a $\phi^4$ coupling on the boundary, whose fixed point is termed as the long-range $\phi^4$ fixed point. 
At this fixed point, the correlation function up to $\mathcal O(\epsilon^2)$ reads
\begin{eqnarray}
\label{eq:2_pt_bulk_free_boundary_phi4_d_3_summary}
    &&\langle\phi(x)\phi(x')\rangle=\frac{1}{\sqrt{4zz'}}\left(\frac{1}{\xi^{1/2}}+\frac{1+\tilde{\lambda}_0^{(2)}\epsilon^2}{(1+\xi)^{1/2}}\right)\,,
\end{eqnarray}
where the OPE coefficient $\tilde{\lambda}_0^{(2)}=-\frac{4(N+2)}{(N+8)^2}$~\cite{prochazka2020composite}. 
Note that the $\mathcal O(\epsilon)$ corrections are vanishing, and the OPE and BOE coefficients at $\mathcal O(\epsilon^2)$  are given by 
\begin{equation} 
\label{eq:OPE_BOE_d_3_boundary_phi_4_image_symm_summary}
\begin{split}
& \tilde{\lambda}_0^{(2)}=-\frac{4(N+2)}{(N+8)^2}\,, \quad \tilde{\lambda}_{n\ge1}^{(2)} = 0\,,\\
& \tilde{\mu}_0^{(2)}=-4\tilde{\mu}_1^{(2)}=\tilde{\lambda}_0^{(2)}\,, \quad \tilde{\mu}_{m\geq2}^{(2)}=0\,.
\end{split}
\end{equation}

%The corresponding data are shown in Table~\ref{tab:3_epsilon_epsilon}.

\begin{table}[t]
  \centering
  {\renewcommand{\arraystretch}{1.3}
  \begin{tabular}{lccccr} % l=left, c=center, r=right
    \toprule
    Fixed point & $\Delta_\phi^{(1)}$ & $\Delta_{0}^{(1)}$ & $\hat{\Delta}_{0}^{(1)}$ ($\hat{\Delta}_{1}^{(1)}$) & $\tilde{\lambda}_{0}^{(1)}$ & other data \\
    \midrule
    special  & 0 & 0 & $\hat{\Delta}_{0}^{(1)} = -\frac{(N+2)(N+4)}{16(3N+22)}$  & 0 & Eq.~\eqref{eq:OPE_BOE_d_3_phi_6_free} \\
    %long-range $\phi^4$   & 0 & 0 & 0 & $0^*$ &  Eq.~\eqref{eq:OPE_BOE_d_3_boundary_phi_4_image_symm}  \\
    % boundary $\phi^4$   &  0 & 0 & $\hat{\Delta}_{0}^{(1)}$ & $\tilde{\lambda}_{0}^{(1)}$  & Eq.~\eqref{eq:OPE_BOE_d_3_phi_6_free}   \\
    image (N)   & $\Delta_\phi^{(1)}$    & $2\Delta_\phi^{(1)}$ & $\hat{\Delta}_{0}^{(1)} = \Delta_\phi^{(1)}$ & $0$ &  Eq.~\eqref{eq:OPE_BOE_d_3_phi_6_image_symm_N}    \\
    ordinary  & 0 & 0 & $\hat{\Delta}_{1}^{(1)} = \frac{(N+2)(N+4)}{8(3N+22)}$ & $\tilde{\lambda}_{0}^{(1)}$ & Eq.~\eqref{eq:OPE_BOE_d_3_phi_6_free_Dirichlet} \\
    image (D)   & $\Delta_\phi^{(1)}$    & $2\Delta_{\phi}^{(1)}$ & $\hat{\Delta}_{1}^{(1)} =  \Delta_\phi^{(1)}$ & 0 &  Eq.~\eqref{eq:OPE_BOE_d_3_phi_6_image_symm_D}   \\
    \bottomrule
  \end{tabular}}
  \caption{Conformal data for BCFTs in the $d=3-\epsilon$ expansion at order $\mathcal{O}(\epsilon)$.
  We use $\hat{\Delta}_{0}^{(1)}$ ($\hat{\Delta}_{1}^{(1)}$) to denote the input anomalous dimension of the boundary primary operator $\hat{\phi}$ ($\partial\hat{\phi}$) with Neumann (Dirichlet) boundary conditions.
  Image (N) and image (D) correspond to the fixed points with the image symmetry and Neumann and Dirichlet boundary conditions.
  % For the long-range $\phi^4$ theory, we use $0^*$ to indicate that all corrections at the order $\mathcal{O}(\epsilon)$ vanish, and the only contribution at the order $\mathcal{O}(\epsilon^2)$ is $\tilde{\lambda}_{0}^{(2)}$.
  }
  \label{tab:3_epsilon_epsilon}
\end{table}

Finally, although the concrete models discussed above do not respect the image symmetry, we explore the image-symmetric fixed point by imposing the image symmetry as a constraint on the general $\mathcal{O}(\epsilon)$ solution, starting from both the zeroth order Neumann and Dirichlet solutions.
The image symmetry imposes an additional constraint among the independent inputs in the general solution, Eq.~\eqref{eq:OPE_d_3_epsilon_Neumann_summary}.  
The results are also presented in Table~\ref{tab:3_epsilon_epsilon}.

\subsubsection{RG investigation for the $O(N)$ theory with cubic boundary
interactions}
\label{sec:Boundary phase transition with bulk-phi4/boundary-phi3 interaction}

The concrete BCFT example in $d=4-\epsilon$, presented above, is defined by the following action
%~\cite{diehl1991surface}
\begin{eqnarray}
\label{eq:bulk_phi4_boundary_phi3}
    S&=&\int_{\mathbb{R}_+^d} {\rm d}^{d-1}r\ {\rm d}z\left[ \frac{1}{2}(\nabla \phi_i)^2 %+\frac{\tau_0}{2}\phi_i^2
    +\frac{u_0}{4!}(\phi_i^2)^2\right.\\ %-h_0\phi_i
    &&\qquad\qquad\qquad \left.+\delta(z)\left(\frac{c_0}{2}\phi_i^2%-h_{1,0}\phi_i
    +\frac{w_0}{3!} d_{ijk}\phi_i\phi_j\phi_k\right) \right]\,, \nonumber
\end{eqnarray}
where the $N$-component field, denoted as $\phi_i$, $i=1,...,N$, transforms in the usual way under the $O(N)$ group in the $d$-dimensional semi-infinite space $\mathbb{R}_+^d$ with a boundary at $z=0$. 
On the boundary, $\phi_i$ fulfills the $N$-dimensional irreducible representation of the permutation group $S_{N+1}$, and hence the $S_{N+1}$ invariant cubic interaction is present. 
The $u_0$ and $w_0$ are the bulk and boundary interaction strengths, respectively, and $c_0$ is the boundary mass. 
The third rank tensor $d_{ijk}$, capturing the Potts anisotropy on the boundary, is defined in Eq.~\eqref{eq:interaction_dijk}. 
%To obtain the inputs for our analytic bootstrap results, we perform a detailed RG analysis of the bulk-$\phi^4$/boundary-$\phi^3$ theory.
%There are two types of boundary-$\phi^3$ interactions, corresponding to a trivial or nontrivial factor $d_{ijk}$.
%Since the bulk RG flow of the $\phi^4$ interaction is independent of the choice of boundary interaction, we focus on the boundary RG flow.
The corresponding RG equation for the boundary coupling is
\begin{equation}
\label{eq:RG_flow_w_summary}
    \beta(w)=-w\left(\frac{\epsilon}{2}+(\frac{3}{2}C_2-6C_3)u+8C_5 w^2\right)\,,
\end{equation}
where $C_2=\frac{N+2}{3}$,$C_3=\frac{2}{3}$, $C_5=(N+1)^2(N-2)$.  
The anomalous dimension of the boundary operator $\hat{\phi}$ is independent of the choice of boundary interaction, whereas the anomalous dimension of the boundary operator $\hat{\phi}^2$ acquires a nontrivial contribution,
\begin{equation}
\label{eq:eta_c_general_n_summary}
    \eta_c=-\frac{N+2}{3}\left(u+\frac{1-4\pi^2}{6}u^2\right)+8C_0 w^2,
\end{equation}
where $C_0=(N+1)^2(N-1)$. 

The RG equation Eq.~\eqref{eq:RG_flow_w_summary} yields three nontrivial fixed points  corresponding, respectively, to the special transition point, the long-range Potts transition point, and the boundary Potts fixed point:
\begin{eqnarray}
\label{eq:fixed_point_RG_dijk_summary}
    (u^*,w^*)_{\rm sp}&=&\left(\frac{3\epsilon}{N+8},0\right),\\
    (u^*,w^*)_{\rm lrP}&=&\left(0,\frac{1}{4}\sqrt{\frac{\epsilon}{(N+1)^2(2-N)}}\right)\,, \nonumber \\
    (u^*,w^*)_{\rm bP}&=&\left(\frac{3\epsilon}{N+8},\sqrt{\frac{(5-2N)\epsilon}{8(N+8)(N+1)^2(N-2)}}\right)\, . \nonumber
\end{eqnarray}
When $N<\frac{5}{2}$, the special fixed point is stable and governs the critical phenomena separating the ordinary and extraordinary transition. 
Notably, for $N>\frac{5}{2}$ the boundary Potts fixed point collides with the special fixed point and subsequently moves into the complex plane, rendering the special fixed point unstable. 
The resulting stable boundary Potts fixed point with an imaginary coupling $w^*$ corresponds to a non-unitary BCFT.  
We therefore conjecture that, for real values of the microscopic parameters, the transition between the ordinary and extraordinary boundary phases becomes first-order-like on the surface.
Moreover, we also derive the OPE coefficient $\tilde{\lambda}_0^{(1)}$, which fixes the input for the bootstrap solution; the results have been summarized above.

\section{Framework of analytic bootstrap for boundary CFT}
\label{sec:Framework of analytic bootstrap for boundary CFT}

In this section, we summarize the analytic bootstrap procedure for boundary CFTs, following Refs.~\cite{liendo2013bootstrap,bissi2019analytic,dey2021analytic}, and highlight certain subtleties together with the strategies used to resolve them.

Consider a BCFT defined on the half-infinite space $\mathbb{R}^d_+=\{x=(r,z)|r\in\mathbb{R}^{d-1},z>0\}$, with a codimension-one boundary at $z=0$.
%For any operator $O$ in an infinite CFT, the one-point function vanishes, and the corresponding two-point function $\langle O(x)O(y)\rangle$ is determined by the separation $x-y$ and the scaling dimension $\Delta_O$.
%The four-point function is determined by a function of the cross-ratios, which depend on the coordinates of four points.
%In the presence of a boundary, the one-point function of a primary field $O(x)$ becomes nontrivial. % because of its image at $\tilde{x}=(r,-z)$. 
In a CFT without boundaries, the four-point correlation function depends on conformally invariant cross-ratios. 
In contrast, in a BCFT, the two-point function plays an analogous role and depends on a single cross-ratio, typically denoted by $\xi$,
\begin{equation}
\label{eq:cross_ratio}
    \xi=\frac{(r-r')^2+(z-z')^2}{4zz'}\,.
\end{equation}
The two-point correlation of a primary field $O$ with scaling dimension $\Delta_O$ can be expressed as 
\begin{equation}
\langle O(x)O(y)\rangle=\frac{F(\xi)}{|x-y|^{2\Delta_O}} \,,
\end{equation}
where $F(\xi)$ is fixed by conformal symmetry, as detailed below.

There are two equivalent ways to express the two-point function in a BCFT via operator expansions.
One way is to use the bulk operator product expansion (OPE).
The bulk OPE for the primary field $O$ gives
\begin{equation}
\label{eq:bulk_ope}
    O(x) O(x')=\frac{1}{|x-x'|^{2\Delta_O}}+\sum_n \lambda_n C[x-x',\partial_{x'}]O_n(x')\,,
\end{equation}
where the sum runs over primary operators $O_n$ appearing in the OPE, $\lambda_n$ are the corresponding OPE coefficients, and $C[x-x',\partial_{x'}]$ is a differential operator fixed by conformal symmetry~\cite{mcavity1995conformal,liendo2013bootstrap}.
Once the two-point function is written in terms of the OPE, we evaluate the one-point function of each bulk operator $O_n(x)$ in the presence of the boundary, namely,
\begin{equation}
\label{eq:expectation_value_bulk_O}
    \langle O_n(x)\rangle=\frac{a_n}{(2z)^{\Delta_n}} \,,
\end{equation}
where $a_n$ depends on the boundary condition or boundary interaction as well as the normalization of $O_n$. 
%In Eq.~\eqref{eq:bulk_ope}, the normalization of $\phi$ is fixed by the first term on the right-hand side.
Substituting Eq.~\eqref{eq:expectation_value_bulk_O} into Eq.~\eqref{eq:bulk_ope} yields
\begin{equation}
\label{eq:bulk_channel_expansion}
    F(\xi)=1+\sum_n \lambda_n a_n f_b(\Delta_n,\xi) \,,
\end{equation}
where $f_b(\Delta_n,\xi)$ is the bulk conformal block~\cite{mcavity1995conformal,liendo2013bootstrap}
\begin{equation}
\label{eq:bulk_conformal_block}
    f_b(\Delta_n,\xi)=\xi^{\Delta_n/2} \mathbin{_2 F_1}\left(\frac{\Delta_n}{2},\frac{\Delta_n}{2},\Delta_n+1-\frac{d}{2},-\xi\right)\,.
\end{equation}
Note that although the normalization of $O_n$ may vary, the combined coefficient $\lambda_n a_{n}$ in Eq.~\eqref{eq:bulk_channel_expansion} is invariant.

The other way to express the two-point function is via the boundary operator expansion (BOE), i.e.,
\begin{equation}
\label{eq:boundary_boe}
    O(x)=\frac{a_O}{(2z)^{\Delta_O}}+\sum_m \mu_m D[z,\partial_r]\hat{O}_m(r)\,,
\end{equation}
where $\hat{O}_m$ denotes the boundary primary operators that appear in the BOE, $\mu_m$ are the corresponding BOE coefficients, and the differential operator $D[z,\partial_r]$ is fixed by the bulk and boundary conformal symmetries.
Taking into account of the two-point function of the boundary primary operator
\begin{equation}
\label{eq:boudnary_2pt_fucntion}
    \langle \hat{O}_m(r)\hat{O}_m(r')\rangle=\frac{1}{|r-r'|^{2\hat{\Delta}_{m}}} \,,
\end{equation}
we obtain an alternative expression for $F(\xi)$:
\begin{equation}
\label{eq:bulk_2pt_boundary_channel}
    F(\xi)=\xi^{\Delta_O}\left(a_O^2+\sum_m \mu_m^2 f_i(\hat{\Delta}_m,\xi)\right) \,,
\end{equation}
where $f_i(\hat{\Delta}_m,\xi)$ denotes the boundary conformal block~\cite{mcavity1995conformal,liendo2013bootstrap}
\begin{equation}
\label{eq:boundary_conformal_block}
    f_i(\hat{\Delta}_m,\xi)=\frac{1}{\xi^{\hat{\Delta}_m}}\mathbin{_2 F_1}\left(\hat{\Delta}_m,\hat{\Delta}_m+1-\frac{d}{2},2\hat{\Delta}_m+2-d,-\frac{1}{\xi}\right) \,.
\end{equation}

%{\color{red} For an extraordinary boundary condition, is the only difference that $a_\phi\neq 0$?
%Remark: the boundary interaction cannot break the $O(N)$ symmetry; otherwise, the analytic bootstrap will not be valid.}

% \begin{figure}
%     \centering
%     \begin{subfigure}{0.8\textwidth}{\includegraphics[width=\linewidth]{bootstrap_eq.pdf}}
%     \end{subfigure}
%     \caption{The branch cut of (a) bulk conformal blocks $f_b(\Delta_n,\xi)$ and (b) boundary conformal blocks $f_i(\hat{\Delta}_m,\xi)$.
%     For the bulk conformal blocks, the branch cut is $\xi<-1$, and for the boundary conformal blocks, the branch cut is $-1<\xi<0$.}
%     \label{fig:bootstrap_eq}
% \end{figure}

Now, by equating Eq.~\eqref{eq:bulk_channel_expansion} and Eq.~\eqref{eq:bulk_2pt_boundary_channel}, we arrive at the crossing symmetry equation,
\begin{equation}
\label{eq:equivalence_bulk_boundary_channel}
    1+\sum_n \lambda_n a_n f_b(\Delta_n,\xi)=\xi^{\Delta_O}\left(a_O^2+\sum_m \mu_m^2 f_i(\hat{\Delta}_m,\xi)\right)\,.
\end{equation}
which is shown in Fig.~\ref{fig:bootstrap_eq} pictorially. 
This relation holds for a general primary field $O$; the indices $n$ and $m$ label the primary operators appearing in the bulk and boundary expansions, with scaling dimensions $\Delta_n$ and $\hat{\Delta}_m$, respectively.
To formulate the analytic bootstrap from the crossing symmetry equation, we consider an $\epsilon=d_0-d$ expansion~\footnote{We consider the dimension $d_0$ to be either $d_0 = 3$ or $d_0=4$ in this paper}, in which all parameters $\lambda_n,a_n,\mu_m,\Delta_n,\hat{\Delta}_m$ are expanded in powers of $\epsilon$.
For notational convenience, we define $\tilde{\lambda}_n=\lambda_n a_n$, $\tilde{\mu}_m=\mu_m^2$, and $\mu_I=a_O^2$.
While the OPE coefficients $\lambda_n$ are independent of boundary conditions or boundary interactions, $\tilde{\lambda}_n$ depend on them through the factor $a_n$. 
In what follows, we refer to $\tilde{\lambda}_n$ as OPE coefficients for brevity, although strictly speaking they are not the boundary-condition independent OPE coefficients.

The analytic bootstrap is based on the analytic properties of the bulk and boundary conformal blocks.
The key ingredient is that the hypergeometric function $\mathbin{_2 F_1}(a,b,c,z)$ has a branch cut for $z>1$.
Therefore, for integer $\hat{\Delta}_m$, the boundary conformal block in Eq.~\eqref{eq:boundary_conformal_block} has a branch cut along $\xi\in[-1,0]$.
And for $\Delta_n/2\in\mathbb{Z}$, the bulk conformal block in Eq.~\eqref{eq:bulk_conformal_block} has a branch cut along $\xi\in(-\infty,-1]$
\footnote{Note that for noninteger $\Delta_n/2$ and $\hat{\Delta}_m$, both bulk and boundary conformal blocks have a branch cut for $\xi<0$ due to the prefactors multiplying the hypergeometric function in Eqs.~\eqref{eq:bulk_conformal_block} and~\eqref{eq:boundary_conformal_block}.}.
It will be clear in the following how one can take advantage of the discontinuity, but before that, we first set up the discussion. 

We consider the $O(N)$ model and start the $\epsilon$-expansion from the Gaussian fixed point with $\Delta_\phi=\frac{d-2}{2}$. 
In general, we can consider general composite operators $O^{(k)}$, where $O^{(2m)}\sim(\sum_i \phi_i^2)^m$ and $O^{(2m+1)}\sim(\sum_i \phi_i^2)^m\phi_j$. 
In Sec.~\ref{sec:General operator expansion}, we discuss the operator expansion for these operators $O^{(k)}$, which shows that the zeroth order expansion has already contained infinitely many conformal blocks in both the bulk and boundary channels. 
One exception is when $k=1$. 
Also, since the two-point correlator $\langle \phi_i(x) \phi_i(y) \rangle$ plays the central role in the analysis, we focus primarily on it in what follows.
We will neglect the subindex for notational simplicity.

For $O=\phi$, the bulk and boundary primary operators that appear in the operator expansions are $O_n \sim\phi\partial^{2n}\phi$ and $\hat O_m \sim \partial^m \hat\phi$, with scaling dimensions $\Delta_n=d-2+2n$ for $n=0,1,\cdots$ and $\hat{\Delta}_m=\frac{d-2}{2}+m$ for $m=0,1,\cdots$, respectively~\footnote{Note that for $d=4$ with a bulk-$\phi^4$ interaction, the equation of motion implies $\sum_i\phi_i\partial^2\phi_i\sim\phi^4$.}.
For a general operator $O^{(k)}$ with $k>1$, the spectrum of operators appearing in the expansions differs; a detailed discussion of the possible primary operators is provided later. 
In this paper, we restrict to boundary interactions that preserve the $S_{N+1}$ or $O(N)$ symmetry, which enforces $a_\phi=0$. 
%More generally, the model may exhibit an enhanced symmetry, e.g., $O(N)$.

%The Gaussian fixed point corresponds to a free theory.
For a Gaussian theory in any dimension $d$, there are two types of boundary conditions, Neumann and Dirichlet, with the corresponding Green's function
\begin{equation}
\label{eq:0th_Greens_function}
    G_0(x,x')=%\langle\phi(x)\phi(x')\rangle=
    \frac{1}{|x-x'|^{2\Delta_\phi}}+\frac{s}{|x-\tilde{x}'|^{2\Delta_\phi}} \,,
\end{equation}
where $s=\pm$ for Neumann and Dirichlet boundary conditions, respectively, and $\tilde{x}'=(r',-z')$ is the image of $x'$.
This two-point function yields the zeroth-order solution of Eq.~\eqref{eq:equivalence_bulk_boundary_channel}
\begin{equation}
\label{eq:0th_solution_general_bootstrap}
\begin{split}
    \Delta_\phi^{(0)}&=\frac{d-2}{2}\,, \qquad\qquad\qquad\Delta_{\phi^2}^{(0)}\equiv\Delta_0^{(0)}=d-2\,, \\
    \hat{\Delta}_{\hat{\phi}}^{(0)N}&\equiv\hat{\Delta}_0^{(0)N}=\frac{d-2}{2}\,,\quad\ \hat{\Delta}_{\partial\hat{\phi}}^{(0)D}\equiv\hat{\Delta}_1^{(0)D}=\frac{d}{2},\\
    \tilde{\lambda}_{\phi^2}^{(0)N}&\equiv\tilde{\lambda}_0^{(0)N}=1\,,\qquad\qquad\tilde{\lambda}_{\phi^2}^{(0)D}\equiv\tilde{\lambda}_0^{(0)D}=-1,\\
    \tilde{\mu}_{{\hat{\phi}}}^{(0)N}&=\tilde{\mu}_0^{(0)N}=2\,, \qquad\qquad\tilde{\mu}_{{\partial\hat{\phi}}}^{(0)D}\equiv\tilde{\mu}_1^{(0)D}=\frac{d-2}{2}.
\end{split}
\end{equation}
All other parameters not listed above vanish. 
The results are exact for any dimension $d$ for the free field theory, and they serve as the starting point for the $\mathcal O(\epsilon)$ expansion for interacting theories. 
To distinguish the correction at different orders, we use $\Delta_n^{(l)}$ and $\hat \Delta_m^{(l)}$ to denote the $\mathcal O(\epsilon^l)$ corrections to the bulk and boundary scaling dimensions, respectively. 
Similarly, we also use $\tilde \lambda_n^{(l)}$ and $\tilde \mu_m^{(l)}$ to denote the $\mathcal O(\epsilon^l)$ corrections to the OPE and BOE coefficients, respectively.

There is a subtlety in evaluating the boundary conformal block under the Neumann boundary condition. 
If we set $\hat{\Delta}_{\hat \phi}^{(0)N}=\frac{d-2}{2}$ in the Neumann boundary condition, the hypergeometric function in the boundary conformal block becomes $\mathbin{_2 F_1}(\frac{d-2}{2},0,0,-\xi^{-1})\equiv1$, which does not reproduce the correct zeroth-order two-point function. 
This is because $b=c=0$ is a singular point for the hypergeometric function, $\mathbin{_2 F_1}(a,b,c,z)$, and the zeroth-order value depends on the way we approach it. 
Hence, to obtain the correct result, we assume a nonzero subleading correction $\hat{\Delta}_{\hat{\phi}}^{N}=\frac{d-2}{2}+\hat{\Delta}_{\hat{\phi}}^{(1)N}\epsilon$, %which gives $\mathbin{_2 F_1}(\frac{d-2}{2}+\hat{\Delta}_{\hat{\phi}}^{(1)N}\epsilon,\hat{\Delta}_{\hat{\phi}}^{(1)N}\epsilon,2\hat{\Delta}_{\hat{\phi}}^{(1)N}\epsilon,-\xi^{-1})$.
and then 
expand the boundary conformal block $f_i(\hat{\Delta}_{\hat{\phi}}^{N},\xi)$ to get 
\begin{equation}
\label{eq:expansion_0th_Neumann}
    f_i(\hat{\Delta}_{\hat{\phi}}^{N},\xi)=\frac{1}{2}\left(\frac{1}{\xi^{\frac{d-2}{2}}}+\frac{1}{(1+\xi)^{\frac{d-2}{2}}}\right)+\mathcal{O}(\epsilon) \,.
\end{equation}
Of course, we will later determine the higher-order corrections to the scaling dimension and the BOE coefficients, where the higher order corrections will be present. 
However, due to the above subtlety, we will treat the BOE coefficient of $\hat{\phi}$ on the boundary separately, which will be clear in the following.
%Accordingly, one should carefully distinguish between these two expressions for $\mathbin{_2 F_1}(\frac{d-2}{2},0,0,-\xi^{-1})$.

With the zeroth-order solution in hand, we can perform a systematic order-by-order $\epsilon$-expansion for the crossing symmetry equation. 
Suppose $F(\xi)$ has been determined up to $\mathcal{O}(\epsilon^{\,l-1})$ with finitely many terms, denoted by $n_{\rm max}$ and $m_{\rm max}$ for the bulk and boundary channels, respectively. 
The next order follows from the procedure below.
First, split the bulk and boundary channels into two parts, $F(\xi)=G_b+H_b$ and $F(\xi)=G_i+H_i$. The sub-index $b$ and $i$ denotes the bulk and boundary channels, respectively. 
In the bulk channel, we define
\begin{equation}
\label{eq:lst_correction_bulk_channel_general}
\begin{split}
    G_b(\xi)=&\ 1+\tilde{\lambda}_0 f_b(\Delta_0,\xi)+\sum_{n=1}^{n_{\rm max}}\tilde{\lambda}_n^{(<l)}f_b(\Delta_n,\xi)\,,\\
    H_b(\xi)=&\ \epsilon^l\sum_{n\geq1}\tilde{\lambda}_n^{(l)}f_b(\Delta_n^{(0)},\xi)\,.
\end{split}
\end{equation}
%where $\lambda_I=1$, $\tilde{\lambda}_{0}=\tilde{\lambda}_{\phi^2}$, and $\Delta_{0}=\Delta_{\phi^2}$.
In $G_b(\xi)$ the scaling dimensions $\Delta_n$ are kept accurate to $\mathcal{O}(\epsilon^{l})$, while the OPE coefficients $\tilde{\lambda}_n^{(<l)}$ are known to $\mathcal{O}(\epsilon^{\,l-1})$.
By contrast, $H_b(\xi)$ uses the bare dimensions $\Delta_n^{(0)}=d-2+2n$, and the coefficients $\tilde{\lambda}_n^{(l)}$ are of order $\mathcal{O}(\epsilon^{l})$.
This separation exploits the analytic structure of the hypergeometric functions appearing in the conformal blocks.
Note that we do not split the conformal block from the operator $\phi^2$ into $H_b(\xi)$; it is included entirely in $G_b(\xi)$ because the hypergeometric function in $f_b(\Delta_{0}^{(0)},\xi)$ has no branch cut for $\xi<-1$. 
Similarly, for the boundary channel, we define
\begin{eqnarray}
\label{eq:lst_correction_boundary_channel_general}
    G_i(\xi)&=&\ \xi^{\Delta_\phi}\left(\tilde{\mu}_0 f_i(\hat{\Delta}_0,\xi)+\sum_{m=1}^{m_{\rm max}}\tilde{\mu}_m^{(<l)}f_i(\hat{\Delta}_m,\xi)\right)\,, \nonumber\\
    H_i(\xi)&=&\ \epsilon^l \xi^{\Delta_\phi^{(0)}}\sum_{m\geq1}\tilde{\mu}_m^{(l)}f_i(\hat{\Delta}_m^{(0)},\xi) \,.
\end{eqnarray}
%where $\mu_I=0$, $\tilde{\mu}_{0}=\tilde{\mu}_{\hat{\phi}}$, $\Delta_{0}=\Delta_{\hat{\phi}}$.
Here $\hat{\Delta}_m^{(0)}=\frac{d-2}{2}+m$ take the bare values, and $\hat{\Delta}_m$ are accurate up to $\mathcal{O}(\epsilon^l)$.
The quantities $\tilde{\mu}_m^{(<l)}$ and $\tilde{\mu}_m^{(l)}$ denote BOE coefficients known up to $\mathcal{O}(\epsilon^{l-1})$ and of order $\mathcal{O}(\epsilon^{l})$, respectively.
We include the entire conformal block $f_i(\hat{\Delta}_{0},\xi)$ in $G_i(\xi)$ because of the subtlety described before. 

\begin{figure}
    \centering
    \subfigure[]{\includegraphics[width=0.22\textwidth]{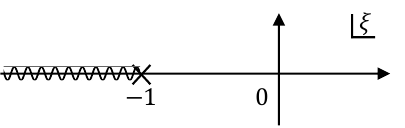}}\ \
    \subfigure[]{\includegraphics[width=0.22\textwidth]{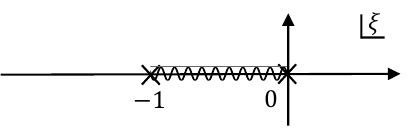}}
    \caption{The branch cut of (a) $H_b(\xi)$, which includes bulk conformal blocks $f_b(\Delta_n^{(0)},\xi)$ for $\Delta_n^{(0)}=d-2+2n$, and (b) $H_i(\xi)$ which includes boundary conformal blocks $f_i(\hat{\Delta}_m^{(0)},\xi)$ and $\xi^{\Delta_\phi^{(0)}}$ for $\hat{\Delta}_m^{(0)} = \frac{d-2}2 +m$.}
    % For the bulk conformal blocks, the branch cut is $\xi<-1$, and for the boundary conformal blocks, the branch cut is $-1<\xi<0$.}
    \label{fig:branchcut}
\end{figure}

\begin{figure*}[t]
\centering
{\setlength{\fboxsep}{6pt}% padding between box and content
 \setlength{\fboxrule}{0.6pt}% border line width
 \fbox{%
\begin{tikzpicture}[
  >=Stealth,
  lab/.style={font=\footnotesize, inner sep=1pt}
]
% Nodes (the first and fourth get a fixed text width to avoid eating arrow length)
\node[align=center, text width=.22\textwidth, inner sep=2pt] (disc)
  {$\underset{\xi<-1}{\mathrm{Disc}}\,H_b
   =\underset{\xi<-1}{\mathrm{Disc}}\,(G_i - G_b)$};

\node[right=16mm of disc] (lam) {$\tilde{\lambda}_n^{(l)}$};
\node[right=23mm of lam]  (Hb)  {$H_b$};
\node[right=27mm of Hb, align=center, inner sep=2pt] (Hi)
  {$H_i = G_b + H_b - G_i$};
\node[right=16mm of Hi]   (mu)  {$\tilde{\mu}_m^{(l)}$};

% Arrows (shorten a tiny bit so tips don’t touch boxes)
\draw[->, shorten >=2pt, shorten <=2pt]
  (disc.east) -- node[lab, above]{Eq.~\eqref{eq:OPE_lambda_n}} (lam.west);
\draw[->, shorten >=2pt, shorten <=2pt]
  (lam.east) -- node[lab, above]{Resummation} (Hb.west);
\draw[->, shorten >=2pt, shorten <=2pt]
  (Hb.east) -- node[lab, above]{Crossing eq.~\eqref{eq:crossing_GH}} (Hi.west);
\draw[->, shorten >=2pt, shorten <=2pt]
  (Hi.east) -- node[lab, above]{Eq.~\eqref{eq:BOE_mu_m}} (mu.west);
\end{tikzpicture}%
}}
\caption{The procedure of solving bootstrap equations and OPE/BOE coefficients.}
\label{fig:bootstrap_pipeline}
\end{figure*}

With the separation outlined above, the crossing equation can be brought into 
\begin{equation} \label{eq:crossing_GH}
    G_i-G_b=H_b-H_i\,.
\end{equation} 
The finite number of terms on the left-hand side then constrains the infinite series on the right-hand side involving the OPE and BOE coefficients.
More importantly, we can simplify the right-hand side by exploiting the analytic properties of the conformal blocks.
It is straightforward to check that the function $H_b(\xi)$ in Eq.~\eqref{eq:lst_correction_bulk_channel_general} has a branch cut for $\xi \in (-\infty, -1]$.
By contrast, the function $H_i(\xi)$ in Eq.~\eqref{eq:lst_correction_boundary_channel_general} has a branch cut for $\xi \in [-1,0]$.
Note that the factor $\xi^{\Delta_\phi^{(0)}}$ in $H_i(\xi)$ would introduce a branch cut for $\xi<0$ if $\Delta_\phi^{(0)} = \frac{d-2}{2}$ is noninteger.
However, the boundary conformal block $f_i(\hat{\Delta}_m,\xi)$ in Eq.~\eqref{eq:boundary_conformal_block} contains an additional factor $\xi^{-\hat{\Delta}_m}$.
With $\hat{\Delta}_m^{(0)}=\frac{d-2}{2}+m$, the noninteger part of $\Delta_\phi^{(0)}$ is canceled.
Thus, $H_i(\xi)$ has a branch cut only for $\xi\in[-1,0]$.
The branch cut of bulk and boundary conformal blocks are shown in Fig.~\ref{fig:branchcut}.

To this end, we define the discontinuity
\begin{equation}
\label{eq:disc_definition}
    \underset{\xi}{\rm Disc}\ f(\xi)=f(\xi+{\rm i}0^+)-f(\xi-{\rm i}0^+)\,.
\end{equation}
Because the branch cuts of $H_b(\xi)$ and $H_i(\xi)$ are disjoint, we can take the discontinuity for $\xi<-1$ to obtain
\begin{equation}
\label{eq:Disc_relation_for_deltaG_deltaH}
    \underset{\xi<-1}{\rm Disc}\ (G_i(\xi)-G_b(\xi))%=\underset{\xi<-1}{\rm Disc}\ (H_b(\xi)-H_i(\xi))
    =\underset{\xi<-1}{\rm Disc}\ H_b(\xi)\,,
\end{equation}
where $H_i(\xi)$ drops out because it has no discontinuity for $\xi<-1$.
The left-hand side of Eq.~\eqref{eq:Disc_relation_for_deltaG_deltaH} contains finitely many terms and thus has a simple discontinuity.
On the right-hand side, we need the discontinuity of the bulk conformal blocks:
\begin{eqnarray}
\label{eq:discontinuity_fb}
    &&\underset{\xi<-1}{\rm Disc}\ f_b(d-2+2n,\xi)\\&=&-2\pi {\rm i}(-1)^n\frac{\Gamma(2n+\frac{d}{2}-1)}{\Gamma(n)\Gamma(n-1+\frac{d}{2})}P^{(\frac{d}{2}-1,0)}_{n-1}\left(-\frac{\xi+2}{\xi}\right) \,, \nonumber
\end{eqnarray}
where $P_n^{(\alpha,\beta)}$ denotes the Jacobi polynomial.
We have used the discontinuity of the hypergeometric function, Eq.~\eqref{eq:discontinuity_2F1}, given in Appendix~\ref{sec:Branch cut of special functions}.
Therefore, using the orthogonality relation of Jacobi polynomials~\cite{NIST:DLMF},
\begin{eqnarray}
\label{eq:orthonormal_Jacobi_polynomials}
\begin{split}
    &\int_{-1}^{1}{\rm d}y (1-y)^\alpha(1+y)^\beta P^{(\alpha,\beta)}_{m}(y)P^{(\alpha,\beta)}_{n}(y)\\=&\delta_{m,n}\frac{2^{\alpha+\beta+1}\Gamma(n+\alpha+1)\Gamma(n+\beta+1)}{(2n+\alpha+\beta+1)\Gamma(n+\alpha+\beta+1)\Gamma(n+1)}\,,
\end{split}
\end{eqnarray}
the OPE coefficients $\tilde{\lambda}_n$ can be written as
\begin{eqnarray}
\label{eq:OPE_lambda_n}
    \tilde{\lambda}_n^{(l)}&=&-\frac{\Gamma(n)\Gamma(n+\frac{d}{2}-1)}{{\rm i}\pi \epsilon^l 2^{\frac{d}{2}+1}(-1)^n \Gamma(2n+\frac{d}{2}-2)}\int_{-1}^{1}{\rm d}y (1-y)^{\frac{d}{2}-1} \nonumber\\
    &&\times P_{n-1}^{(\frac{d}{2}-1,0)}(y) \underset{\xi<-1}{\rm Disc}\ (G_i(\xi)-G_b(\xi))|_{\xi=-\frac{2}{y+1}} \,.
\end{eqnarray}
After obtaining the OPE coefficients, we must resum the bulk conformal blocks in $H_b$ and verify the resulting discontinuity, since summing an infinite series may not commute with taking discontinuities.

Once the OPE coefficients are determined, we can likewise obtain the BOE coefficients.
We first resum all bulk conformal blocks weighted by the OPE coefficients, Eq.~\eqref{eq:OPE_lambda_n}, to obtain a closed-form expression for $H_b(\xi)$ in Eq.~\eqref{eq:lst_correction_bulk_channel_general}.
The procedure is as follows.
First, express the hypergeometric function %$\mathbin{_2 F_1}(a,b,c,z)$ in Eq.~\eqref{eq:lst_correction_bulk_channel_general} using 
by the integral representation given in Eq.~\eqref{eq:integral_rep_2F1} in Appendix~\ref{sec:Expansion of conformal blocks}.
Next, perform the sum over $n\geq1$ inside this integral. %using the coefficients $\tilde{\lambda}_n^{(l)}$.
Finally, evaluate the remaining $t$-integral by expanding the integrand as a power series in $\xi$, integrating term by term, and then resumming the resulting series.

After obtaining the closed-form expression for $H_b$ by resummation, we have $H_i=G_b+H_b-G_i$.
The BOE coefficients can then be derived using the orthonormal condition for hypergeometric functions~\cite{heemskerk2009holography}
\begin{eqnarray}
\label{eq:orthonormal_2F1}
    \oint_{w}\frac{{\rm d}w}{2\pi{\rm i}}&&w^{n-m-1}\mathbin{_2 F_1}\left(1-m,-m-\frac{d-4}{2},2-2m,-w\right) \nonumber\\
    &&\times\mathbin{_2 F_1}\left(n,n+\frac{d-2}{2},2n,-w\right)=\delta_{m,n}\,,  
\end{eqnarray}
which yields
\begin{eqnarray}
\label{eq:BOE_mu_m}
    \tilde{\mu}_m^{(l)}=\underset{w=0}{\rm Res}&&\left[\mathbin{_2 F_1}\left(1-m,-m-\frac{d-4}{2},2-2m,-w\right)\right. \nonumber \\
    &&\qquad\qquad\times\left.w^{-m-1}\frac{1}{\epsilon^l}H_i(\xi)|_{\xi=\frac{1}{w}}\right]\,.
\end{eqnarray} 
We summarize the procedure of solving the bootstrap equation $G_i - G_b = H_b - H_i$ in Fig.~\ref{fig:bootstrap_pipeline}.
% \begin{widetext}
% \begin{equation}
% \label{eq:solving bootstrap equations}
%     \boxed{\underset{\xi<-1}{\rm Disc}\ H_b = \underset{\xi < -1}{\rm Disc}\ (G_i-G_b)   \xrightarrow{{\rm Eq.~\eqref{eq:OPE_lambda_n}}} \\ \tilde{\lambda}_n^{(l)}  \xrightarrow{{\rm resummation}} H_b  \xrightarrow{\rm crossing~eq.~\eqref{eq:crossing_GH}} H_i=G_b+H_b-G_i  \xrightarrow{{\rm Eq.~\eqref{eq:BOE_mu_m}}} \tilde{\mu}_m^{(l)}}
% \end{equation}
% \end{widetext} 

With this procedure, we expand $F(\xi)$ as a series in $\epsilon$, order by order, and stop once infinitely many conformal blocks appear in $F(\xi)$, since higher-order terms can mix operators with degenerate primary fields~\cite{osborn1991weyl,prochazka2020composite}.
A simple example is the zeroth-order result for the $\langle\phi^k \phi^k\rangle$ correlator discussed in Sec.~\ref{sec:General operator expansion}, where degenerate primaries occur when $d$ is taken to be an integer.
Note that, this procedure cannot solve the problem order-by-order within itself. 
Loosely speaking, what the analytic bootstrap outlined above achieves is to determine the OPE and BOE exponents $\tilde\lambda_{n>0}^{(l)}$, $\tilde \mu_m^{(l)}$ at the $l$-th order, with the knowledge of the OPE and BOE exponents at the ($l-1$)-th order as well as the scaling dimension. 
Recall that in the definition of $G$ and $H$, we have included the conformal blocks for $\tilde \lambda_0$ and and $\tilde \mu_0$ in $G$. 
Hence, the analytic bootstrap will not generate $\tilde \lambda_0$ to the next order.
On the other hand, we will use Eq.~\eqref{eq:orthonormal_2F1} for $m = 0$ as a constraint. 
%, we do not use it to compute the BOE coefficient $\tilde \mu_0^{(l)}$, because the $n=0$ case in Eq.~\eqref{eq:orthonormal_2F1} involves $\mathbin{_2 F_1}(0,\frac{d-2}{2},0,z)=1$, which is not the conformal block used here.
%Instead, 
Namely, we treat $\tilde{\mu}_{0}^{(l)}$ as an input in $G_i$ and then compute the BOE data in $H_i$ using Eq.~\eqref{eq:BOE_mu_m}.
For $\tilde{\mu}_m^{(l)}$ with $m=0$ obtained from Eq.~\eqref{eq:BOE_mu_m}, we require it to vanish, since $H_i$ contains no conformal block with $\hat{\Delta}_{0}^{(0)}=\frac{d-2}{2}$.
This condition provides an additional constraint that determines $\tilde{\mu}_0^{(l)}$ in terms of the other inputs, in contrast to $\tilde \lambda_0$. 

Finally, we briefly discuss the image symmetry, which is present in the $\phi^4$-theory under either the Neumann or Dirichlet boundary condition at $d=4-\epsilon$. 
The image symmetry is associated with a transformation that takes one of the coordinates to its image, i.e., it takes $(r,z)$ to $(r,-z)$.
Under this transformation the cross ratio changes as $\xi \rightarrow -(1+\xi)$. 
Then the image symmetry follows from the property of boundary conformal blocks,
\begin{equation}
\label{eq:image_symm_boundary_block}
    f_i(\hat{\Delta}_m,e^{\pm{\rm i}\pi}(\xi+1))=e^{\mp{\rm i}\pi\hat{\Delta}_m}f_i(\hat{\Delta}_m,\xi)\,.
\end{equation}
Therefore, $\hat{\Delta}_m^{(0)}=1+m\in\mathbb{Z}$ in $d=4-\epsilon$ implies that if only boundary conformal blocks with odd (even) $m$ are present, then the $H_i$ satisfies the following symmetry,
\begin{equation}
\label{eq:image_symm_Hi_phi4}
    \left.\frac{H_i(\xi')}{\xi'}\right|_{\xi'=e^{\pm{\rm i}\pi}(\xi+1)}=e^{\mp{\rm i}\pi}(-1)^m \frac{H_i(\xi)}{\xi}\,.
\end{equation}

For the $\phi^4$-theory in $d=4-\epsilon$ under Neumann and Dirichlet boundary conditions, only even and odd $m$ appear in the boundary conformal blocks, respectively. 
%This is consistent with the zeroth-order results, where the Neumann and Dirichlet boundary conditions lead to nonzero boundary primary fields $\hat{\phi}$ and $\partial\hat{\phi}$ with $m=0,1$. 
However, when the boundary interaction is present, the image symmetry is no longer preserved. 
For example, in the bulk-$\phi^4$/boundary-$\phi^3$ theory at $d=4-\epsilon$, the BOE contains both even and odd $m$ at order $\mathcal{O}(\epsilon)$. 
This follows from the boundary equation of motion: the boundary interaction induces a mixed boundary condition where both $\hat{\phi}$ and $\partial\hat{\phi}$ appear in the BOE.

%A simple example is the bulk-free/boundary-$\phi^4$ theory at $d=3-\epsilon$, where the $\mathcal{O}(\epsilon)$ correction to the two-point function $\langle\phi(x)\phi(y)\rangle$ explicitly breaks image symmetry.
The situation is subtle in odd dimensions. 
With $\hat{\Delta}_m=\frac{d-2}{2}+m$ in Eq.~\eqref{eq:image_symm_boundary_block}, if $H_i(\xi)$ contains only even or only odd $m$, then it has the following symmetry
\begin{equation}
\label{eq:image_symm_Hi_odd_d}
    \left.\frac{H_i(\xi')}{\xi'^{\frac{d-2}{2}}}\right|_{\xi'=e^{\pm{\rm i}\pi}(\xi+1)}=e^{\mp{\rm i}\pi(\frac{d-2}{2})}(-1)^m \frac{H_i(\xi)}{\xi^{\frac{d-2}{2}}}\,.
\end{equation}
However, in the $\phi^6$ theory in $d=3-\epsilon$ under either Neumann or Dirichlet boundary condition, we show below that $H_i$ contains both even and odd $m$ conformal blocks even without a boundary interaction.

%Therefore, in the analysis that follows we do not assume image symmetry; instead, we treat it as an additional constraint on the conformal data.

\section{General operator expansion}
\label{sec:General operator expansion}

%With the results in Sec.~\ref{sec:Framework of analytic bootstrap for boundary CFT}, 
%The crossing equation Eq.~\eqref{eq:equivalence_bulk_boundary_channel} is applicable to composite primary operators such as $O^{(k)}\sim\phi^k$, where the spectra $\Delta_n$ and $\hat{\Delta}_m$ are fixed by the choice of $O^{(k)}$. 

In this section, we consider the primary operator $O^{(k)}\sim\phi^k$ in the $O(N)$ theory, and evaluate their OPE and BOE at tree level. 
Concretely, owing to the $O(N)$ symmetry, we have $O^{(2m)}\sim(\sum_{i=1}^N \phi_i^2)^m$ and $O^{(2m+1)}_i\sim(\sum_{i=1}^N \phi_i^2)^m\phi_i$.
We first solve the zeroth‐order correlator $\langle O^{(k)}O^{(k)}\rangle$, then extract the corresponding spectra $\Delta_n$ and $\hat{\Delta}_m$ together with their OPE and BOE coefficients. 
We find that the expansion of $\langle O^{(k)}O^{(k)}\rangle$ already contain infinitely many conformal blocks.
The resulting expressions are valid for general space dimension $d$. 
This serves as the starting point for the analytic bootstrap.
In particular, it leads to Eq.~\eqref{eq:0th_solution_general_bootstrap} for $k=1$.

\subsection{Correlation function $\langle O^{(k)}O^{(k)}\rangle$}
\label{sec:Correlation function O(k)O(k)}

We derive the correlation function $\langle O^{(k)}O^{(k)}\rangle$ for the free theories with $O(N)$ symmetry.
We define the operator $O^{(k)}$ as
\begin{equation}
\label{eq:O_k_definition}
\begin{split}
    O^{(2m)}&=f_{2m,N}:\left(\sum_{i=1}^N \phi_i^2\right)^m: \,,\\ O^{(2m+1)}_i&=f_{2m+1,N}:\left(\sum_{i=1}^N \phi_i^2\right)^m\phi_i: \,, 
\end{split}
\end{equation}
where $f_{k,N}$ denotes the normalization factor to be chosen to satisfy
\begin{equation}
\label{eq:O_k_definition_bulk_correlation}
    \langle O^{(k)}(x)O^{(k)}(y)\rangle_b=\frac{1}{|x-y|^{2\Delta_{O^{(k)}}}}\,,
\end{equation}
and $\langle\cdot\rangle_b$ denotes the expectation value in the bulk without a boundary. 
In the Appendix~\ref{append:normalization}, the normalization coefficients are obtained as
\begin{eqnarray}
    f_{2m,N}&=&\left(\frac{\Gamma(m+N/2)\Gamma(m+1)}{\Gamma(N/2)} \right)^{-1/2}2^{-m}\,, \\ f_{2m+1,N}&=&\left(\frac{\Gamma(m+N/2+1)\Gamma(m+1)}{\Gamma(N/2+1)} \right)^{-1/2}2^{-m}\,. \nonumber
\end{eqnarray}

In the following, we write $\phi^2=\sum_{i=1}^N \phi_i^2$ for brevity. 
We now compute the correlation function with boundary conditions and denote $\langle\cdot\rangle_s$ with $s=\pm$ for the Neumann and Dirichlet boundary conditions.
Applying the Wick theorem for free theories, the correlation function for $O^{(k)}$ reads,
\begin{equation}
\label{eq:correlation_with_boundary}
\begin{split}
    \langle O^{(k)}(x)O^{(k)}(y)\rangle_s=&\left[\frac{1}{|x-y|^{2\Delta_\phi}}+s\frac{1}{|x-\tilde{y}|^{2\Delta_\phi}}\right]^{k}\\
    &+\langle O^{(k)}(x)\rangle_s\langle O^{(k)}(y)\rangle_s\,,
\end{split}
\end{equation}
where $\tilde{y}= (r_y,-z_y)$ with $y=(r_y,z_y)$.
The one-point function $\langle O^{(k)}(x)\rangle_s$ vanishes for odd $k$ due to the $O(N)$ symmetry.
For $k=2m$, consider the generating function $\langle:e^{t\phi^2}:\rangle_s$, which yields
\begin{eqnarray}
\label{eq:onepoint_O}
    &&\langle:e^{t\phi^2}:\rangle_s= \prod_{i=1}^N \langle:e^{t\phi^2_i}:\rangle_s=\left(\sum_{n=0}^\infty\frac{t^n}{n!}\langle:(\phi_i^2)^n :\rangle_s\right)^N \nonumber \\
    && =\left(\sum_{n=0}^\infty\frac{t^n}{n!}(2n-1)!!\langle:\phi_i^2:\rangle_s^n\right)^N =\left[1-2t\langle:\phi_i^2:\rangle_s\right]^{-N/2} \nonumber 
    \\
    && =\sum_{n=0}^\infty\frac{(\frac{N}{2})_n}{n!}\left(2t\langle:\phi_i^2:\rangle_s\right)^n\,.
\end{eqnarray}
We also have $\langle:e^{t\phi^2}:\rangle_s=\sum_{n=0}^\infty\frac{t^n}{n!}\langle:(\phi^2)^n:\rangle_s$.
Matching these two expressions, we obtain
\begin{equation}
\label{eq:onepoint_O_results}
\begin{split}
    \langle O^{(2m)}\rangle_s=&f_{2m,N}\left(\frac{N}{2}\right)_m (2s)^m \frac{1}{(2z)^{2m\Delta_\phi}}\\
    =&\sqrt{\frac{\Gamma(m+N/2)}{\Gamma(N/2)\Gamma(m+1)}}\frac{s^m}{(2z)^{2m\Delta_\phi}}\,.
\end{split}
\end{equation}

Therefore, writing $\langle O^{(k)}(x)O^{(k)}(y)\rangle_s=\frac{F_s^{(k)}(\xi)}{|x-y|^{2\Delta_{O^{(k)}}}}$, we obtain
\begin{equation}
\label{eq:F_phik}
    F_s^{(k)}(\xi)=\left(1+s\left(\frac{\xi}{\xi+1}\right)^{\Delta_\phi}\right)^k + \alpha_{k,N}\xi^{k\Delta_\phi}\,,
\end{equation} 
where $\alpha_{2m+1,N} = 0 $ and  $\alpha_{2m,N}=\frac{\Gamma(m+N/2)}{\Gamma(N/2)\Gamma(m+1)}$.

\subsection{Operator expansion for $O^{(k)}$}
\label{sec:Scaling dimension and operator expansion for general operators}

We expand $F_s^{(k)}(\xi)$ in Eq.~\eqref{eq:F_phik} in terms of conformal blocks in the bulk and boundary channels using Eq.~\eqref{eq:equivalence_bulk_boundary_channel}. 

We begin with the bulk channel.
For the second term in Eq.~\eqref{eq:F_phik}, we expand the bulk conformal block in Eq.~\eqref{eq:bulk_conformal_block} in powers of $\xi$. 
To obtain a noninteger power $\xi^a$, the scaling dimensions in the conformal blocks must take the values $\Delta_n=2a+2n$, with $n=0,1,\cdots$:
\begin{eqnarray}
\label{eq:bulk_channel_second_term_expansion}
    \xi^a&=&\sum_{n=0}^\infty c_{a,n} f_b(2a+2n,\xi) \\
    &=&\sum_{n=0}^\infty c_{a,n} \xi^{a+n} \sum_{m=0}^\infty\frac{(a+n)_m(a+n)_m}{(2a+2n+1-\frac{d}{2})_m m!}(-\xi)^m\,, \nonumber
\end{eqnarray}
where the second line uses the series expansion of the hypergeometric function.
The coefficients $c_{a,n}$ are then fixed by the linear equations
\begin{equation}
\label{eq:bulk_second_c_equations}
    \delta_{l,0}=\sum_{n=0}^l c_{a,n}\frac{(-1)^{l-n}(a+n)_{l-n}(a+n)_{l-n}}{(2a+2n+1-\frac{d}{2})_{l-n} (l-n)!}\,.
\end{equation}
Solving these linear equations yields
\begin{equation}
\label{eq:bulk_second_c_solution}
    c_{a,n}=\frac{\Gamma(a+n)^2\Gamma(2a+n-\frac{d}{2})}{\Gamma(a)^2\Gamma(n+1)\Gamma(2a+2n-\frac{d}{2})}\,.
\end{equation}

For the first term in Eq.~\eqref{eq:F_phik}, we can first consider the expansion for $\left(\frac{\xi}{1+\xi} \right)^a$, which serves as the building block. 
On the one hand, $\left(\frac{\xi}{1+\xi} \right)^a=\xi^a\sum_{l=0}^\infty\frac{(a)_l}{l!}(-1)^l\xi^l$.
On the other hand, the expansion in terms of the bulk conformal block reads $\left(\frac{\xi}{1+\xi} \right)^a=\sum_{n=0}^\infty c_{a,n}' f_b(2a+2n,\xi)$. 
%\begin{equation}
%\label{eq:bulk_channel_first_term_expansion}
%\begin{split}
%    &\left(\frac{\xi}{1+\xi} \right)^a=\sum_{n=0}^\infty c_{a,n}' f_b(2a+2n,\xi)\,.%\\
    %=&\sum_{n=0}^\infty c_{a,n}' \xi^{a+n} \sum_{m=0}^\infty\frac{(a+n)_m(a+n)_m}{(2a+2n+1-\frac{d}{2})_m m!}(-\xi)^m\,.
%\end{split}
%\end{equation}
%Hence, the associated linear equations are
%\begin{equation}
%\label{eq:bulk_first_c_equations}
%    \frac{(a)_l}{l!}(-1)^l=\sum_{n=0}^l c_{a,n}'\frac{(-1)^{l-n}(a+n)_{l-n}(a+n)_{l-n}}{(2a+2n+1-\frac{d}{2})_{l-n} (l-n)!}\,.
%\end{equation}
We can further expand the bulk conformal blocks as power series of $\xi$, and match the coefficients of these two expansion to give
\begin{equation}
\label{eq:bulk_first_c_solution}
    c_{a,n}'=\frac{(-1)^n\Gamma(a+n)\Gamma(a+n+1-\frac{d}{2})\Gamma(2a+n-\frac{d}{2})}{\Gamma(a)\Gamma(a+1-\frac{d}{2})\Gamma(n+1)\Gamma(2a+2n-\frac{d}{2})}.
\end{equation}

Now, with these two results, Eqs.~\eqref{eq:bulk_second_c_solution} and \eqref{eq:bulk_first_c_solution}, 
%As an illustration, consider $O=O^{(2)}$. 
%There are four terms in Eq.~\eqref{eq:F_phik}, where $1$ and $2s\bigl(\frac{\xi}{\xi+1}\bigr)^{\Delta_\phi}$ correspond to $1$ and $2s\, f_b(2\Delta_\phi,\xi)$ with $\tilde{\lambda}_{2\Delta_\phi}=2s$.
%The term $\bigl(\frac{\xi}{\xi+1}\bigr)^{2\Delta_\phi}$ yields $\tilde{\lambda}_{4\Delta_\phi+2n}=c_{2\Delta_\phi,n}'$.
%The final term gives $\tilde{\lambda}_{4\Delta_\phi+2n}=\alpha_{2,N}c_{2\Delta_\phi,n}$ with $\alpha_{2,N}=\frac{N}{2}$.
%This agrees with Ref.~\cite{liendo2013bootstrap}. 
%For general $k$, the last term of Eq.~\eqref{eq:F_phik} contributes $\tilde{\lambda}_{2k\Delta_\phi+2n}=\alpha_{k,N}c_{k\Delta_\phi,n}$. 
%The first part of Eq.~\eqref{eq:F_phik} is $\sum_{l=0}^k \frac{\Gamma(k+1)}{\Gamma(l+1)\Gamma(k-l+1)}s^l \left(\frac{\xi}{1+\xi}\right)^{l\Delta_\phi}$, in which each term contributes $\tilde{\lambda}_{2l\Delta_\phi+2n}=\frac{\Gamma(k+1)}{\Gamma(l+1)\Gamma(k-l+1)}s^l c_{l\Delta_\phi,n}'$.
it is straightforward to obtain the OPE coefficients for $\langle O^{(k)} O^{(k)} \rangle$ 
\begin{eqnarray}
\label{eq:OPE_phi_k}    \tilde{\lambda}_{2l\Delta_\phi+2n}&=&\frac{\Gamma(k+1)}{\Gamma(l+1)\Gamma(k-l+1)}s^l c_{l\Delta_\phi,n}' \nonumber \\
    && +\delta_{l,k} \alpha_{k,N}c_{k\Delta_\phi,n},
\end{eqnarray}
with $l=0,1,\cdots,k$ and $n=0,1,\cdots$. 
The expansion reveals $k$ distinct families of conformal blocks corresponding to the scaling dimension $\Delta=2l\Delta_\phi+2n$, where $l=1,\cdots,k$. 
Among each family, the lowest member is $:\phi^{2l}:$ and higher $n$ are the scalar primaries $:\phi^l \square^n\phi^l:$.

The expressions hold for any $d$.
In particular, for $\Delta_\phi=\frac{d-2}{2}$ with an integer $d$, if $d$ is even then all families collapse to a single series with $\Delta=d-2+2n$, and if $d$ is odd then two families remain, with $\Delta=d-2+2n$ and $\Delta=2d-4+2n$.

Next, we evaluate the BOE coefficients. 
The second term $\alpha_{k,N}\xi^{k\Delta_\phi}$ in the correlator gives $\Delta_O=k\Delta_\phi$ and $a_O^2=\alpha_{k,N}$ in Eq.~\eqref{eq:bulk_2pt_boundary_channel}. 
In what follows, we focus on the first part in Eq.~\eqref{eq:F_phik}.
%We expand the boundary conformal blocks Eq.~\eqref{eq:boundary_conformal_block} 
%To obtain the BOE, we expand it in powers of $\xi^{-1}$:
%Then we expand each term of the form 
Consider the expansion $(1+\xi^{-1})^{-a}=\sum_{l=0}^\infty \frac{(a)_l}{l!}(-1)^l(\xi^{-1})^l$ in terms of the boundary conformal block:
\begin{equation}
\label{eq:boudnary_first_expansion}
\begin{split}
    (1+\xi^{-1})^{-a}=\xi^{k\Delta_\phi}\sum_{n=0}^\infty d_{a,k,n} f_i(k\Delta_\phi+n,\xi)\,. %\\
    %=&\sum_{n=0}^\infty d_{a,k,n}\xi^{-n}\mathbin{_2 F_1}\left(k\Delta_\phi+n,k\Delta_\phi+n+1-\frac{d}{2},2k\Delta_\phi+2n+2-d,-\xi^{-1}\right)\\
    %=&\sum_{n=0}^\infty d_{a,k,n}\xi^{-n}\sum_{m=0}^{\infty}\frac{(k\Delta_\phi+n)_m(k\Delta_\phi+n+1-\frac{d}{2})_m}{(2k\Delta_\phi+2n+2-d)_m m!}(-\xi^{-1})^m \,,
\end{split}
\end{equation}
%and the associated linear equations for the expansion coefficients are
%\begin{equation}
%\label{eq:boundary_first_d_equations}
%    \frac{(a)_l}{l!}(-1)^l=\sum_{n=0}^l d_{a,k,n}\frac{(-1)^{l-n}(k\Delta_\phi+n)_{l-n}(k\Delta_\phi+n+1-\frac{d}{2})_{l-n}}{(2k\Delta_\phi+2n+2-d)_{l-n} (l-n)!}\,.
%\end{equation}
%In general, solving $d_{a,k,n}$ in a closed form for arbitrary $a$, $k$, and $n$ is difficult. 
To proceed, we use a generalization of Eq.~\eqref{eq:orthonormal_2F1}\footnote{Here we do not present a full proof. 
Instead, we verify the identity numerically. 
An analytic proof may be obtained by expanding both hypergeometric functions in series and isolating the \(w^{-1}\) term in the integrand, and the coefficient of this term reproduces the \(\delta\)-function.},
\begin{equation}
\label{eq:orthonormal_2F1_general}
\begin{split}
    &\oint_{w}\frac{{\rm d}w}{2\pi{\rm i}}\mathbin{_2 F_1}\left(1-f_{\Delta_\phi}^{(k+1,m)},1-f_{\Delta_\phi}^{(k,m)},2-2f_{\Delta_\phi}^{(k,m)},-w\right)\\
    &\times w^{n-m-1}\mathbin{_2 F_1}\left(f_{\Delta_\phi}^{(k,n)}+\Delta_\phi,f_{\Delta_\phi}^{(k,n)},2f_{\Delta_\phi}^{(k,n)},-w\right)=\delta_{m,n}\,,
\end{split}
\end{equation}
where we define $f_{\Delta}^{(k,n)}=(k-1)\Delta+n$. 
The above orthogonal relation reduces to Eq.~\eqref{eq:orthonormal_2F1} when $k=1$.
To simplify the notation, we define $\bar{f}_{\Delta}^{(k,n)}=1-f_{\Delta}^{(k,n)}$.
This identity allows us to extract $d_{a,k,n}$ as
\begin{widetext}
\begin{eqnarray}
\label{eq:BOE_mu_m_general}
    d_{a,k,m} =
    %=&\underset{w=0}{\rm Res}\left[\frac{w^{-m-1}}{(1+w)^{a}}\mathbin{_2 F_1}\left(\bar{f}_{\Delta_\phi}^{(k+1,m)},\bar{f}_{\Delta_\phi}^{(k,m)},2\bar{f}_{\Delta_\phi}^{(k,m)},-w\right)\right]\\
    \begin{cases}  \frac{(-1)^m(a)_m}{\Gamma(m+1)}\mathbin{_3 F_2}\left(\{\bar{f}_{\Delta_\phi}^{(k+1,m)},\bar{f}_{\Delta_\phi}^{(k,m)},-m\},\{2\bar{f}_{\Delta_\phi}^{(k,m)},1-a-m\},1\right)\,,
    \quad &a>0\\
    \\\frac{4^{\bar{f}_{\Delta_\phi}^{(k,n)}}\sqrt{\pi}\Gamma(2f_{\Delta_\phi}^{(k,n)}-1-n)\Gamma(f_{\Delta_\phi}^{(k+1,n)})}{\Gamma((k-1)\Delta_\phi)\Gamma(k\Delta_\phi)\Gamma(n+1)\Gamma(f_{\Delta_\phi}^{(k,n)}-\frac{1}{2})}\,, \quad &a=0
    \end{cases}
\end{eqnarray}

This leads to the BOE coefficients by noticing $
(1+s(1+\xi^{-1})^{-\Delta_\phi})^k=\sum_{l=0}^k \frac{\Gamma(k+1)}{\Gamma(l+1)\Gamma(k-l+1)}s^l (1+\xi^{-1})^{-l\Delta_\phi}$: 
\begin{equation}
\label{eq:BOE_general_free}
    \tilde{\mu}_{k\Delta_\phi+n}=\sum_{l=0}^{k}\frac{\Gamma(k+1)}{\Gamma(l+1)\Gamma(k-l+1)}s^l d_{l\Delta_\phi,k,n}\,.
\end{equation}
In contrast to the bulk channel, the boundary channel contains only a single series with $\hat{\Delta}=k\Delta_\phi+n$ and $n=0,1,\cdots$, corresponding to the boundary primaries $:\partial^n \hat \phi^k:$.
\end{widetext}

%From the BOE coefficients with $k=2$, we find $\tilde{\mu}_{2\Delta_\phi+n}=0$ for any odd $n$. 
We can prove the following statement for the BOE.
Under the Neumann boundary condition, $\tilde{\mu}_{k\Delta_\phi+n}=0$ for all odd $n$.
Under the Dirichlet boundary condition, $\tilde{\mu}_{k\Delta_\phi+n}=0$ for odd $n+k$. 
In particular, for $k=1$, only $\tilde{\mu}_{\Delta_\phi}$ is nonzero under the Neumann condition, and only $\tilde{\mu}_{\Delta_\phi+1}$ is nonzero for the Dirichlet condition. 
In general, the BOE coefficients have the following property
\begin{eqnarray}
\label{eq:BOE_property}
    &&d_{a,k,m} \nonumber\\
    &=&\oint_{w}\frac{{\rm d}w}{2\pi{\rm i}}\frac{w^{-m-1}}{(1+w)^{a}}\mathbin{_2 F_1}\left(\bar{f}_{\Delta_\phi}^{(k+1,m)},\bar{f}_{\Delta_\phi}^{(k,m)},2\bar{f}_{\Delta_\phi}^{(k,m)},-w\right) \nonumber\\
    &=&\oint_{u}\frac{{\rm d}u}{2\pi{\rm i}}\frac{u^{-m-1}(-1)^m}{(1+u)^{k\Delta_\phi-a}}\mathbin{_2 F_1}\left(\bar{f}_{\Delta_\phi}^{(k+1,m)},\bar{f}_{\Delta_\phi}^{(k,m)},2\bar{f}_{\Delta_\phi}^{(k,m)},-u\right)\nonumber \\
    &=&(-1)^m d_{k\Delta_\phi-a,k,m}\,,
\end{eqnarray}
where in the second line we set $w=-\frac{u}{1+u}$. 
Therefore the BOE coefficients satisfy
\begin{eqnarray}
\label{eq:BOE_property_result}
    \tilde{\mu}_{k\Delta_\phi+n}&=&\sum_{m=0}^{k}\frac{\Gamma(k+1)}{\Gamma(k-m+1)\Gamma(m+1)}s^{k-m} d_{(k-m)\Delta_\phi,k,n} \nonumber \\
    &=&\sum_{m=0}^{k}\frac{\Gamma(k+1)}{\Gamma(k-m+1)\Gamma(m+1)}s^{k-m} (-1)^n d_{m\Delta_\phi,k,n} \nonumber\\
    &=&\frac{1+s^{k} (-1)^n}{2}\tilde{\mu}_{k\Delta_\phi+n}\,.
\end{eqnarray}
Thus, for $s=1$, $\tilde{\mu}_{k\Delta_\phi+n}=0$ for odd $n$. For $s=-1$, $\tilde{\mu}_{k\Delta_\phi+n}=0$ for odd $n+k$.
It is easy to check that when $k=1$ this result is equivalent to the image symmetry in Eqs.~\eqref{eq:image_symm_Hi_phi4} and \eqref{eq:image_symm_Hi_odd_d}.
%For the correlation function in Eq.~\eqref{eq:F_phik}, we have 
%\begin{equation}
%\label{eq:image_symmetry_free}
%    \left.\frac{F_s^{(k)}(\xi')}{\xi'^{k\Delta_\phi}}\right|_{\xi'=e^{\pm{\rm i}\pi}(\xi+1)}= e^{\mp{\rm i}\pi k\Delta_\phi}s^k\frac{F_s^{(k)}(\xi)}{\xi^{k\Delta_\phi}}\,.
%\end{equation}
%Therefore, for $k=1$, $s=1$ or $s=-1$ corresponds to boundary conformal blocks $\hat{\Delta}=\Delta_\phi+n$ with even or odd $n$, which means $s^k=(-1)^n$, matching Eq.~\eqref{eq:image_symm_Hi_odd_d}. 
%The same holds for all odd $k$. For even $k$, both $s=\pm1$ lead to the same set of boundary blocks $\hat{\Delta}=\Delta_\phi+n$ with even $n$, consistent with the $k=2$ result for $\langle\phi^2\phi^2\rangle_s$.
%The structure of the free theory correlators in Eq.~\eqref{eq:image_symmetry_free} and the pattern of nonzero BOE coefficients agree with the property of boundary conformal blocks in Eq.~\eqref{eq:image_symm_boundary_block}.

\section{$d=4-\epsilon$ expansion}
\label{sec:d 4-epsilon expansion}

In this section, we consider the $O(N)$ model in $d=4-\epsilon$ and use the analytic bootstrap to derive anomalous dimensions as well as OPE and BOE coefficients up to $\mathcal{O}(\epsilon^2)$ in $F(\xi)$ for $\langle \phi \phi \rangle$. 
While the Wilson-Fisher fixed point has been investigated using the analytic bootstrap~\cite{liendo2013bootstrap,bissi2019analytic,dey2021analytic}, we emphasize the analysis of the $O(N)$ model with interacting boundaries. 
Note that determining $F(\xi)$ to order $\mathcal{O}(\epsilon^l)$ does not in general fix all conformal data to the same order.
A simple example is that, when $F(\xi)$ is computed to $\mathcal{O}(\epsilon)$, the conformal blocks that do not appear in $G_{b}$ and $G_i$ in the zeroth order will be present in $H_b$ and $H_i$ with their OPE and BOE coefficients at $\mathcal O(\epsilon)$, while their scaling dimensions remain at the bare values.

\subsection{$\mathcal{O}(\epsilon)$ expansion}
\label{sec:O(epsilon) expansion}

An $O(N)$ theory in a semi-infinite space can have different boundary conditions or boundary interactions. 
In the presence of only a boundary mass term $\int{\rm d}^{d-1}x\, c\phi^2$, the ordinary fixed point corresponds to the limit $c\rightarrow\infty$, while the special fixed point has $c\rightarrow 0$ at tree level. 
More generally, the boundary theories can host interactions. 
The analytic bootstrap in $4- \epsilon$ expansion works for any interacting boundary theories that are marginal at $\epsilon = 0$ and connected to the Gaussian theory. 
For example, a cubic $S_{N+1}$ symmetric interaction $\sim \int{\rm d}^{d-1}x\, d_{ijk} \phi_i \phi_j \phi_k$ is marginal on the boundary. 
A perturbative calculation for such interactions will be present in Sec.~\ref{sec:RG of the bulk-phi4/boundary-phi3 theory}.

We shall begin with the zeroth-order solution under the Neumann boundary condition, and explore the solution to the bootstrap equation in the $\epsilon$ expansion. 
%Using the solution in the Neumann boundary condition in Eq.~\eqref{eq:0th_solution_general_bootstrap}, we first evaluate Eq.~\eqref{eq:Disc_relation_for_deltaG_deltaH}.
%Fixing $\lambda_I=1,\  \mu_I=0$, and 
We expand around the solution in the Neumann boundary condition in Eq.~\eqref{eq:0th_solution_general_bootstrap}
$\Delta_{0}=d-2+\Delta_{0}^{(1)}\epsilon$, $ \Delta_\phi=\frac{d-2}{2}+\Delta_\phi^{(1)}\epsilon$, $ \hat{\Delta}_{0}=\frac{d-2}{2}+\hat{\Delta}_{0}^{(1)}\epsilon$ and $\tilde{\lambda}_{0}=1+\tilde{\lambda}_{0}^{(1)}\epsilon,\ \tilde{\mu}_0=2+\tilde{\mu}_0^{(1)}\epsilon$.
Following the procedure in Fig.~\ref{fig:bootstrap_pipeline}, we first obtain the discontinuity,
\begin{equation}
\label{eq:Disc_d_4_epsilon}
    \underset{\xi<-1}{\rm Disc}\ (G_i(\xi)-G_b(\xi))=-{\rm i}\pi\epsilon\frac{A_1+A_2\xi}{1+\xi}\,,
\end{equation}
with
\begin{equation}
\label{eq:parameters_A_d_4_epsilon}
    A_1=2\hat{\Delta}_{0}^{(1)}-2\Delta_\phi^{(1)}\,,~ A_2=4\hat{\Delta}_{0}^{(1)}-4\Delta_\phi^{(1)}+\Delta_{0}^{(1)}\,.
\end{equation}
Substituting Eq.~\eqref{eq:Disc_d_4_epsilon} into Eq.~\eqref{eq:OPE_lambda_n}, the OPE coefficients are
\begin{equation}
\label{eq:OPE_d_4_epsilon}
    \tilde{\lambda}_n^{(1)}=-\frac{\Gamma(n)\Gamma(n+1)}{8(-1)^n \Gamma(2n)}\begin{cases}
        (2A_1-4A_2) \,,& n=1 \,,\\[0.7ex]
        \frac{4}{n}(A_1-A_2)\,, & n>1 \,.
    \end{cases}
\end{equation}
We then compute $H_b$ by summing over $n$.
Treating the $A_1$ and $A_2$ terms separately yields
\begin{eqnarray}
\label{eq:sum_n_Hb_d_4_epsilon}
    H_b(\xi)=\epsilon\left(A_1\frac{\xi-\log{(1+\xi)}}{2(1+\xi)}-A_2\frac{\xi\log{(1+\xi)}}{2(1+\xi)}\right) \,.
\end{eqnarray}
We can check that $H_b(\xi)$ gives the discontinuity %${\rm Disc}_{\xi<-1} \ H_b(\xi)=\epsilon\!\left(A_1\frac{-2\pi{\rm i}}{2(1+\xi)}-A_2\frac{2\pi{\rm i}\xi}{2(1+\xi)}\right)$, 
agreeing with Eq.~\eqref{eq:Disc_d_4_epsilon}.

Next, $H_i$ is obtained via $H_i=G_b+H_b-G_i$:
\begin{eqnarray}
\label{eq:difference_Hi_xi}
   H_i(\xi)&=&\epsilon\left(B_1 \frac{1}{\xi+1}+B_2 \frac{\xi}{\xi+1}+B_3 \frac{\log{\xi}}{\xi+1}+B_4 \frac{\xi\log{\xi}}{\xi+1}\right. \nonumber \\
    &&\left.+B_5 \frac{\log{(1+\xi)}}{\xi+1}+B_6 \frac{\xi\log{(1+\xi})}{\xi+1}\right) \,,
\end{eqnarray}
where
\begin{eqnarray}
\label{eq:parameters_B_d_4_epsilon}
    && B_1 =-\frac{1}{2}\tilde{\mu}_0^{(1)}\,,~  B_2=-\Delta_\phi^{(1)}+\tilde{\lambda}_0^{(1)}+\hat{\Delta}_0^{(1)}-\tilde{\mu}_0^{(1)}\,,\\ 
    && B_3=-B_5=-\Delta_\phi^{(1)}\,,
    B_4=-B_6=-2\Delta_\phi^{(1)}+\frac{1}{2}\Delta_0^{(1)}+\hat{\Delta}_0^{(1)} \,. \nonumber
\end{eqnarray}
Substituting this into Eq.~\eqref{eq:BOE_mu_m} and projecting onto the boundary conformal blocks, we obtain the BOE coefficients $\tilde{\mu}_m^{(1)}$.
In particular, when $m=0$, we have 
\begin{eqnarray}
    \tilde{\mu}_{0,c}^{(1)}=\hat{\Delta}_{0}^{(1)}-\Delta_\phi^{(1)}+\tilde{\lambda}_{0}^{(1)}-\tilde{\mu}_{0}^{(1)}\,.
\end{eqnarray}
As we described before, we have grouped the entire $m=0$ block in $G_i$, which indicates that $\tilde{\mu}_{0,c}=0$ should be treated as a constraint, indicated by the subscript, $c$.  
This constraint fixes $\tilde{\mu}_0$ to be $\tilde{\mu}_{0}^{(1)}=\hat{\Delta}_{0}^{(1)}-\Delta_\phi^{(1)}+\tilde{\lambda}_{0}^{(1)}$. 
Therefore, the BOE coefficients are
\begin{equation}
\label{eq:BOE_d_4_epsilon}
\begin{split}
    \tilde{\mu}_{0}^{(1)}&=\hat{\Delta}_{0}^{(1)}-\Delta_\phi^{(1)}+\tilde{\lambda}_{0}^{(1)}\,, \\
    \tilde{\mu}_{1}^{(1)}&=\frac{1}{2}(-3\hat{\Delta}_{0}^{(1)}+5\Delta_\phi^{(1)}-\tilde{\lambda}_{0}^{(1)}-\Delta_{0}^{(1)})\,, \\
    \tilde{\mu}_{m\geq2}^{(1)}&=B_3'\cdot \frac{(-4)^{1-m}\sqrt{\pi}\Gamma(m-1)}{\Gamma(m-\frac{1}{2})}\\
    &\quad+B_4'\cdot \frac{4^{1-m}(1+(-1)^m)\sqrt{\pi}\Gamma(m-1)}{\Gamma(m-\frac{1}{2})} \,,
\end{split}
\end{equation}
where 
\begin{eqnarray}
\label{eq:parameters_Bp_d_4_epsilon}
    B_3'=&-\hat{\Delta}_{0}^{(1)}+2\Delta_\phi^{(1)}-\frac{1}{2}\Delta_{0}^{(1)},\qquad
    B_4'=\Delta_\phi^{(1)}\,.
\end{eqnarray}

In general, we expect the operations of taking discontinuities, performing series expansions, and resumming to commute.
To avoid any ambiguity, we resum the conformal blocks in $H_i(\xi)$ using the BOE coefficients in Eq.~\eqref{eq:BOE_d_4_epsilon} and check whether the result reproduces $H_i(\xi)$ in Eq.~\eqref{eq:difference_Hi_xi}.
After a lengthy but straightforward calculation, we indeed obtain the same $H_i(\xi)$ by summing all boundary conformal blocks $f_i(\hat{\Delta}_m,\xi)$ for $m\geq1$ with their corresponding BOE coefficients.
We have done similar checks for all cases studied in this paper, and found that they are all consistent. 
Hence, we omit the explicit mentioning for the other cases in the rest of the paper.

The OPE and BOE coefficients at $\mathcal O(\epsilon)$ are determined by four parameters, $\Delta_\phi^{(1)}$, $\Delta_0^{(1)}$, $\hat \Delta_0^{(1)}$ and $\tilde \lambda_0^{(1)}$. 
We now discuss the image symmetry, Eq.~\eqref{eq:image_symm_Hi_phi4}, in light of the results above.
%In $d=4-\epsilon$, the image symmetry implies Eq.~\eqref{eq:image_symm_Hi_phi4}. 
Under the Neumann boundary condition, it implies that all boundary conformal blocks have even $m$.  %which yields $\frac{H_i(\xi')}{\xi'}|_{\xi'=e^{\pm{\rm i}\pi}(\xi+1)}=-\frac{H_i(\xi)}{\xi}$.
Imposing this constraint on Eq.~\eqref{eq:difference_Hi_xi} leads to relations among the $B_i$: $B_1=B_2=0$ and $B_3=B_4=-B_5=-B_6$, which in turn gives the relations among the conformal data
\begin{equation}
\label{eq:relation_conformal_data_image_symmetry}
    \hat{\Delta}_0^{(1)}=-\frac{1}{2}\Delta_0^{(1)}+\Delta_\phi^{(1)}\,,\qquad \tilde{\lambda}_0^{(1)}=\frac{1}{2}\Delta_{0}^{(1)}\,.
    %\qquad \tilde{\mu}_0^{(1)}=0\,.
\end{equation}
Recall that the OPE and BOE coefficients at $\mathcal O(\epsilon)$ are determined by four inputs, $\Delta_\phi^{(1)}$, $\Delta_0^{(1)}$, $\hat{\Delta}_0^{(1)}$, $\tilde{\lambda}_{0}^{(1)}$. 
The image symmetry reduces it into two bulk data $\Delta_\phi^{(1)}$, $\Delta_0^{(1)}$. 
However, we will see that when the boundary interaction is present, the image symmetry is no longer preserved. 

Without boundary interactions, the Wilson-Fisher theory at the special fixed point respects the image symmetry~\cite{liendo2013bootstrap,bissi2019analytic,dey2021analytic}. 
Using $\Delta_\phi^{(1)}=0$ and $ \Delta_{0}^{(1)}=2\alpha$, we obtain the OPE and BOE coefficients from Eqs.~\eqref{eq:OPE_d_4_epsilon} and \eqref{eq:BOE_d_4_epsilon},
\begin{equation}
\label{eq:bulk_phi4_boundary_free_output}
    \tilde{\lambda}_1^{(1)}=\frac{\alpha}{2}\,,\quad \tilde{\lambda}_{n\geq2}^{(1)}=0\,,\quad \tilde{\mu}_{m\geq0}^{(1)}=0\,.
\end{equation}
From the RG calculation for the critical $O(N)$ theory, we have the following at the well-known Wilson-Fisher fixed point~\cite{wilson1973quantum}:
\begin{equation}
\label{eq:RG_phi4_free}
    \alpha=\frac{1}{2}\frac{N+2}{N+8}.
\end{equation}

While the result for the special transition without boundary interactions was well-known previously, we now extend the analysis to interacting boundaries. 
First, we consider a Gaussian theory in the bulk supplemented by a boundary cubic interaction that respects $S_{N+1}$ symmetry. 
%analyze the long-range Potts/Yang-Lee fixed point with a free bulk and a boundary-$\phi^3$ interaction. 
We will investigate the theory explicitly using perturbative RG calculation in Sec.~\ref{sec:RG of the bulk-phi4/boundary-phi3 theory}, where we identify its fixed point, referred to as the long-range Potts fixed point. 
Also, the bootstrap analysis is valid for the long-range Yang-Lee fixed point for $N=1$~\cite{diehl1991surface}.

Because the boundary cannot renormalize the bulk, the anomalous dimensions of bulk operators vanish to all orders.
Notably, the anomalous dimension of the boundary operator $\hat{\phi}$ is also zero at the long-range Potts/Yang-Lee fixed point, because, with a free bulk theory, the two-point function of $\hat{\phi}$ on the boundary has the nonanalytic form $\frac{1}{|p|}$, which cannot be renormalized by local boundary interactions. 
Therefore, the only nontrivial conformal data are the OPE and BOE coefficients.
With the input $\tilde{\lambda}_{0}^{(1)}=\alpha'$, the remaining OPE and BOE coefficients are~\footnote{If we include composite operators on the boundary, they may acquire nontrivial anomalous dimensions~\cite{prochazka2020composite}.}
\begin{equation}
\label{eq:bulk_free_boundary_phi3_output}
    \tilde{\lambda}_{n\geq1}^{(1)}=0\,,\quad \tilde{\mu}_0^{(1)}=-2\tilde{\mu}_1^{(1)}=\alpha'\,,\quad \tilde{\mu}_{m\geq2}^{(1)}=0\,.
\end{equation}
Since both even and odd $m$ appear in the BOE, it does not preserve image symmetry. 
%Plugging these BOE coefficients into $F(\xi)$, we obtain the two-point function
%\begin{eqnarray}
%\label{eq:2_pt_bulk_free_boundary_phi3_d_4}
    %\langle\phi(x)\phi(x')\rangle&=&\frac{\xi^{-1}}{2z\cdot 2z'}\xi\left[(2+\alpha'\epsilon)f_i(1,\xi)-\frac{\alpha'}{2}\epsilon f_i(2,\xi)\right] \nonumber \\
    %&=&\frac{1}{4zz'}\left(\frac{1}{\xi}+\frac{1+\alpha'\epsilon}{1+\xi}\right)\,.
%\end{eqnarray}
%where we use the conformal block $f_i(\frac{d-2}{2},\xi)=\frac{1}{2}(\frac{1}{\xi}+\frac{1}{\xi+1})+\mathcal{O}(\epsilon)$.
%This result is consistent with Ref.~\cite{prochazka2020composite}. 
We will determine $\alpha'$ from the RG calculation in Sec.~\ref{sec:RG of the bulk-phi4/boundary-phi3 theory}.

Finally, we turn to the Wilson-Fisher fixed point in the presence of interacting boundaries. 
Again, we will investigate this theory in Sec.~\ref{sec:RG of the bulk-phi4/boundary-phi3 theory}, where we identify its fixed point, referred to as the boundary Potts/Yang-Lee fixed point. 
Since the boundary cannot renormalize the bulk, the anomalous dimensions of bulk operators are the same as the Wilson-Fisher fixed point.
The RG calculations further show that, up to $\mathcal{O}(\epsilon)$, the anomalous dimension of the boundary operator $\hat{\phi}$ is also unchanged.
Hence, to order $\mathcal{O}(\epsilon)$, the only new input is $\tilde{\lambda}_{0}^{(1)}$.
Defining $\tilde{\lambda}_{0}^{(1)}=\beta$, the conformal data for the boundary Potts/Yang-Lee fixed point is determined by two independent inputs, $\alpha$ and $\beta$, as 
\begin{equation}
\label{eq:bulk_phi4_boundary_phi3_output}
\begin{split}
    &\Delta_\phi^{(1)}=0\,,\qquad \Delta_{0}^{(1)}=2\alpha\,,\qquad \hat{\Delta}_{0}^{(1)} = -\alpha\,, \\
    &\tilde{\lambda}_{0}^{(1)}=\beta,\qquad \tilde{\lambda}_1^{(1)}=\frac{\alpha}{2}\,,\qquad \tilde{\lambda}_{n\geq2}^{(1)}=0\,,\\
    &\tilde{\mu}_{0}^{(1)}=-2\tilde{\mu}_{1}^{(1)}=-\alpha+\beta\,,\qquad \tilde{\mu}_{m\geq2}^{(1)}=0\,.
\end{split}
\end{equation}
Again, with nonzero $\tilde{\mu}_{0}^{(1)}$ and $\tilde{\mu}_{1}^{(1)}$, image symmetry is not preserved.

Note that all theories we have considered so far, including the special fixed point, the long-range Potts/Yang-Lee fixed point, and the boundary Potts/Yang-Lee fixed point, have finitely many nonvanishing OPE and BOE coefficients. 
They can be unified by two independent inputs, $\Delta_0^{(1)}$ and $\tilde\lambda_0^{(1)}$
as summarized in Table~\ref{tab:4_epsilon_epsilon1}. 
Hence, we can continue the analytic bootstrap to the next order.

\subsection{$\mathcal{O}(\epsilon^2)$ expansion}
\label{sec:O(epsilon2) expansion}

In this section we use the $\mathcal{O}(\epsilon)$ bootstrap solution, which contains finitely many conformal blocks in $F(\xi)$, to derive the $\mathcal{O}(\epsilon^2)$ corrections. 
Since the solution in Eq.~\eqref{eq:bulk_phi4_boundary_phi3_output} with two independent inputs, which can be taken as, $\Delta_0^{(1)}$ and $\tilde\lambda_0^{(1)}$, is the most general one in our setup, we use it as the starting point and later specialize its parameters to recover all other cases.
% \textcolor{red}{might be better to use $\Delta_0^{(1)}$ and $\tilde\lambda_0^{(1)}$ for the general cases. and use $\alpha$ for the bulk-$\phi^4$/boundary-free, $\alpha'$ for the bulk-free, and $\alpha$, $\beta$ for bulk-$\phi^4$/boundary-$\phi^3$. namely $\alpha$ $\beta$ are reserved for explicit numbers.}
% \textcolor{blue}{Corrected.}

%Using Eqs.~\eqref{eq:lst_correction_bulk_channel_general} and \eqref{eq:lst_correction_boundary_channel_general}, we evaluate the difference $G_i-G_b$.
The first step is to get the discontinuity of $G_i - G_b$ as shown in the general procedure Fig.~\ref{fig:bootstrap_pipeline}.
% \textcolor{red}{write out the explicit expansion coefficients for completeness: Consider the expansion}
Consider the expansion 
$\Delta_\phi^{(\leq2)}$, $\Delta_{0}^{(\leq2)}$, $\Delta_{1}^{(\leq1)}$, $\Delta_{n\geq2}^{(0)}$, $\tilde{\lambda}_{m}^{(\leq2)}$, $\hat{\Delta}_{0}^{(\leq2)}$, $\hat{\Delta}_{1}^{(\leq1)}$, $\hat{\Delta}_{m\geq2}^{(0)}$, and $\tilde{\mu}_n^{(\leq2)}$, 
where the superscripts indicate the orders of expansion.
The zeroth-order results are $\Delta_\phi^{(0)}=\frac{d-2}{2},\,\Delta_{n}^{(0)}=d-2+2n,\,\hat{\Delta}_{m}^{(0)}=\frac{d-2}{2}+m$, while the first-order results are   $\Delta_\phi^{(1)}=0,\,\hat{\Delta}_{0}^{(1)}=-\frac{1}{2}\Delta_0^{(1)},\,\tilde{\lambda}_{1}^{(1)}=\frac{1}{4}\Delta_0^{(1)},\, \tilde{\mu}_{0}^{(1)}=-2\tilde{\mu}_{1}^{(1)}=-\frac{1}{2}\Delta_0^{(1)}+\tilde\lambda_0^{(1)}$.
%Here we base on the most general solution Eq.~\eqref{eq:bulk_phi4_boundary_phi3_output}, but do not plug $\alpha$ and $\beta$ into the solution to express the conformal data.
% Later, we will find that the final input $\Delta_0^{(1)},\tilde\lambda_0^{(1)}$ and $\Delta_\phi^{(2)},\Delta_0^{(2)},\Delta_1^{(1)},\tilde{\lambda}_0^{(2)},\hat{\Delta}_0^{(2)},\hat{\Delta}_1^{(1)}$.
To obtain the $\mathcal{O}(\epsilon^2)$ expansion in $G_i$ and $G_b$, we need the expansion of the hypergeometric function $\mathbin{_2 F_1}(a,b,c,x)$, given in Appendix~\ref{sec:Expansion of conformal blocks}.
From this we find the discontinuity
\begin{eqnarray}
\label{eq:Disc_d_4_epsilon2}
    &&\underset{\xi<-1}{\rm Disc}\ (G_i(\xi)-G_b(\xi))={\rm i}\pi\epsilon^2\left(A_1 \frac{1}{1+\xi}+A_2\log{(-1-\xi)}\right. \nonumber \\
    &&\qquad\quad\left.+A_3\log{(-\xi)}+A_4\frac{\xi}{1+\xi}+A_5\frac{\xi\log{(-\xi)}}{1+\xi}\right)\, ,
\end{eqnarray}
where
\begin{eqnarray}
\label{eq:parameters_A_d_4_epsilon2}
    A_1 &=&\frac{1}{4}\Delta_0^{(1)}-\frac{1}{4}(\Delta_0^{(1)})^2+\frac{1}{2}\Delta_0^{(1)}\tilde\lambda_0^{(1)}-\frac{3}{2}\Delta_0^{(1)} \Delta_1^{(1)} \nonumber \\
    &&+2\Delta_\phi^{(2)}-2\hat{\Delta}_{0}^{(2)}+(-\frac{1}{2}\Delta_0^{(1)}+\tilde\lambda_0^{(1)})\hat{\Delta}_1^{(1)}\,, \nonumber \\ A_2&=&\frac{1}{2}\Delta_0^{(1)}(\Delta_0^{(1)}-\Delta_1^{(1)})\,, \qquad
    A_3=\frac{1}{2}\Delta_0^{(1)} \Delta_1^{(1)}\,, \nonumber\\
    A_4&=&\frac{\Delta_0^{(1)}}{4}-\frac{(\Delta_0^{(1)})^2}{2}-\Delta_0^{(2)}-\Delta_0^{(1)} \Delta_1^{(1)}+4\Delta_\phi^{(2)}-4\hat{\Delta}_0^{(2)}\,, \nonumber\\
    A_5&=&\frac{1}{2}(\Delta_0^{(1)})^2\,. 
\end{eqnarray}

Plugging Eq.~\eqref{eq:Disc_d_4_epsilon2} into Eq.~\eqref{eq:OPE_lambda_n}, we obtain the OPE coefficients
\begin{equation}
\label{eq:OPE_d_4_epsilon2}
    \tilde{\lambda}_n^{(2)}=\begin{cases}
        \frac{1}{8}(-2A_1+2A_2+3A_3+4A_4+4A_5)\,,& n=1\,,\\[2.7ex]
        \frac{\Gamma(n)\Gamma(n+1)}{8(-1)^n\Gamma(2n)}\left(\frac{4}{n}A_1+\frac{4(-1)^{n}((-1)^n+n)}{n(n^2-1)}A_2\right.\\
        \qquad+\left.\frac{4(-1)^{n}}{n^2-1}A_3-\frac{4}{n}A_4+\frac{4(-1)^{n}}{n^2}A_5\right)\,,& n>1\,.
    \end{cases}
\end{equation}
We then compute $H_b$ by summing the infinite set of conformal blocks to arrive at
\begin{equation}
\label{eq:sum_n_Hb_d_4_epsilon2}
\begin{split}
    H_b(\xi)=\epsilon^2&\left[A_1\frac{-\xi+\log{(1+\xi)}}{2(1+\xi)}+A_2\frac{\log{(1+\xi)^2}}{4}\right.\\
    &+A_3\left(\frac{-\xi}{2(1+\xi)}-\frac{1}{2}{\rm Li}_2(-\xi)\right)\\
    &\left.+A_4\frac{\xi\ \log{(1+\xi)}}{2(1+\xi)}+A_5\frac{-\xi\ {\rm Li}_2(-\xi)}{2(1+\xi)}\right]\,.
\end{split}
\end{equation}
Next, $H_i=G_b+H_b-G_i$ takes the form
\begin{eqnarray}
\label{eq:difference_Hi_xi_epsilon2}
    && H_i(\xi)=\epsilon^2\left(B_1 \frac{1}{\xi+1}+B_2 \frac{\log{\xi}}{\xi+1}+B_3 \frac{\log{(\xi+1)}}{\xi+1}\right. \nonumber\\
    &&+B_4 \frac{\log{\xi}\log{(1+\xi)}}{\xi+1}+B_5 \frac{\log{(1+\xi)}^2}{\xi+1} \nonumber\\
    &&+B_6 \frac{{\rm Li}_2(-\xi)}{\xi+1}+B_7 \frac{\xi\log{\xi}}{\xi+1}+B_8 \frac{\xi(\log{\xi})^2}{\xi+1}\\
    && +B_9 \frac{\xi\log{\xi}\log{(\xi+1)}}{\xi+1}+B_{10} \frac{\xi\log{(\xi+1)}}{\xi+1}\nonumber\\
    &&\left.+B_{11} \frac{\xi\log{(1+\xi)}^2}{\xi+1}+B_{12} \frac{\xi}{\xi+1}+B_{13}\frac{\xi {\rm Li}_2(-\xi)}{\xi+1} \right)\,, \nonumber
\end{eqnarray}
where the coefficients are
\begin{eqnarray}
\label{eq:parameters_B_d_4_epsilon2}
    B_1&=&-\frac{\Delta_0^{(1)}}{8}+\frac{\pi^2}{24}(1-\Delta_0^{(1)})\Delta_0^{(1)} \nonumber\\
    &&\qquad\qquad+\frac{\tilde\lambda_0^{(1)}}{4}-\frac{\hat{\Delta}_1^{(1)}}{2}(\frac{1}{2}\Delta_0^{(1)}-\tilde\lambda_0^{(1)})-\frac{1}{2}\tilde{\mu}_0^{(2)},\nonumber\\
    B_2&=&-B_3=-\Delta_\phi^{(2)}\,,\nonumber\\ B_4&=&-2B_5=B_6=B_9=-2B_{11}=\frac{\Delta_0^{(1)}}{4}(-\Delta_0^{(1)}+\Delta_1^{(1)})\,, \nonumber\\
    B_7&=&-B_{10}=\frac{1}{4}\Delta_0^{(1)}(\frac{1}{2}\Delta_0^{(1)}+\tilde\lambda_0^{(1)})-2\Delta_\phi^{(2)}+\frac{1}{2}\Delta_0^{(2)} \nonumber\\
    &&\qquad\quad-\frac{1}{4}\Delta_0^{(1)}\Delta_1^{(1)}+\hat{\Delta}_0^{(2)}+\frac{1}{2}(\tilde\lambda_0^{(1)}-\frac{1}{2}\Delta_0^{(1)})\hat{\Delta}_1^{(1)}\,,
    \nonumber\\
    B_8&=&\frac{1}{8}\Delta_0^{(1)}(\Delta_0^{(1)}-1)\,,\nonumber\\
    B_{12}&=&-\frac{1}{4}\Delta_0^{(1)}(1-\frac{1}{2}\Delta_0^{(1)}+\tilde\lambda_0^{(1)})+\frac{1}{4}\Delta_0^{(1)}\Delta_1^{(1)}-\Delta_\phi^{(2)} \nonumber\\
    &&\qquad+\tilde{\lambda}_0^{(2)}+\hat{\Delta}_0^{(2)}+\frac{1}{2}(\frac{1}{2}\Delta_0^{(1)}-\tilde\lambda_0^{(1)})\hat{\Delta}_1^{(1)}-\tilde{\mu}_0^{(2)}\,, \nonumber\\
    B_{13}&=&-\frac{1}{4}\Delta_0^{(1)}(1-\Delta_1^{(1)})\,.
\end{eqnarray}
Finally, inserting Eq.~\eqref{eq:difference_Hi_xi_epsilon2} into Eq.~\eqref{eq:BOE_mu_m} yields the BOE coefficients
\begin{equation}
\label{eq:BOE_d_4_epsilon2}
\begin{split}
    \tilde{\mu}_m^{(2)}=\begin{cases}
        B_1'\,,& m=0,\\[0.7ex]
        B_2'-B_4'+B_5'\,,& m=1\,,\\[2.7ex]
        \frac{4^{1-m}\sqrt{\pi}\Gamma(m-1)}{(-1)^{m-1}\Gamma(m-\frac{1}{2})}\left(B_2'+\frac{2(1+(-1)^m)}{m(m-1)}B_3'\right.\\
        \left.\qquad\ +\frac{(-1)^{m-1}}{m(m-1)}B_4'-(1+(-1)^m)B_6'\right),& m>1,
    \end{cases}
\end{split}
\end{equation}
where
\begin{eqnarray}
\label{eq:parameters_Bp_d_4_epsilon2}
    && B_1'=\frac{1}{2}(\frac{1}{2}\Delta_0^{(1)}-\tilde\lambda_0^{(1)})(\frac{1}{2}\Delta_0^{(1)}+\hat{\Delta}_1^{(1)})+\hat{\Delta}_0^{(2)} \nonumber\\
    &&\qquad\qquad+\frac{1}{4}\Delta_0^{(1)}(\frac{\pi^2}{6}-1)(1-\Delta_1^{(1)})-\Delta_\phi^{(2)}+\tilde{\lambda}_0^{(2)}\,, \nonumber\\
    &&B_2'=-\frac{1}{4}\Delta_0^{(1)}(\frac{1}{2}\Delta_0^{(1)}+\tilde\lambda_0^{(1)})-\frac{1}{2}\Delta_0^{(2)}+\frac{1}{4}\Delta_0^{(1)}\Delta_1^{(1)} \nonumber\\
    &&\qquad\qquad+2\Delta_\phi^{(2)}-\hat{\Delta}_0^{(2)}+\frac{1}{2}(\frac{1}{2}\Delta_0^{(1)}-\tilde\lambda_0^{(1)})\hat{\Delta}_1^{(1)}\,, \nonumber\\
    &&B_3'=\frac{1}{8}\Delta_0^{(1)}(\Delta_0^{(1)}-\Delta_1^{(1)})\,, \quad B_4'=\frac{1}{4}\Delta_0^{(1)}(1-\Delta_1^{(1)})\,,\nonumber\\
    &&B_5'=\frac{3}{8}\hat{\Delta}_1^{(1)}(2\tilde\lambda_0^{(1)}-\Delta_0^{(1)})+\frac{1}{2}(\Delta_\phi^{(2)}-\tilde\lambda_0^{(2)}-\hat{\Delta}_0^{(2)}) \nonumber\\
    &&+\frac{\tilde\lambda_0^{(1)}}{8}(2+\Delta_0^{(1)})-\frac{\Delta_0^{(1)}\Delta_1^{(1)}}{8}(1+\frac{\pi^2}{6})+\frac{\Delta_0^{(1)}}{8}(\frac{\pi^2}{6}-\frac{\Delta_0^{(1)}}{2})\,, \nonumber\\
    &&B_6'=\Delta_\phi^{(2)}\,.
\end{eqnarray}
Here, we have imposed the constraint that the coefficient of the lowest boundary conformal block ($m=0$) inferred from $H_i$ vanishes. 
This condition fixes the physical BOE coefficient $\tilde{\mu}_{0}^{(2)}$ and determines other parameters $B_i'$.
Therefore, the independent inputs are  $\Delta_\phi^{(2)},\Delta_0^{(\le2)},\Delta_1^{(1)},\tilde{\lambda}_0^{(\le2)},\hat{\Delta}_0^{(2)},\hat{\Delta}_1^{(1)}$.

Similarly, we apply image symmetry to the $\mathcal{O}(\epsilon^2)$ results obtained above, which imposes the %following 
constraints on $B_i$ and,
%begin{equation}
%\label{eq:image_symm_epsilon2_constraint_B}
%\begin{split}
%    B_3&=-B_2,\qquad\qquad\qquad\quad
%    B_4=-2B_5=B_6=-2B_8=B_9=-2B_{11},\\
%    B_7&=B_{10}=B_{13}=0,\qquad\ B_{12}=2B_1+\frac{2\pi^2}{3}B_8.
%\end{split}
%\end{equation}
%These constraints 
consequently implies the relations among the independent inputs,
\begin{equation}
\label{eq:relation_conformal_data_image_symmetry_epsilon2}
\begin{split}
    \Delta_1^{(1)}&=1\,,\\
    \hat{\Delta}_0^{(2)}&=\frac{1}{2}\left(-\frac{1}{2}\Delta_0^{(1)}(-1+\frac{1}{2}\Delta_0^{(1)}+\tilde\lambda_0^{(1)})-\Delta_0^{(2)}\right.\\
    &\qquad\qquad\left.+4\Delta_\phi^{(2)}+(\frac{1}{2}\Delta_0^{(1)}-\tilde\lambda_0^{(1)})\hat{\Delta}_1^{(1)}\right)\,,\\
    \tilde{\lambda}_0^{(2)}&=\frac{1}{2}\left(\Delta_0^{(1)}(-1+\tilde\lambda_0^{(1)})+\tilde\lambda_0^{(1)}+\Delta_0^{(2)}-2\Delta_\phi^{(2)}\right.\\
    &\qquad\qquad\left.-4(\frac{1}{2}\Delta_0^{(1)}-\tilde\lambda_0^{(1)})\hat{\Delta}_1^{(1)}\right)\,.
\end{split}
\end{equation}
The special fixed point of the Wilson-Fisher fixed point preserves the image symmetry.
We include the result~\cite{bissi2019analytic} here for completeness: 
%we can simplify it, with the additional input $\Delta_0^{(1)}=2\alpha$ and $\tilde\lambda_0^{(1)}=\alpha$, to
\begin{equation}
\label{eq:relation_conformal_data_two_image_symmetry_epsilon2}
\begin{split}
    \hat{\Delta}_0^{(2)}&=\frac{1}{2}\left(\alpha-2\alpha^2-\Delta_0^{(2)}+4\Delta_\phi^{(2)}\right)\,,\\
    \tilde{\lambda}_0^{(2)}&=\frac{1}{2}\left(-\alpha+2\alpha^2+\Delta_0^{(2)}-2\Delta_\phi^{(2)}\right)\,,
\end{split}
\end{equation}
%which agree with the $\mathcal{O}(\epsilon^2)$ results for the bulk-$\phi^4$/boundary-free theory~\cite{bissi2019analytic}
%Then, substituting Eq.~\eqref{eq:relation_conformal_data_two_image_symmetry_epsilon2} into $\tilde{\mu}_0^{(2)}$ in Eq.~\eqref{eq:BOE_d_4_epsilon2}, we obtain $\tilde{\mu}_0^{(2)}=0$. %consistent with the bulk-$\phi^4$/boundary-free result.
%All other OPE and BOE coefficients are listed here:
% \begin{equation}
% \label{eq:bulk_phi4_bpundary_free_epsilon2_OPE_BOE}
% \begin{split}
%     \tilde{\lambda}_1^{(2)}=&\frac{1}{4}(\alpha(-3+8\alpha)+\Delta_0^{(2)}-6\Delta_\phi^{(2)})\,,\\
%     \tilde{\lambda}_{n\geq2}^{(2)}=&\frac{(-1)^n\Gamma(n)\Gamma(n+1)}{n(n^2-1)\Gamma(2n+1)}\left(2\alpha^2(-1)^n(2n^2-1)+\alpha (2\alpha-1)n+2n(n^2-1)\Delta_\phi^{(2)}\right)\,,\\
%     %\tilde{\mu}_1^{(2)}=&0,\\
%     \tilde{\mu}_{m>0}^{(2)}=&\frac{2^{1-2m}(1+(-1)^m)\sqrt{\pi}\Gamma(m-1)}{m(m-1)\Gamma(m-\frac{1}{2})}\left(\alpha-2\alpha^2+2(m-1)m\Delta_\phi^{(2)} \right)\,,\\
% \end{split}
% \end{equation}
as well as the OPE and BOE coefficients,
\begin{equation}
\label{eq:bulk_phi4_bpundary_free_epsilon2_OPE_BOE}
\begin{split}
    &\tilde{\lambda}_n^{(2)}=\begin{cases}
        \frac{1}{4}(\alpha(-3+8\alpha)+\Delta_0^{(2)}-6\Delta_\phi^{(2)}),& n=1,\\[0.7ex]
        \frac{(-1)^n\Gamma(n)\Gamma(n+1)}{n(n^2-1)\Gamma(2n+1)}\left(2\alpha^2(-1)^n(2n^2-1)\right.\\
        \left.\qquad\ +\alpha (2\alpha-1)n+2n(n^2-1)\Delta_\phi^{(2)}\right),& n\geq2,
    \end{cases}\\
    &\tilde{\mu}_{m>0}^{(2)}=\frac{2^{1-2m}\sqrt{\pi}\Gamma(m-1)}{m(m-1)\Gamma(m-\frac{1}{2})}\left(\alpha-2\alpha^2+2(m-1)m\Delta_\phi^{(2)} \right)\,.\\
\end{split}
\end{equation} 
The BOE coefficients above are valid only for even $m$, and $\tilde{\mu}_{m>0}^{(2)}=0$ for odd $m$, as a result of the image symmetry. 
Also, the bulk scaling dimensions for the Wilson-Fisher fixed point are 
\begin{equation}
\label{eq:WF_2}
\begin{split}
&\Delta_1^{(1)}=1\,,  \quad \Delta_\phi^{(2)}=\frac{N+2}{4(N+8)^2}\,,  \\ 
&\Delta_0^{(2)}=\frac{(N+8)(13N+44)}{2(N+8)^3}\,.
\end{split}
\end{equation}

We now discuss interacting boundaries, including the long-range Potts/Yang-Lee fixed point, and the boundary Potts/Yang-Lee fixed point. 
At the long-range Potts/Yang-Lee fixed point, we set to zero the anomalous dimensions of all bulk operators and of the boundary operator $\hat{\phi}$, while keeping nontrivial $\tilde{\lambda}_0^{(2)}$ and $\hat{\Delta}_1^{(1)}$.
With $\Delta_0^{(1)}=0$ and $\tilde\lambda_0^{(1)}=\alpha'$, the OPE and BOE coefficients simplify to
\begin{equation}
\label{eq:bulk_free_boundary_phi3_epsilon2_confromal_data}
\begin{split}
    &\tilde{\lambda}_n^{(2)}=\begin{cases}
        -\frac{1}{4}\alpha'\hat{\Delta}_1^{(1)},&\qquad\qquad\ n=1,\\[0.7ex]
        \frac{\Gamma(n)^2}{2(-1)^n \Gamma(2n)}\alpha'\hat{\Delta}_1^{(1)},&\qquad\qquad\ n\geq2,
    \end{cases}\\
    &\tilde{\mu}_m^{(2)}=\begin{cases}
        -\frac{1}{2}\alpha'\hat{\Delta}_1^{(1)}+\tilde{\lambda}_0^{(2)},&\quad m=0,\\[0.7ex]
        \frac{1}{4}\alpha'\hat{\Delta}_1^{(1)}+\frac{1}{4}\alpha'-\frac{1}{2}\tilde{\lambda}_0^{(2)},&\quad m=1,\\[0.7ex]
        -\frac{(-4)^{1-m}\sqrt{\pi}\Gamma(m-1)}{2\Gamma(m-\frac{1}{2})}\alpha'\hat{\Delta}_1^{(1)},&\quad m\geq2,
    \end{cases}
\end{split}
\end{equation}
The image symmetry is absent as both even and odd $m$ appear in the BOE.
Here, $\hat{\Delta}_1^{(1)}$ and $\tilde{\lambda}_0^{(2)}$ are the two only unknown inputs.

At the boundary Potts/Yang-Lee fixed point, the anomalous dimensions of bulk operators, namely $\Delta_\phi^{(2)}$, $\Delta_0^{(2)}$, and $\Delta_1^{(1)}$, remain the same as the Wilson-Fisher point given in Eq.~\eqref{eq:WF_2}, while the anomalous dimensions of boundary operators, $\hat{\Delta}_0^{(2)}$ and $\hat{\Delta}_1^{(1)}$, and the OPE coefficient $\tilde{\lambda}_0^{(2)}$ can differ.
Additional calculations are therefore required.
It would be interesting to evaluate them by the perturbative RG calculation in the future.
%Even with only a finite set of RG data, the analytic bootstrap developed above produces infinitely many OPE and BOE coefficients.
%In this setting, the only new unknown inputs are $\hat{\Delta}_0^{(2)}$, $\hat{\Delta}_1^{(1)}$, and $\tilde{\lambda}_0^{(2)}$.

\subsection{Perturbative RG calculation}
\label{sec:RG of the bulk-phi4/boundary-phi3 theory}

In this section, we perform the RG study for the $O(N)$ BCFT supplemented by a boundary interaction that preserves $S_{N+1}$ symmetry. 
We classify distinct fixed points in $d=4-\epsilon$. 
We also evaluate the OPE coefficient $\tilde{\lambda}_0^{(1)}$, which provides all necessary information for the bootstrap solution at $\mathcal O(\epsilon)$ order.

\subsubsection{RG calculation for the $O(N)$ theory with cubic boundary interactions}
\label{sec:RG calculation}

We consider an $O(N)$ theory in a $d$-dimensional semi-infinite space $\mathbb{R}_+^d$ with a boundary at $z=0$.
In the bulk, the $N$-component field, denoted as $\phi_i$, $i=1,...,N$, transforms in the usual way under the $O(N)$ group.  
On the boundary, we consider an $S_{N+1}$ invariant cubic interaction, where $\phi_i$ fulfills the $N$-dimensional irreducible representation of the permutation group $S_{N+1}$.
The theory is governed by the following action,
%~\cite{diehl1991surface}
\begin{eqnarray}
\label{eq:d_4_bulk_phi4_boundary_phi3}
    S&=&\int_{\mathbb{R}_+^d} {\rm d}^{d-1}r\ {\rm d}z\left[ \frac{1}{2}(\nabla \phi_i)^2 %+\frac{\tau_0}{2}\phi_i^2
    +\frac{u_0}{4!}(\phi_i^2)^2\right.\\ %-h_0\phi_i
    &&\qquad\qquad\qquad \left.+\delta(z)\left(\frac{c_0}{2}\phi_i^2%-h_{1,0}\phi_i
    +\frac{w_0}{3!} d_{ijk}\phi_i\phi_j\phi_k\right) \right]\,, \nonumber
\end{eqnarray}
where $u_0$ and $w_0$ is the bulk and boundary interaction strength, respectively, and $c_0$ is the boundary mass. 
%We consider the $\phi_i$ transform in the bulk under the $O(N)$ group in usual way, more crucially, it fulfills the $N$ dimensional irreducible representation of the permutation group $S_{N+1}$ in the boundary. 
%There are two classes of boundary interactions. 
%One sets $d_{ijk}=1$ with a single scalar field $\phi$, which connects to the usual Lee-Yang fixed point.
The cubic boundary interaction is characterized by a third rank tensor, $d_{ijk}$ that satisfies~\cite{de1981critical}
\begin{equation}
\label{eq:interaction_dijk}
    d_{ijk}=\sum_{\alpha=1}^{N+1}e_i^\alpha e_j^\alpha e_k^\alpha\,,
\end{equation}
where $N+1$ vectors $e^\alpha_i$ in $N$ dimensions obey
\begin{equation}
\label{eq:condition_e}
\begin{split}
    &\sum_{\alpha=1}^{N+1}e_i^\alpha=0\,,\qquad \sum_{\alpha=1}^{N+1}e_i^\alpha e_j^\alpha=(N+1)\delta_{ij}\,,\\
    &\sum_{i=1}^N e_i^\alpha e_i^\beta=(N+1 )\delta^{\alpha\beta}-1\,.
\end{split}
\end{equation}
This construction is related to the $(N+1)$-state Potts model~\footnote{For $N=1$, the cubic tensor $d_{ijk}$ vanishes, and the Potts model is identical to the Ising model. 
%Thus, the leading allowed interaction is quartic, which is consistent with Ising criticality being governed by a $\phi^4$ (not $\phi^3$) fixed point. 
Consequently, the Yang-Lee edge singularity is not obtained by simply setting $N=1$ in the Potts $\phi^3$ theory.}. 
The theory Eq.~\eqref{eq:d_4_bulk_phi4_boundary_phi3}, can be realized in an $O(N)$ vector lattice model with a boundary field $e_i^{\alpha}$ that reduces the symmetry to $S_{N+1}$, as will be given in Eq.~\eqref{eq:lattice_model}.

\begin{figure}
    \centering
    \subfigure[]{\includegraphics[width=0.33\textwidth]{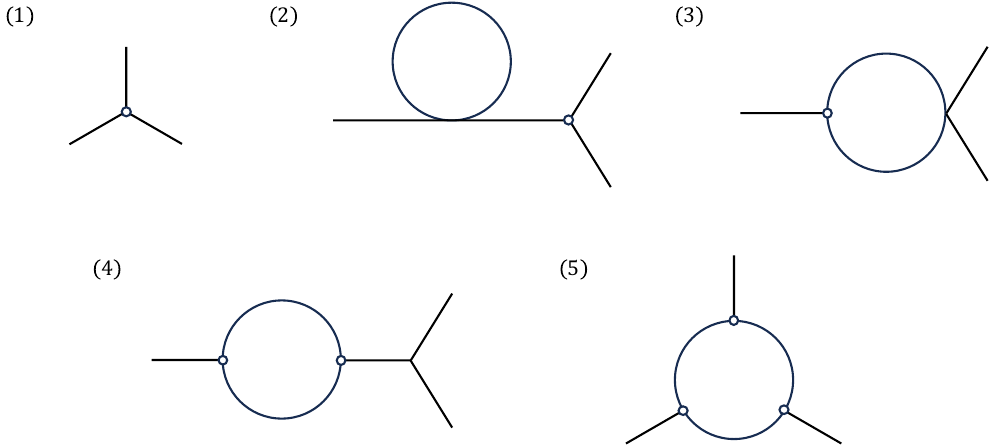}}\qquad
    \subfigure[]{\includegraphics[width=0.105\textwidth]{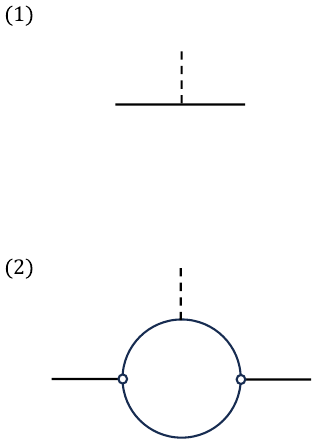}}
    \caption{Feynman diagrams for (a) $\langle\phi_i\phi_j\phi_k\rangle$ and (b) $\langle\phi_i\phi_l\phi^2\rangle$, where simple vertices and empty circles correspond to the bulk $\phi^4$-type vertex and the boundary qubic interaction vertex. 
    A vertex with a dashed line indicates the boundary operator $\phi^2$.}
    \label{fig:Feynman_diagram_w0}
\end{figure}

%Near the fixed point we treat $u_0$ and $w_0$ as tuning parameters.
The RG equation of the bulk coupling $u_0$ is well known~\cite{Domb1986PHASE,amit2005field}:
\begin{equation}
\label{eq:RG_phi4_d4_u0}
    \beta(u)=-\epsilon u+\frac{N+8}{3}u^2-\frac{3N+14}{3}u^3+\mathcal{O}(u^4)\,,
\end{equation}
which yields the Wilson-Fisher fixed point 
\begin{equation}
\label{eq:fixed_point_u}
    u^*=\frac{3}{N+8}\epsilon+\frac{9(3N+14)}{(N+8)^3}\epsilon^2+\mathcal{O}(\epsilon^3)\,.
\end{equation}
In what follows, we compute the RG flow of the boundary coupling parameter $w_0$.
Under the Neumann boundary condition, where we set $c_0=0$, the Green's function is
\begin{equation}
\label{eq:phi4_neumann_GF_momentum}
    \langle\phi_i(p,z_1)\phi_i(-p,z_2)\rangle_0=\frac{1}{2p}\left(e^{-p|z_1-z_2|}+e^{-p(z_1+z_2)}\right)\,.
\end{equation}

Consider the three-point correlation function $\langle\phi_i(p_1,z_1)\phi_j(p_1,z_2)\phi_j(p_3,z_3)\rangle$ with $\sum_i p_i=0$, the relevant Feynman diagrams are shown in Fig.~\ref{fig:Feynman_diagram_w0} (a). 
The diagram (4) is finite, and the divergence from diagram (2), (3), (5) are listed below:
% \textcolor{red}{$S_d$ not depend on $d$?} \textcolor{blue}{Yes, $A_d$ is similar to usual $S_d$, but $S_d$ is just used to match the convention in other reference.}
%Diagram (2) gives
% \begin{equation}
% \label{eq:Feynman_diagram_integral_2}
% \begin{split}
%     I_2=&\frac{e^{-p_2 z_2}}{p_2}\frac{e^{-p_3 z_3}}{p_3}\int \frac{{\rm d}^{d-1}k}{(2\pi)^{d-1}}{\rm d}z \frac{1}{2p_1}(e^{-p_1|z-z_1|}+e^{-p_1(z+z_1)})\frac{e^{-p_1 z}}{p_1}\frac{1+e^{-2kz}}{2k}\\=&\frac{e^{-p_2 z_2}}{p_2}\frac{e^{-p_3 z_3}}{p_3}\int {\rm d}z \frac{1}{2p_1}(e^{-p_1|z-z_1|}+e^{-p_1(z+z_1)})\frac{e^{-p_1 z}}{p_1}S_4\ z^{2-d}
%     =-\frac{S_4}{\epsilon} \prod_{i=1}^3 \frac{1}{p_i}e^{-p_i z_i}\,,
% \end{split}
% \end{equation}
\begin{equation}
\label{eq:Feynman_diagram_integral_2}
\begin{split}
    I_2&=-\frac{S_d}{\epsilon} \prod_{i=1}^3 \frac{1}{p_i}e^{-p_i z_i}\,, \\
    I_3&=\frac{4S_d}{\epsilon} \prod_{i=1}^3 \frac{1}{p_i}e^{-p_i z_i} \,, \\
    I_5&=\frac{8S_d}{\epsilon} \prod_{i=1}^3 \frac{1}{p_i}e^{-p_i z_i}\,.
\end{split}
\end{equation}
where $S_d=\frac{1}{(4\pi)^{d/2}}$. %, with $S_4=\frac{1}{16\pi^2}$. 
%With the nontrivial $d_{ijk}$ introduced above, 
Each diagram also acquires additional factors from the tensor $d_{ijk}$.
The factors $C_i$ are given by
\begin{eqnarray}
\label{eq:factor_dijk_N}
    &&  C_2\delta_{kj}=\sum_l F_{kllj}\,, \quad \quad\qquad C_2=\frac{N+2}{3}\,, \nonumber\\
    && C_3 d_{ijk}=\sum_{lm}d_{ilm}F_{lmjk}\,, \qquad C_3=\frac{2}{3}\,,\\
    && C_5 d_{ijk}=\sum_{lmn}d_{ilm}d_{jmn}d_{knl}\,, \quad C_5 = (N+1)^2(N-2)\,, \nonumber
\end{eqnarray}
%\begin{equation}
%\label{eq:factor_dijk_N}
%\begin{split}
    %& C_2=\frac{\sum_l F_{kllj}}{\delta_{kj}}=\frac{N+2}{3}\,,\quad
    %C_3=\frac{\sum_{lm}d_{ilm}F_{lmjk}}{d_{ijk}}=\frac{2}{3}\,,\\
    %&C_5=\frac{\sum_{lmn}d_{ilm}d_{jmn}d_{knl}}{d_{ijk}}=(N+1)^2(N-2)\,,
%\end{split}
%\end{equation}
where $F_{ijkl}=\frac{1}{3}(\delta_{ij}\delta_{kl}+\delta_{ik}\delta_{jl}+\delta_{il}\delta_{jk})$ denotes the coupling structure of the bulk interaction.
Therefore, the corresponding one-loop three-point function becomes
\begin{eqnarray}
\label{eq:correlation_function_d_1_add_factor}
    &&\langle\phi_i(p_1,z_1)\phi_j(p_1,z_2)\phi_j(p_3,z_3)\rangle_{\rm bare}=-w_0\left(\prod_{i=1}^3 \frac{1}{p_i}e^{-p_i z_i}\right) \nonumber\\
    &&\qquad\quad\times\left(1+u_0\frac{S_4}{\epsilon}(\frac{3}{2}C_2-6C_3)+w_0^2\frac{8S_4}{\epsilon}C_5\right)\,.
\end{eqnarray}
Note that the RG factor $Z_\phi$ is trivial at one loop. 
We define $Z_w \mu^{\epsilon/2} w=\sqrt{S_d} w_0$ and $u_0 S_d=\mu^\epsilon Z_u u$ with $Z_u=1+\frac{N+8}{3\epsilon}u+(\frac{(N+8)^2}{9\epsilon^2}-\frac{3N+14}{6\epsilon})u^2+\mathcal{O}(u^3)$.
Cancelling the divergence in the three-point function yields %$\langle\phi_i(p_1,z_1)\phi_j(p_1,z_2)\phi_j(p_3,z_3)\rangle_{\rm bare}$ yields
\begin{equation}
\label{eq:Zw_phi4_phi3}
    Z_w=1-u\left(\frac{3}{2}C_2-6C_3\right)\epsilon^{-1}-w^2\cdot8C_5\epsilon^{-1}\,,
\end{equation}
and the corresponding RG equation for $w$
\begin{equation}
\label{eq:RG_flow_w}
    \beta(w)=-w\left(\frac{\epsilon}{2}+(\frac{3}{2}C_2-6C_3)u+8C_5 w^2\right)\,.
\end{equation}
%Setting $C_2=C_3=C_5=1$ reproduces the known result in Ref.~\cite{diehl1991surface}, and nontrivial $C_i$ give the RG flow for the bulk $\phi^4$ theory with a different boundary interaction.

Before we discuss the fixed points, it's worth noting that by setting $C_2=C_3=C_5=1$, which corresponds to a single scalar field $\phi$ with $d_{iii}=1$, our results
reduce to a boundary Yang-Lee type interaction explored in Ref.~\cite{diehl1991surface}, which gives three nontrivial fixed points, the special fixed point $(u^*,w^*)=(\frac{\epsilon}{3},0)$, the long-range Yang-Lee fixed point $(u^*,w^*)=(0,\pm\frac{\rm i}{4}\sqrt{\epsilon})$, and a new fixed point (boundary Yang-Lee fixed point) $(u^*,w^*)=(\frac{\epsilon}{3},\pm\sqrt{\frac{\epsilon}{8}})$~\footnote{The special fixed point is stable; the Yang-Lee fixed point is attractive along the boundary coupling but unstable against the bulk coupling; and the new fixed point is a saddle with one attractive and one repulsive direction.}.

Now, combining Eqs.~\eqref{eq:RG_phi4_d4_u0} (truncated at $\mathcal{O}(u^2)$) and \eqref{eq:RG_flow_w}, we analyze the RG flow as a function of $N$.
Setting $\beta(u)=\beta(w)=0$ yields four distinct fixed points.
Since for any fixed point $(u^*,w^*)$, the pair $(u^*,-w^*)$ is equivalent, we restrict to $w>0$. 
The fixed points are 
\begin{eqnarray}
\label{eq:fixed_point_RG_dijk}
    (u^*,w^*)_{\rm G}&=&\left(0,0\right)\,, \\
    (u^*,w^*)_{\rm sp}&=&\left(\frac{3\epsilon}{N+8},0\right)\,, \nonumber\\
    (u^*,w^*)_{\rm lrP}&=&\left(0,\frac{1}{4}\sqrt{\frac{\epsilon}{(N+1)^2(2-N)}}\right)\,, \nonumber\\
    (u^*,w^*)_{\rm bP}&=&\left(\frac{3\epsilon}{N+8},\sqrt{\frac{(5-2N)\epsilon}{8(N+8)(N+1)^2(N-2)}}\right)\,. \nonumber
\end{eqnarray}
The first one, $(u^*,w^*)_{\rm G}$, is the Gaussian fixed point.
The second one, $(u^*,w^*)_{\rm sp}$, denotes the special transition point without boundary interactions.
The third one, $(u^*,w^*)_{\rm lrP}$, corresponds to a long-range Potts fixed point with a free bulk.
The last one, $(u^*,w^*)_{\rm bP}$, is a boundary Potts fixed point with nonvanishing bulk and boundary interactions. 
These results imply two critical values of $N$, namely $N_{c,1}=2$ and $N_{c,2}=\frac{5}{2}$.
In particular, for $N=2$ the coefficient $C_5$ in Eq.~\eqref{eq:RG_flow_w} vanishes. 
Consequently, any nontrivial fixed point $w^*$, if present, can only be resolved at the next order, i.e., at two loops.
In the following, we plot the RG flow for both real and imaginary $w$ at representative values of $N$. %, since for some $N$ the fixed points in $w$ are imaginary.

\begin{figure}
    \centering
    \subfigure[]{\includegraphics[width=0.23\textwidth]{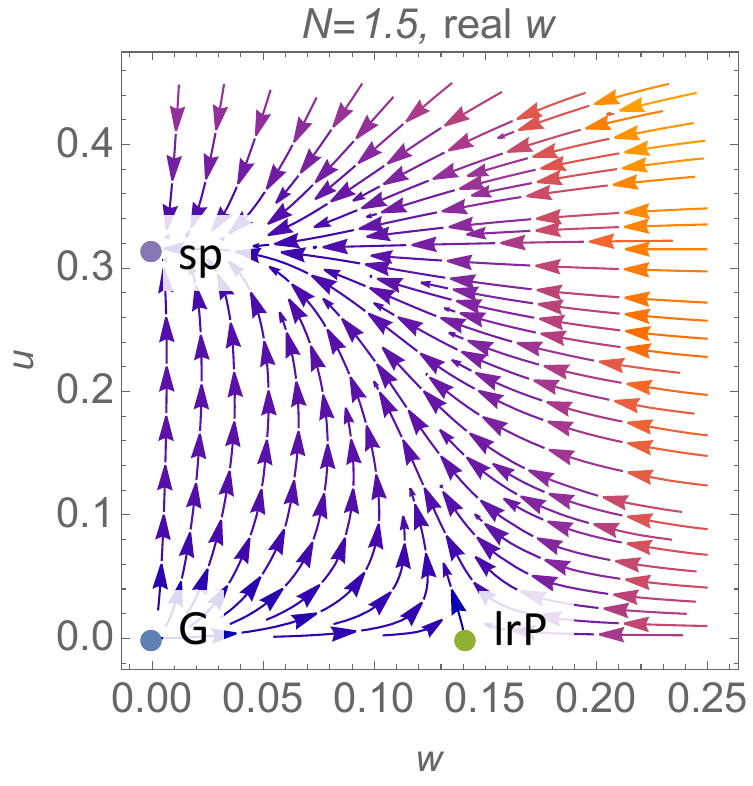}}
    \subfigure[]{\includegraphics[width=0.23\textwidth]{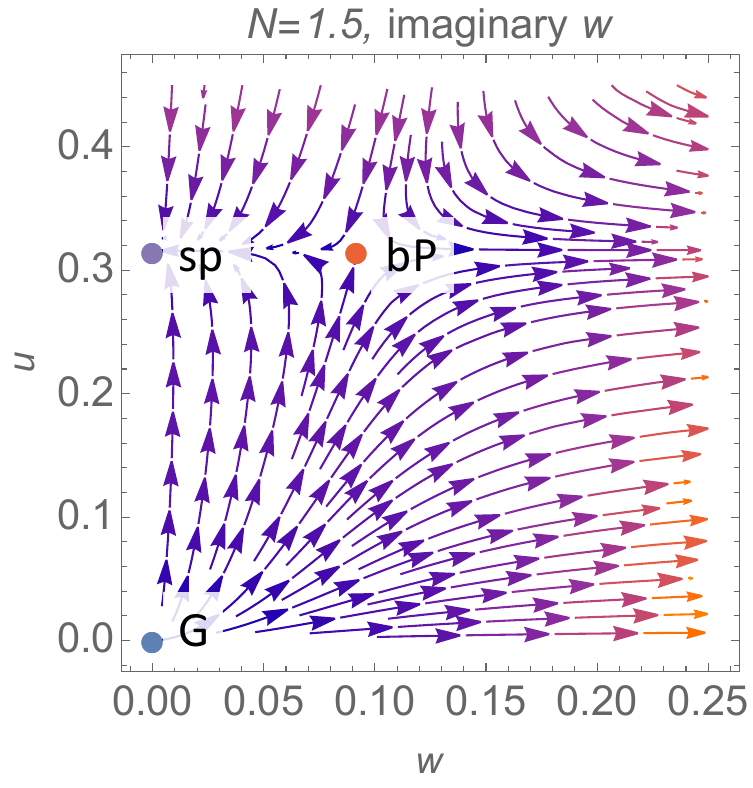}}
    \subfigure[]{\includegraphics[width=0.23\textwidth]{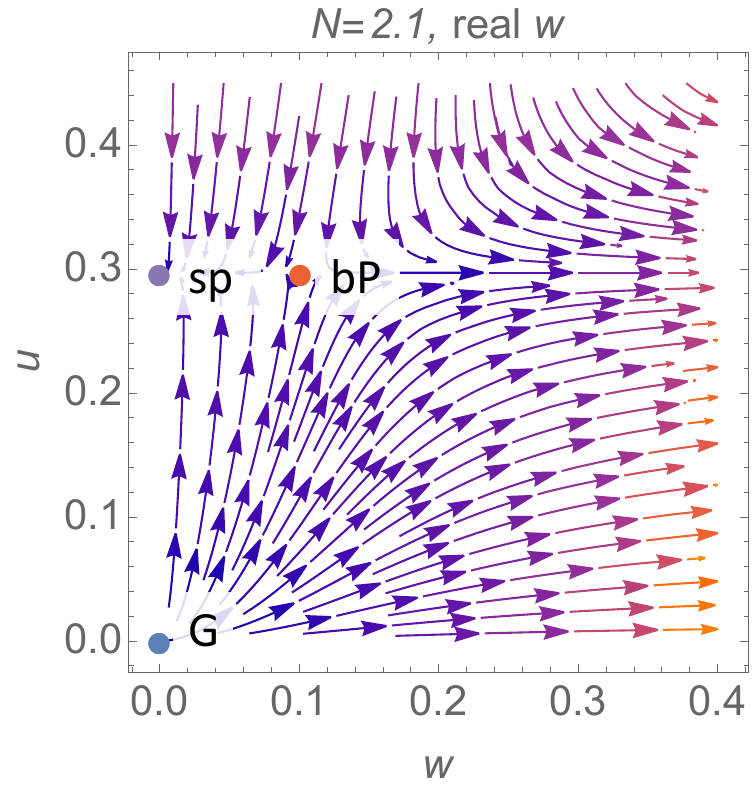}}
    \subfigure[]{\includegraphics[width=0.23\textwidth]{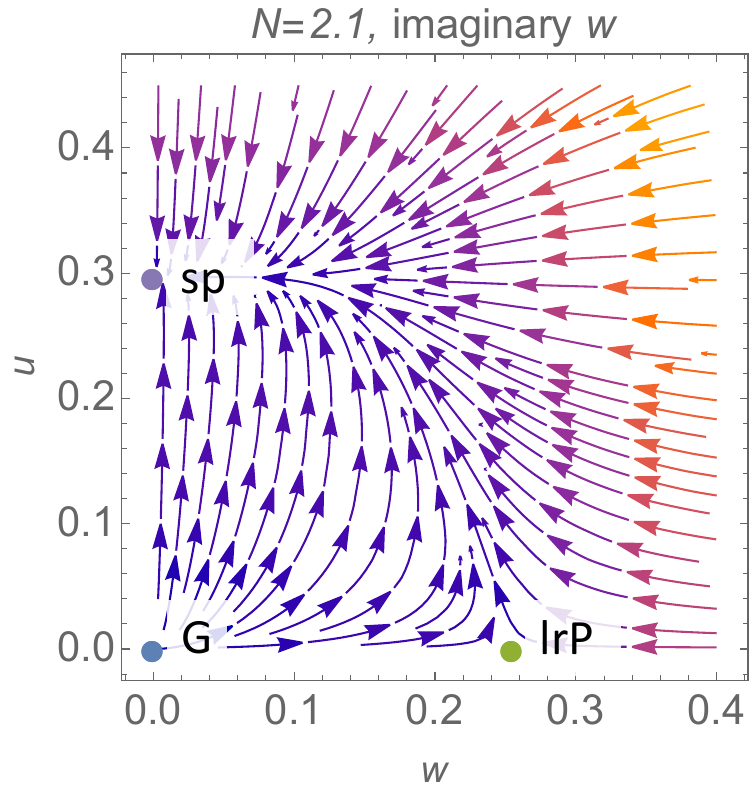}}
    \subfigure[]{\includegraphics[width=0.23\textwidth]{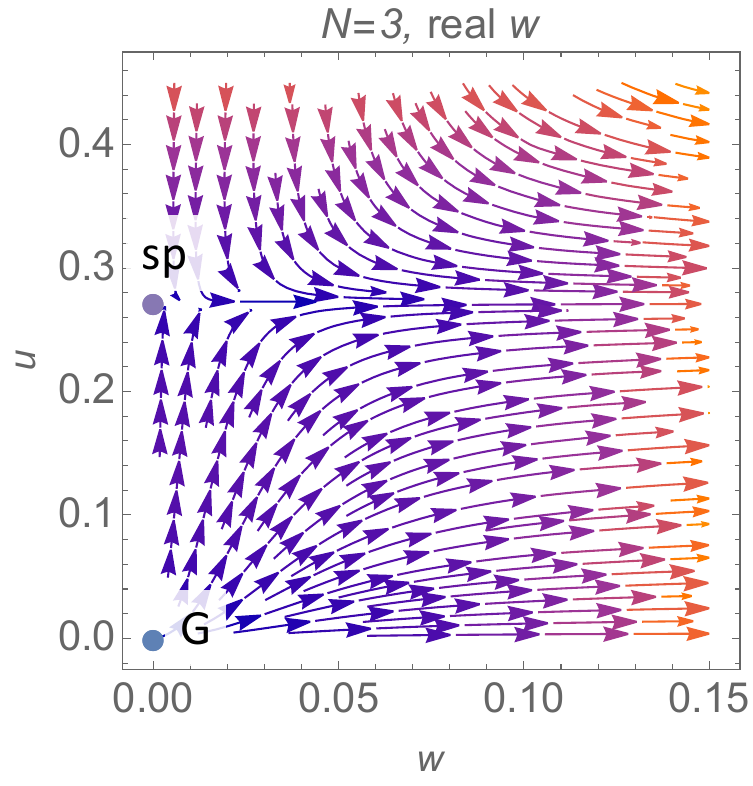}}
    \subfigure[]{\includegraphics[width=0.23\textwidth]{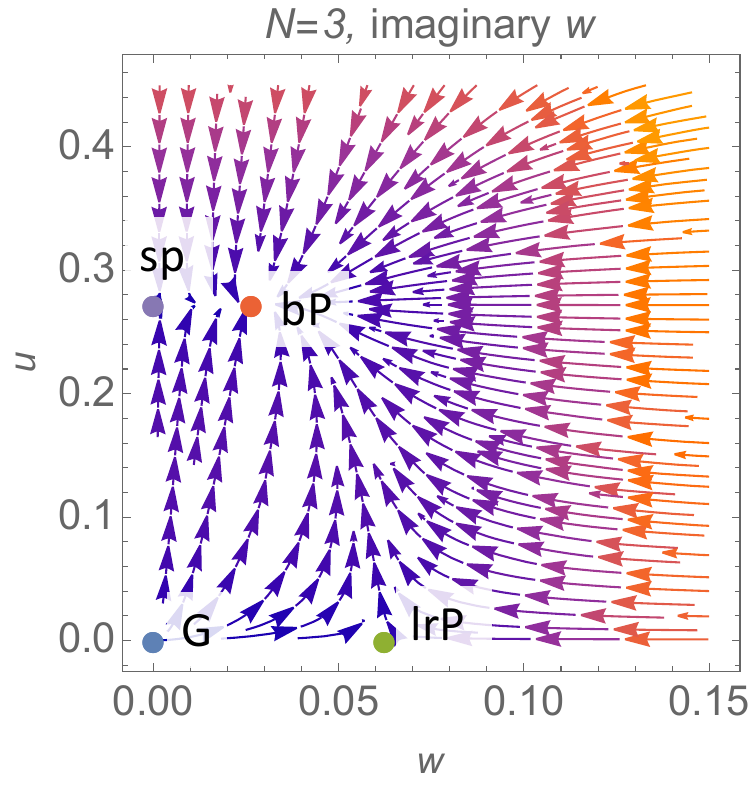}}
    \caption{RG flow phase diagrams for $N=1.5,\,2.1,\,3$; the left column shows real $w$, and the right column shows imaginary $w$.
    We set $\epsilon=1$. 
    Gaussian (G), special (sp), long-range Potts (lrP), and boundary Potts (bP) fixed points are indicated by blue, purple, green, and red dots, respectively.}
    \label{fig:RG_flow}
\end{figure}

As shown in Fig.~\ref{fig:RG_flow}, we plot the RG flow with $\epsilon=1$ for $N=1.5,\,2.1,\,3$ for both real and imaginary $w$, and indicate Gaussian, special, long-range Potts, and boundary Potts fixed points with blue, purple, green, and red dots. 
(i) For $N<N_{c,1}$, shown in panels (a) and (b), there is one real stable fixed point $(u^*,w^*)_{\rm sp}$, two real unstable fixed points $(u^*,w^*)_{\rm G}$ and $(u^*,w^*)_{\rm lrP}$, and one imaginary unstable fixed point $(u^*,w^*)_{\rm bP}$. 
(ii) For $N_{c,1}<N<N_{c,2}$, shown in panels (c) and (d), there is one real stable fixed point $(u^*,w^*)_{\rm sp}$, two real unstable fixed points $(u^*,w^*)_{\rm G}$ and $(u^*,w^*)_{\rm bP}$, and one imaginary unstable fixed point $(u^*,w^*)_{\rm lrP}$. 
(iii) For $N>N_{c,2}$, shown in panels (e) and (f), there are two real unstable fixed points $(u^*,w^*)_{\rm G}$ and $(u^*,w^*)_{\rm sp}$, one imaginary unstable fixed point $(u^*,w^*)_{\rm lrP}$, and one imaginary stable fixed point $(u^*,w^*)_{\rm bP}$.
We summarize the results in Table~\ref{tab:phase diagram}. 
The special fixed point is stable only for $N<N_{c,2}$. 
The boundary Potts fixed point $(u^*,w^*)_{\rm bP}$ is unstable for $N<N_{c,1}$ with imaginary $w^*$ and for $N_{c,1}<N<N_{c,2}$ with real $w^*$, while it becomes stable for $N>N_{c,2}$ with imaginary $w^*$.
It is interesting to also note that while the long-range Potts fixed point is unstable once the bulk interaction is turned on, it remains stable under the boundary interaction, with either real or imaginary $w^*$. 

\begin{figure}
    \centering
    \subfigure[]{\includegraphics[width=0.42\textwidth]{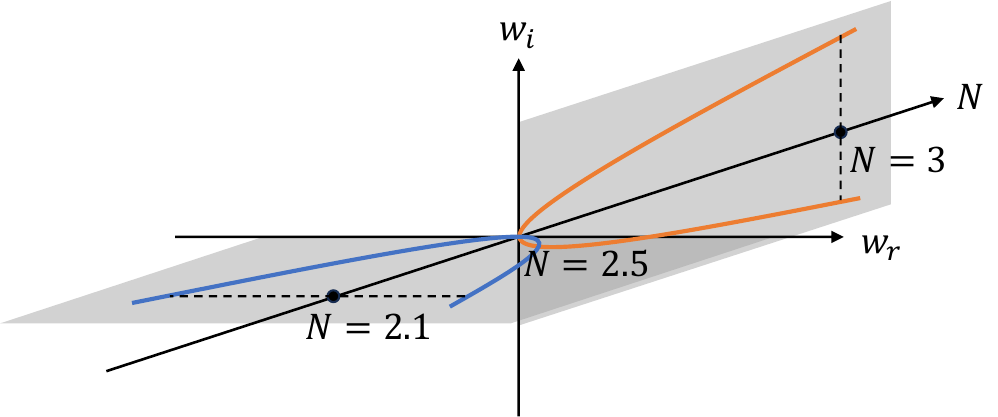}}
    \subfigure[]{\includegraphics[width=0.23\textwidth]{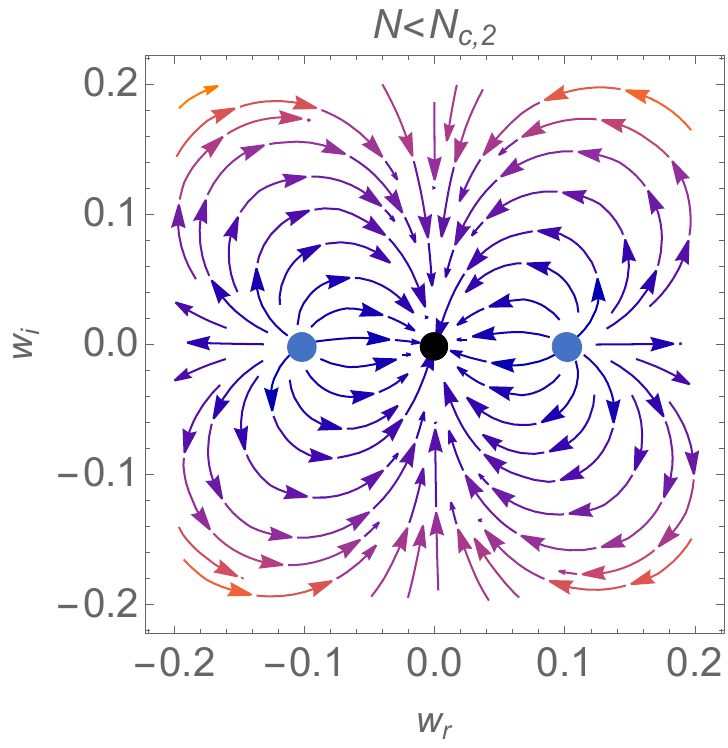}}
    \subfigure[]{\includegraphics[width=0.239\textwidth]{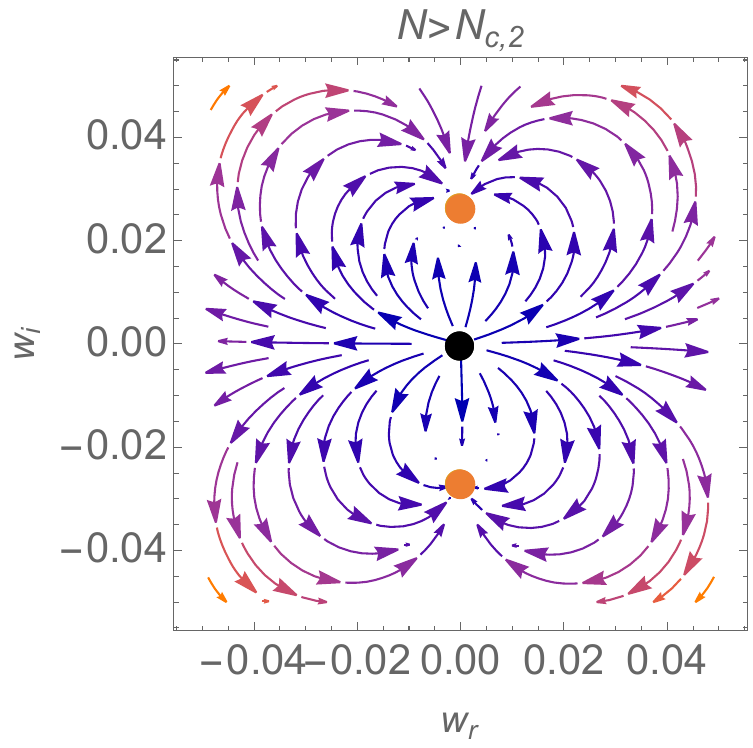}}
    \caption{(a) Collision and subsequent separation of the special and boundary Potts fixed points near $N = N_{c,2} = 2.5$. 
    (b), (c) RG flow phase diagrams for $N = 2.1$ and $N = 3.0$, respectively, in the complex plane of $w = w_r + \mathrm{i} w_i$, at fixed $u = u_{\mathrm{sp}}^* = \dfrac{3\epsilon}{N + 8}$ with $\epsilon = 1$. 
    The special fixed point is indicated by black dots, while the boundary Potts fixed point for $N < N_{c,2}$ ($N > N_{c,2}$) is shown as blue (orange) dots.}
    \label{fig:RG_flow_along_N}
\end{figure}

When $N < N_{c,2}$, the special transition remains stable and controls the critical behavior separating the ordinary and extraordinary transitions. 
At this special boundary fixed point, an emergent $O(N)$ symmetry appears, while the Potts anisotropy is dangerously irrelevant.
As $N$ increases beyond
$N_{c,2}$, the boundary Potts fixed point collides with the special fixed point and subsequently moves into the complex plane, rendering the special fixed point unstable. 
The resulting stable boundary Potts fixed point possesses an imaginary coupling $w^*$ corresponding to a non-unitary BCFT, as shown in Fig.~\ref{fig:RG_flow_along_N}.
At this point, the model also features a composite symmetry of $i \rightarrow -i $ and $\phi \rightarrow - \phi$. 
We therefore conjecture that, for real values of the microscopic parameters, the transition between the ordinary and extraordinary boundary phases becomes first-order–like on the surface.
Importantly, this first-order character is confined to the surface, while the bulk remains governed by a unitary critical theory.

\begin{table}[t]
  \centering
  {\renewcommand{\arraystretch}{1.3}
  \begin{tabular}{l|ccr} % l=left, c=center, r=right
    \toprule
     & stable fixed point & unstable fixed point \\\hline
    % \midrule
    \multirow{2}*{$N<N_{c,1}$}  & special & long-range Potts \\ \cline{2-3}
     & N/A & boundary Potts \\\hline
    \multirow{2}*{$N_{c,1}<N<N_{c,2}$}   & special & boundary Potts \\ \cline{2-3}
      &  N/A & long-range Potts \\\hline
    \multirow{2}*{$N>N_{c,2}$}   & N/A & special \\ \cline{2-3}
      & boundary Potts & long-range Potts \\
    \bottomrule
  \end{tabular}}
  \caption{The stable and unstable fixed points for the $O(N)$ theory with boundary $S_{N+1}$ invariant coupling.
  For each case of $N$, the first and second lines are the fixed points with real and imaginal $w$. 
  The Gaussian fixed point is unstable and is not included in the table. }
  \label{tab:phase diagram}
\end{table}

Here are several remarks.
(i) Non-unitary BCFTs can be realized in open quantum systems by subjecting only the boundary to Lindbladian dynamics, while the bulk evolves unitarily~\cite{Prosen2011Exact,Wang2023Scale}.
(ii) In contrast to standard numerical bootstrap methods, which rely on reflection positivity and the positivity of OPE/BOE coefficients and therefore struggle with non-unitary theories, our analytic bootstrap framework does not require positivity and can be applied directly to solve such cases.

We now calculate the surface critical exponents. 
The boundary operator $\hat{\phi}$ has the same anomalous dimension with and without boundary interaction at one loop~\cite{diehl1991surface}. 
We next compute the anomalous dimension of the boundary composite operator $\frac{1}{2}\phi^2=\frac{1}{2}\sum_{i=1}^N \phi_i^2$, which acquires a nontrivial contribution from the boundary interaction. 
The boundary contribution is shown in Fig.~\ref{fig:Feynman_diagram_w0}~(b). 
The symmetry factor for this diagram is 
\begin{equation}
    C_0 \delta_{il}=\sum_{j,k}d_{ijk}d_{ljk}\,,\quad C_0=(N+1)^2(N-1)\,.
\end{equation} 
Combining this with the bulk contribution, the corresponding normalization factor $Z_c$ is
\begin{equation}
\label{eq:RG_phi2_Z_c}
\begin{split}
    Z_c=&1+\frac{N+2}{3\epsilon}u+\left[\frac{(N+2)(N+5)}{9\epsilon^2}\right.\\
    &\qquad\left.+\frac{(N+2)(1-4\pi^2)}{36\epsilon}\right]u^2-C_0 w^2\frac{8}{\epsilon}\,,
\end{split}
\end{equation}
leading to the corresponding anomalous dimension~\footnote{Setting $C_5'=1$ for the Yang-Lee type interaction, this reproduces the results in Ref.~\cite{diehl1991surface}.}:
\begin{equation}
\label{eq:eta_c_general_n}
    \eta_c=-\frac{N+2}{3}\left(u+\frac{1-4\pi^2}{6}u^2\right)+8C_0 w^2\,.
\end{equation}
Then with the fixed point values given in Eq.~\eqref{eq:fixed_point_RG_dijk}, we can get the anomlous dimensions for each of these universality class. 

\subsubsection{BOE coefficients at the long-range Potts/Yang-Lee fixed point}
\label{sec:BOE coefficients for bulk-free/boundary-phi3 theories}

%With the RG results above, 
We evaluate the OPE coefficient $\tilde{\lambda}_0^{(1)}$ at the long-range Potts/Yang-Lee fixed point. 
%For a boundary interaction of a scalar field with $d_{ijk}=1$, the boundary Lee-Yang critical point is $(u^*,w^*)=(0, \pm{\rm i}\frac{1}{4}\sqrt{\epsilon})$. 
%For the vector field $\phi_i$ with $d_{ijk}$ given in Eq.~\eqref{eq:interaction_dijk}, the critical point is $(u^*,w^*)=(u^*,w^*)_{\rm bP}$ in Eq.~\eqref{eq:fixed_point_RG_dijk}. 
We compute the Green's function to the first order. 
The boundary contribution is similar to the diagram (5) in Fig.~\ref{fig:Feynman_diagram_w0}~(a) with one of the external line removed, and gives 
%given by the Feynman diagram shown in Fig.~\ref{fig:Feynman_diagram_w0}~(b), which evaluates to
% \begin{equation}
% \label{eq:2_pt_bulk_free_boundary_phi3}
% \begin{split}
%     I'=&\frac{e^{-p z_1}}{p} \frac{e^{-p z_2}}{p}\int \frac{{\rm d}^{d-1}k}{(2\pi)^{d-1}}\frac{1}{k|p-k|}=\frac{e^{-p (z_1+z_2)}}{p^2} \int {\rm d}\alpha \frac{\Gamma(1)}{\Gamma(\frac{1}{2})^2} \int \frac{{\rm d}^{d-1}k}{(2\pi)^{d-1}}\frac{[\alpha(1-\alpha)]^{-1/2}}{\alpha(p-k)^2+(1-\alpha)k^2}\\
%     =&\frac{e^{-p (z_1+z_2)}}{p} \frac{\Gamma(1)}{\Gamma(\frac{1}{2})^2} \frac{A_{d-1}}{(2\pi)^{d-1}}\frac{\Gamma(-\frac{d-3}{2})\Gamma(\frac{d-1}{2})}{2\Gamma(1)}=\frac{e^{-p (z_1+z_2)}}{p}\left(-\frac{1}{4\pi^2}\right)\,.
% \end{split}
% \end{equation}
\begin{equation}
\label{eq:2_pt_bulk_free_boundary_phi3}
\begin{split}
    I'=\frac{e^{-p (z_1+z_2)}}{p}\left(-\frac{1}{4\pi^2}\right)\,.
\end{split}
\end{equation}
Therefore, the total Green's function simplifies to
\begin{equation}
\label{eq:correction_2_pt_bulk_free_boundary_phi3}
\begin{split}
    \langle\phi_i\phi_j\rangle=\delta_{ij}\cdot\frac{1}{2p}\left(e^{-p|z_1-z_2|}+(1+\alpha'\epsilon)e^{-p(z_1+z_2)}\right)\,,
\end{split}
\end{equation}
where $\alpha'\epsilon=-\frac{C_5'}{4\pi^2}w^2$ and $C_5' \delta_{ij}=\sum_{kl}d_{ikl}d_{jkl}$, $C_5'=(N+1)^2(N-1)$~\footnote{Note that $C_5'=1$ for the boundary Yang-Lee type interaction.}.
%\begin{eqnarray}
%    && \alpha'\epsilon=-\frac{C_5'}{4\pi^2}w^2\,,\\
%    && C_5' \delta_{ij}=\sum_{kl}d_{ikl}d_{jkl}\,, \quad C_5'=(N+1)^2(N-1)\,. \nonumber
%\end{eqnarray}
%$\alpha'\epsilon=-\frac{C_5'}{4\pi^2}w^2$, and $C_0 \delta_{ij}=\sum_{kl}d_{ikl}d_{jkl}$, $C_0=(N+1)^2(N-1)$ from $d_{ijk}$ in Eq.~\eqref{eq:interaction_dijk}
By substituting $w=w^*$, the OPE coefficient 
\begin{eqnarray}
\alpha'=-\frac{N-1}{64\pi^2(2-N)}\,, \quad \alpha'=-\frac{1}{64\pi^2}\,,
\end{eqnarray}
for the long-range Potts fixed point and the long-range Yang-Lee fixed point, respectively. 
Therefore, the OPE coefficient $\tilde{\lambda}_0^{(1)}=\alpha'$.

\subsubsection{BOE coefficients at the boundary Potts/Yang-Lee fixed point}

For the BOE coefficient in the boundary Potts fixed point, $\tilde\lambda_0$, we first note that 
\begin{eqnarray}
    \tilde\lambda_0 = \lambda_0 a_0 = \lambda_0^{(0)}  a_0^{(0)} +  (\lambda_0^{(1)} a_0^{(0)} + \lambda_0^{(0)} a_0^{(1)}) \epsilon + \mathcal O(\epsilon^2)\,, \nonumber\\
\end{eqnarray}
which indicates $\tilde\lambda_0^{(1)} = \lambda_0^{(1)} a_0^{(0)} + \lambda_0^{(0)} a_0^{(1)}$. 
$\lambda_0$ is the bulk conformal data, that will not be affected by the boundary interactions. 
For the long-range Potts fixed point, $\lambda_0=\lambda_0^{(0)}$ while $a_0^{(1)}\neq0$, which gives $\tilde\lambda_0^{(1)}=\lambda_0^{(0)}a_{0,i}^{(1)}\equiv \alpha'$.
As shown previously, this contribution comes solely from the boundary interaction vertex, and the subscript $i$ in $a_{0,i}^{(1)}$ labels the boundary piece. 
Conversely, at the special transition without boundary interactions, we have $\tilde{\lambda}_{0}^{(1)}=\lambda_0^{(0)}a_{0,b}^{(1)}+\lambda_0^{(1)}a_0^{(0)}=\alpha$, where $a_{0,b}^{(1)}$ encodes the bulk-interaction correction and $\lambda_0^{(1)}$ is the bulk OPE correction. 

At the boundary Potts model with both bulk and boundary interactions present, the OPE correction becomes 
\begin{eqnarray}
\tilde{\lambda}_{0}^{(1)}&=&\lambda_0^{(0)}(\tilde{a}_{0,i}^{(1)}+a_{0,b}^{(1)})+\lambda_0^{(1)}a_0^{(0)} \nonumber \\
&=& \alpha + \lambda_0^{(0)}\tilde{a}_{0,i}^{(1)} \,.
\end{eqnarray}
Compared with the long-range Potts fixed point, it is easy to see that $\lambda_0^{(0)}\tilde{a}_{0,i}^{(1)}$ has the same form as $\lambda_0^{(0)} a_{0,i}^{(1)}$, but with different fixed point value, $w^*$.
This is because the Feynman diagram will be the same as in the long-range fixed point at one-loop order. 
Therefore, we have \begin{eqnarray} \beta &\equiv& \tilde{\lambda}_0^{(1)} = \alpha+\tilde{\alpha}'\, , \\
 \tilde{\alpha}'&=&-\frac{(N+1)^2(N-1)}{4\pi^2}\frac{(w^*_{\rm bP})^2}{\epsilon}=-\frac{(N-1)(5-2N)}{32\pi^2(N+8)(N-2)}\,,\nonumber
\end{eqnarray}
for the boundary Potts fixed point while $\tilde{\alpha}'=-\frac{1}{32\pi^2}$ for the boundary Yang-Lee fixed point.

\section{$d=3-\epsilon$ expansion}
\label{sec:d 3-epsilon expansion}

In this section we focus on the analytic bootstrap for $\langle\phi(x)\phi(y)\rangle$ in the $O(N)$ theory in $d=3-\epsilon$, with interacting boundaries. 
We fix the zeroth-order solution of the bootstrap equation by the Neumann boundary condition for boundary interactions, and it can have two nontrivial fixed points, the long-range fixed point with a free bulk and an interacting boundary with $\phi^4$ type couplings, and the special fixed point with both nonvanishing interactions in the bulk and the boundary~\cite{eisenriegler1988surface}. 
The special fixed point characterizes the critical phenomena separating the ordinary and extraordinary boundary phases~\cite{sun2025boundary} for the tricritical $O(N)$ model. 
The boundary $\phi^4$ coupling is necessarily generated under the RG flow, even if it is initially set to zero~\cite{eisenriegler1988surface}. 

In addition to the long-range fixed point and the special fixed point with interacting boundaries, we also consider the ordinary fixed point under the Dirichlet boundary condition for the tricritical $O(N)$ model. 
In the following, we solve the bootstrap equation to order $\mathcal{O}(\epsilon)$, which already produces infinitely many conformal blocks in $F(\xi)$.
Unlike the situation in $d=4-\epsilon$, we find that, all solutions do not respect the image symmetry.
%Therefore, we do not treat image symmetry as a general constraint in $d=3-\epsilon$.

\subsection{$\mathcal{O}(\epsilon)$ expansion with interacting boundaries}
\label{sec:O(epsilon) expansion with the effective Neumann boundary condition}

We begin with the zeroth-order solution under the Neumann boundary condition and follow the procedure in Fig.~\ref{fig:bootstrap_pipeline}.
Using the parameters in Eq.~\eqref{eq:0th_solution_general_bootstrap}, we compute the discontinuity of $G_i(\xi)-G_b(\xi)$, 
\begin{equation}
\label{eq:Disc_d_3_epsilon}
    \underset{\xi<-1}{\rm Disc}\ (G_i(\xi)-G_b(\xi))=-{\rm i}\pi\epsilon\left(A_1+A_2\sqrt{\frac{\xi}{1+\xi}}\right)\,,
\end{equation}
with 
\begin{equation}
\label{eq:parameters_A_d_3_epsilon}
    A_1=2\hat{\Delta}_{0}^{(1)}-2\Delta_\phi^{(1)}\,,\quad A_2=2\hat{\Delta}_{0}^{(1)}-2\Delta_\phi^{(1)}+\Delta_{0}^{(1)}\,.
\end{equation}
Substituting Eq.~\eqref{eq:Disc_d_3_epsilon} into Eq.~\eqref{eq:OPE_lambda_n}, we obtain the OPE coefficients 
\begin{equation}
\label{eq:OPE_d_3_epsilon}
    \tilde{\lambda}_n^{(1)}=\frac{\Gamma(n)\Gamma(n+\frac{1}{2})}{2^{\frac{5}{2}}(-1)^n \Gamma(2n-\frac{1}{2})}\begin{cases}
        (\frac{4\sqrt{2}}{3}A_1+2\sqrt{2}A_2)\,,& n=1\,,\\[0.7ex]
        2\sqrt{\frac{2}{\pi}}\frac{\Gamma(n-\frac{1}{2})}{\Gamma(n+1)}A_2\,,& n>1\,.
    \end{cases}
\end{equation}
Summing the conformal blocks in $H_b$ yields
\begin{equation}
\label{eq:sum_n_Hb_d_3_epsilon}
\begin{split}
    H_b(\xi)=&\epsilon\left(A_1\left(\sqrt{\frac{\xi}{1+\xi}}-{\rm arcsinh}{\sqrt{\xi}}\right)\right.\\
    &\qquad\qquad\left.-A_2\ \frac{1}{2}\sqrt{\frac{\xi}{1+\xi}}\log{(1+\xi)}\right)\,.
\end{split}
\end{equation}
%We have verified that it can reproduce the discontinuity in Eq.~\eqref{eq:Disc_d_3_epsilon}. 
Next we obtain $H_i=G_b+H_b-G_i$, 
\begin{eqnarray}
\label{eq:difference_Hi_xi_d_3}
    H_i(\xi)&=&\epsilon\left(B_1 \sqrt{\frac{\xi}{\xi+1}}\ {\rm arcsinh}{\sqrt{\xi}}+B_2\ {\rm arcsinh}{\sqrt{\xi}}\right. \nonumber\\
    &&\quad+B_3 \sqrt{\frac{\xi}{\xi+1}}+B_4+B_5 \sqrt{\frac{\xi}{\xi+1}}\log{\xi} \nonumber \\
    &&\quad\left.+B_6 \log{\xi}+B_7\sqrt{\frac{\xi}{1+\xi}}\log{(1+\xi)}\right) \,,
\end{eqnarray}
where
\begin{eqnarray}
\label{eq:parameters_B_d_3_epsilon}
    B_1&=&2\hat{\Delta}_0^{(1)}\,, \qquad \qquad \qquad B_2=-2B_6=2\Delta_\phi^{(1)}\,,\nonumber\\ B_3&=&-2\Delta_\phi^{(1)}+\tilde{\lambda}_{0}^{(1)}-(\log{4}-2)\hat{\Delta}_0^{(1)}-\frac{1}{2}\tilde{\mu}_0^{(1)}\,,\nonumber\\
    B_4&=&-\log{4}\cdot \hat{\Delta}_0^{(1)}-\frac{1}{2}\tilde{\mu}_0^{(1)}\,, \quad
    B_5=-\Delta_\phi^{(1)}+\frac{1}{2}\Delta_0^{(1)}\,, \nonumber\\
    B_7&=&-\hat{\Delta}_0^{(1)}+\Delta_\phi^{(1)}-\frac{1}{2}\Delta_0^{(1)}\,.
\end{eqnarray}
Inserting $H_i(\xi)$ into Eq.~\eqref{eq:BOE_mu_m} yields the BOE coefficients
\begin{equation}
\label{eq:BOE_d_3_epsilon}
\begin{split}
    \tilde{\mu}_m^{(1)}=\begin{cases}
        B_2',& m=0,\\[2.ex]
        -\frac{1}{2}B_1'+B_3'+\frac{1-\log{4}}{4}B_4'+\frac{1}{4}B_5',& m=1,\\[2.ex]
        -\frac{2^{-2m}}{m(m-1)}\left[2(1+(-1)^m(2m-1))B_3'\right.\\
        \left.\qquad\qquad\qquad\quad+B_4'-(2m-1)B_5'\right],& m>1,
    \end{cases}
\end{split}
\end{equation}
where 
\begin{equation}
\label{eq:parameters_Bp_d_3_epsilon}
\begin{split}
    B_1'=&-(1+\log{2})\Delta_\phi^{(1)}+\frac{\tilde{\lambda}_{0}^{(1)}}{2}+(1-\log{2})\hat{\Delta}_{0}^{(1)}\,,\\
    B_2'=&(-2+\log{4})(\Delta_\phi^{(1)}-\hat{\Delta}_{0}^{(1)})+\tilde{\lambda}_{0}^{(1)}\,,\\
    B_3'=&-\hat{\Delta}_{0}^{(1)}+\Delta_\phi^{(1)}-\frac{1}{2}\Delta_{0}^{(1)}\,,\\
    B_4'=&2\hat{\Delta}_0^{(1)}\,,\qquad\qquad
    B_5'=2\Delta_\phi^{(1)}\,.
\end{split}
\end{equation}
Here, we have imposed the constraint that the coefficient of the lowest boundary conformal block ($m=0$) inferred from $H_i$ vanishes. 
The independent inputs include $\Delta_\phi^{(1)},\Delta_0^{(1)},\hat{\Delta}_0^{(1)}$ and $\tilde\lambda_0^{(1)}$.

% \begin{equation}
% \label{eq:BOE_d_3_epsilon}
% \begin{split}
%     \tilde{\mu}_{0,c}^{(1)}&=B_1'+B_2'+\log{2}\ (B_4'+B_5')\,,\\
%     \tilde{\mu}_1^{(1)}&=-\frac{1}{2}B_1'+B_3'+\frac{1-\log{4}}{4}B_4'+\frac{1}{4}B_5'\,,\\
%     \tilde{\mu}_{m\geq2}^{(1)}&=-\frac{2^{-2m}}{m(m-1)}\left[2(1+(-1)^m(2m-1))B_3'+B_4'-(2m-1)B_5'\right]\,,
% \end{split}
% \end{equation}
% where 
% \begin{equation}
% \label{eq:parameters_Bp_d_3_epsilon}
% \begin{split}
%     B_1'=&-2\Delta_\phi^{(1)}+\tilde{\lambda}_{0}^{(1)}+(2-\log{4})  \hat{\Delta}_{0}^{(1)}-\frac{1}{2}\tilde{\mu}_{0}^{(1)}\,,\qquad B_2'=-\log{4}\ \hat{\Delta}_{0}^{(1)}-\frac{1}{2}\tilde{\mu}_{0}^{(1)}\,,\\
%     B_3'=&-\hat{\Delta}_{0}^{(1)}+\Delta_\phi^{(1)}-\frac{1}{2}\Delta_{0}^{(1)}\,,\qquad\qquad\qquad\quad\quad\quad\ \
%     B_4'=2\hat{\Delta}_0^{(1)}\,,\qquad\qquad\qquad B_5'=2\Delta_\phi^{(1)}\,.
% \end{split}
% \end{equation}
% The constraint $\tilde{\mu}_{0,c}^{(1)}=0$ implies
% \begin{equation}
% \label{eq:constraint_mu_0_d_3}
%     \tilde{\mu}_{0}^{(1)}=(-2+\log{4})(\Delta_\phi^{(1)}-\hat{\Delta}_{0}^{(1)})+\tilde{\lambda}_{0}^{(1)}\,.
% \end{equation}
% The next BOE coefficient is $\tilde{\mu}_{1}^{(1)}=\frac{1}{4}[-2\Delta_{0}^{(1)}+(8+\log{4})\Delta_\phi^{(1)}-(4+\log{4})\hat{\Delta}_{0}^{(1)}-\tilde{\lambda}_{0}^{(1)}]$ and the others can be obtained by substituting Eq.~\eqref{eq:constraint_mu_0_d_3}. 

We apply the image symmetry constraint to the results above.
%For $d=3-\epsilon$, image symmetry leads to Eq.~\eqref{eq:image_symm_Hi_odd_d}, and 
Under the Neumann boundary condition, it can hold only if all $m$ are even, which implies $\frac{H_i(\xi')}{\xi'}|_{\xi'=e^{\pm{\rm i}\pi}(\xi+1)}=e^{\mp{\rm i}\pi/2} \frac{H_i(\xi)}{\xi}$.
Taking $\xi'=e^{{\rm i}\pi}(\xi+1)$, the constraint on $H_i(\xi)$ gives $B_1=B_2=-2B_6=-2B_7$, $B_3=B_4$, and $B_5=0$.
Using Eq.~\eqref{eq:parameters_B_d_3_epsilon}, we obtain
\begin{equation}
\label{eq:image_symmetry_phi6_d_3}
    \Delta_\phi^{(1)}=\hat{\Delta}_{0}^{(1)}\,,\qquad \Delta_0^{(1)}=2\hat{\Delta}_{0}^{(1)}\,,\qquad \tilde{\lambda}_0^{(1)}=0\,.
\end{equation}
Substituting these into the expressions above, the OPE and BOE coefficients constrained by image symmetry are
\begin{equation}
\label{eq:OPE_BOE_d_3_phi_6_image_symm_N}
\begin{split}
    \tilde{\lambda}_1^{(1)}=-\Delta_\phi^{(1)}\,,\ \tilde{\lambda}_{n\geq2}^{(1)}=\frac{(-1)^n\Gamma(n-1/2)\Gamma(n+1/2)}{n\sqrt{\pi}\Gamma(2n-1/2)}\Delta_\phi^{(1)}\,,\\
    \tilde{\mu}_0^{(1)}=\tilde{\mu}_1^{(1)}=0\,,\ \tilde{\mu}_{m\geq2}^{(1)}=\frac{2^{1-2m}(2m-1)(1+(-1)^m)}{m(m-1)}\Delta_\phi^{(1)}\,,
\end{split}
\end{equation}
with a single paramete $\Delta_\phi^{(1)}$. 
It is natural to ask whether there exists a BCFT that possesses the image symmetry. %and hence obeys the anomalous-dimension constraints in Eq.~\eqref{eq:image_symmetry_phi6_d_3}.
But for the theory we discuss below, it turns out that the image symmetry is not preserved. 

We now discuss fixed points with boundary $\phi^4$ couplings, including the long-range $\phi^4$ fixed point with a free bulk and the special fixed point of the tricritical $O(N)$ model.
At the long-range $\phi^4$ fixed point, the scaling dimensions are all trivial, $\Delta_\phi=\Delta_0=\hat{\Delta}_0=0$, due to the same reason as discussed in the long-range Potts fixed point.
Actually, at order $\mathcal{O}(\epsilon)$, the OPE coefficient $\tilde{\lambda}_0^{(1)}=0$ is also trivial.

While the $\mathcal O(\epsilon)$ order is trivial, we can discuss the next order. 
The correlation function at $\mathcal O(\epsilon^2)$ reads
\begin{eqnarray}
\label{eq:2_pt_bulk_free_boundary_phi4_d_3}
    &&\langle\phi(x)\phi(x')\rangle=\frac{1}{\sqrt{4zz'}}\left(\frac{1}{\xi^{1/2}}+\frac{1+\tilde{\lambda}_0^{(2)}\epsilon^2}{(1+\xi)^{1/2}}\right)\\
    &=&\frac{\xi^{-1/2}}{(2z\cdot 2z')^{1/2}}\xi^{1/2}\left[(2+\tilde{\mu}_0^{(2)}\epsilon^2)f_i(\frac{1}{2},\xi)+\tilde{\mu}_1^{(2)}\epsilon^2 f_i(\frac{3}{2},\xi)\right]\,. \nonumber 
\end{eqnarray}
The first line is given by the Feynman diagram calculation, with the OPE coefficient $\tilde{\lambda}_0^{(2)}=-\frac{4(N+2)}{(N+8)^2}$~\cite{prochazka2020composite}. 
Then, by matching two sides in the second equation, we can obtain the BOE coefficients. 
In summary, the OPE and BOE coefficients for the long-range $\phi^4$ model are given by 
\begin{equation} 
\label{eq:OPE_BOE_d_3_boundary_phi_4_image_symm}
\begin{split}
& \tilde{\lambda}_0^{(2)}=-\frac{4(N+2)}{(N+8)^2}\,, \quad \tilde{\lambda}_{n\ge1}^{(2)} = 0\,,\\
& \tilde{\mu}_0^{(2)}=-4\tilde{\mu}_1^{(2)}=\tilde{\lambda}_0^{(2)}\,, \quad \tilde{\mu}_{m\geq2}^{(2)}=0\,.
\end{split}
\end{equation}
%$\tilde{\mu}_0^{(2)}=-4\tilde{\mu}_1^{(2)}=\tilde{\lambda}_0^{(2)}$ and $\tilde{\mu}_{m\geq2}^{(2)}=0$ %which are consistent with the order $\mathcal{O}(\epsilon)$ results in Eq.~\eqref{eq:OPE_BOE_d_3_boundary_phi_4_image_symm}.

As we mentioned in the beginning of this section, the special fixed point characterizes the critical phenomena separating the ordinary and extraordinary boundary phases~\cite{sun2025boundary} for the tricritical $O(N)$ model. 
The boundary $\phi^4$ coupling is necessarily generated under the RG flow, even if it is initially set to zero~\cite{eisenriegler1988surface}. 
Hence, it is unavoidable that the special fixed point of the tricritical $O(N)$ model at $d=3-\epsilon$ features interacting boundaries. 
At the special fixed point, the RG calculation gives $\Delta_\phi^{(1)}=\Delta_0^{(1)}=0$ and 
\begin{equation}
\label{eq:d_3_phi_6_Neumann_special_epsilon}
    \hat{\Delta}_0^{(1)}=-\frac{(N+2)(N+4)}{16(3N+22)} \,.
\end{equation}

Next, we argue that the remaining unknown parameter $\tilde{\lambda}_0^{(1)}=0$ at the special fixed point. 
Similar to the boundary Potts fixed point at $4-\epsilon$, the $\mathcal{O}(\epsilon)$ correction becomes $\tilde{\lambda}_{0}^{(1)}=\lambda_0^{(0)}(a_{0,i}^{(1)}+a_{0,b}^{(1)})+\lambda_0^{(1)}a_0^{(0)}$, with both bulk and boundary interactions present. %including the bulk and boundary interaction correction.
%we have $\tilde{\lambda}_{0}^{(1)}=(\lambda_0 a_0)^{(1)}$, where $\lambda_0$ is the bulk OPE coefficient for $\phi\times\phi\to\phi^2$ and $a_0$ is the normalization factor of $\langle\phi^2\rangle$.
%We write $\lambda_0=\lambda_0^{(0)}+\lambda_0^{(1)}\epsilon$ and $a_0=1+a_0^{(1)}\epsilon$, with the convention $a_0^{(0)}=1$ for the normalization of $\phi^2$.
%For a free bulk with a boundary interaction, $\lambda_0=\lambda_0^{(0)}$ while $a_0^{(1)}\neq0$.
From  Eq.~\eqref{eq:2_pt_bulk_free_boundary_phi4_d_3}, we know that $a_{0,i}^{(1)}=0$. % the contribution of the boundary interaction also vanished.
Moreover, due to the $\phi^6$ interaction in the bulk, there is no contribution at the one-loop order for $a_{0,b}^{(1)}$ and $\lambda_0^{(1)}$, which leads to $\tilde{\lambda}_{0}^{(1)}=0$. 
Hence, the corresponding OPE and BOE coefficients are
\begin{equation}
\label{eq:OPE_BOE_d_3_phi_6_free}
\begin{split}
    &\tilde{\lambda}_n^{(1)}=\begin{cases}
        -\frac{5}{3}\hat{\Delta}_0^{(1)},&\qquad n=1,\\[0.7ex]
        \frac{(-1)^n \Gamma(n-\frac{1}{2})\Gamma(n+\frac{1}{2})}{n\sqrt{\pi}\Gamma(2n-\frac{1}{2})}\hat{\Delta}_0^{(1)},&\qquad n>1.
    \end{cases}\\[0.5ex]
    &\tilde{\mu}_m^{(1)}=\begin{cases}
        (2-\log{4})\hat{\Delta}_0^{(1)},&\quad m=0,\\[0.7ex]
        -(1+\frac12 \log{2})\hat{\Delta}_0^{(1)},&\quad m=1,\\[0.7ex]
        \frac{(-1)^m 2^{1-2m}(2m-1)}{m(m-1)}\hat{\Delta}_0^{(1)},&\quad m>1,
    \end{cases}
\end{split}
\end{equation}
This shows explicitly that both even and odd boundary conformal blocks appear with nonzero BOE coefficients.

Finally, let's discuss the $\mathcal O(\epsilon^{1/2})$ correction for the tricritical $O(N)$ model. 
It has been shown in Ref.~\cite{eisenriegler1988surface} that the fixed point value of the special transition has the expansion of  $\sqrt{\epsilon}$. 
Hence, this would motivate an expansion in terms of $\sqrt \epsilon$ instead of $\epsilon$. 
%To present a uniform derivation, 
We can begin an $\mathcal{O}(\epsilon^{1/2})$ expansion by setting $d=3-\tilde{\epsilon}^{\,2}$ and expand all relevant conformal data to $\mathcal{O}(\tilde{\epsilon})$. 
Repeating the derivation outlined above, we find that the structure of the result is unchanged.
The BOE and OPE coefficients retain the same forms as in Eqs.~\eqref{eq:OPE_d_3_epsilon} and \eqref{eq:BOE_d_3_epsilon}, now with coefficients $\tilde{\lambda}_n^{(1/2)}$ and $\tilde{\mu}_m^{(1/2)}$.
The input anomalous dimensions are also corresponding to the order $\mathcal{O}(\epsilon^{1/2})$ correction.

It follows that the image-symmetry relations, Eqs.~\eqref{eq:image_symmetry_phi6_d_3} and \eqref{eq:OPE_BOE_d_3_phi_6_image_symm_N}, continue to hold at $\mathcal{O}(\epsilon^{1/2})$ with the obvious replacements.
Applied to the models discussed above, the long-range $\phi^4$ theory exhibits no nontrivial corrections at $\mathcal{O}(\epsilon^{1/2})$.
For an interacting bulk, we focus on the fixed point with a boundary interaction.
At this order the anomalous dimensions of the bulk and boundary primaries vanish, $\Delta_\phi^{(1/2)}=\Delta_{0}^{(1/2)}=\hat{\Delta}_{0}^{(1/2)}=0$, so the only potentially nonzero new coefficient is $\tilde{\lambda}_{0}^{(1/2)}$.
By the same argument used above for $\tilde{\lambda}_{0}^{(1)}$, we expect $\tilde{\lambda}_{0}^{(1/2)}$ to vanish as well.
Consequently, there are no $\mathcal{O}(\epsilon^{1/2})$ corrections to the conformal data considered here~\footnote{Note, however, that composite boundary operators such as $\hat{\phi}^2$ can acquire nontrivial $\mathcal{O}(\epsilon^{1/2})$ anomalous dimensions~\cite{eisenriegler1988surface}.}.
Motivated by this, we proceed to the next order and determine the $\mathcal{O}(\epsilon)$ corrections, as shown above.

%It is natural to ask whether there exists a BCFT that possesses image symmetry and hence obeys the anomalous-dimension constraints in Eq.~\eqref{eq:image_symmetry_phi6_d_3}.
%At least for the $\phi^6$ interaction, the trivial anomalous dimension of bulk fields at order $\mathcal{O}(\epsilon)$ rules out image symmetry.

\subsection{$\mathcal{O}(\epsilon)$ expansion in the Dirichlet boundary condition}
\label{sec:O(epsilon) expansion with the Dirichlet boundary condition}

In this section, we bootstrap the tricritical $O(N)$ theory under the Dirichlet boundary condition in $d= 3- \epsilon$, i.e., the ordinary transition. 
%Although the model is simple, we will see that it does not respect image symmetry.
%Note that the constraint discussed above is valid only for the effective Neumann boundary condition.

As in the previous calculation, we first analyze the discontinuity of $G_i(\xi)-G_b(\xi)$.
Using the zeroth-order solution in Eq.~\eqref{eq:0th_solution_general_bootstrap}, we set
$\tilde{\lambda}_0=-1+\tilde{\lambda}_0^{(1)}\epsilon$, $\ \tilde{\mu}_0=\tilde{\mu}_0^{(1)}\epsilon$, $\ \tilde{\mu}_1=\frac{1}{2}+\tilde{\mu}_1^{(1)}\epsilon$, and
$\Delta_\phi=\frac{d-2}{2}+\Delta_\phi^{(1)}\epsilon$, $\ \Delta_0=d-2+\Delta_0^{(1)}\epsilon$, $\ \hat{\Delta}_0=\frac{d-2}{2}+\hat{\Delta}_0^{(1)}\epsilon$, $\ \hat{\Delta}_1=\frac{d}{2}+\hat{\Delta}_1^{(1)}\epsilon$.
Because all $\mathcal{O}(\epsilon)$ contributions to the BOE coefficients are moved into $H_i(\xi)$, we take $\tilde{\mu}_1^{(0)}=\frac{1}{2}$ rather than $\tilde{\mu}_1^{(0)}=\frac{d-2}{2}$.
The resulting discontinuity has the same structure as Eq.~\eqref{eq:Disc_d_3_epsilon}, but with different parameters
\begin{equation}
\label{eq:parameters_A_d_3_epsilon_Dirichlet}
    A_1=2\hat{\Delta}_{1}^{(1)}-2\Delta_\phi^{(1)}\,,\  A_2=-2\hat{\Delta}_{1}^{(1)}+2\Delta_\phi^{(1)}-\Delta_{0}^{(1)}\,.
\end{equation}
Also, the corresponding OPE coefficients and $H_b(\xi)$ take the same forms as in Eqs.~\eqref{eq:OPE_d_3_epsilon} and \eqref{eq:sum_n_Hb_d_3_epsilon}, with $A_i$ given by Eq.~\eqref{eq:parameters_A_d_3_epsilon_Dirichlet}.

Plugging these into $H_i=G_b+H_b-G_i$, we obtain $H_i(\xi)$ with the same form as Eq.~\eqref{eq:difference_Hi_xi_d_3}, but with different coefficients $B_i$ given by
\begin{eqnarray}
\label{eq:parameters_B_d_3_epsilon_Dirichlet}
    B_1&=&-2\hat{\Delta}_1^{(1)}\,,\quad B_2=-2B_6=2\Delta_\phi^{(1)},\\ 
    B_3 &=&1-2\Delta_\phi^{(1)}+\tilde{\lambda}_{0}^{(1)}+(\log{4}+2)\hat{\Delta}_1^{(1)}-\frac{1}{2}\tilde{\mu}_0^{(1)}\,, \nonumber\\
    B_4&=&-1-\log{4}\cdot \hat{\Delta}_1^{(1)}-\frac{1}{2}\tilde{\mu}_0^{(1)}\,, \nonumber\\
    B_5&=& \Delta_\phi^{(1)}-\frac{1}{2}\Delta_0^{(1)}\,,~~ B_7=\hat{\Delta}_1^{(1)}-\Delta_\phi^{(1)}+\frac{1}{2}\Delta_0^{(1)}\,. \nonumber 
\end{eqnarray}
The BOE coefficients keep the form of Eq.~\eqref{eq:BOE_d_3_epsilon}, with modified $B_i'$ given by
\begin{eqnarray}
\label{eq:parameters_Bp_d_3_epsilon_Dirichlet}
    B_1'&=&1+\frac{\tilde{\lambda}_{0}^{(1)}}{2}+(1+3\log{2})  \hat{\Delta}_{1}^{(1)}-(1+\log{2})\Delta_\phi^{(1)}\,, \nonumber\\
    B_2'&=&(-2+\log{4})(\Delta_\phi^{(1)}-\hat{\Delta}_{1}^{(1)})+\tilde{\lambda}_{0}^{(1)}\, ,\\
    B_3'&=&\hat{\Delta}_{1}^{(1)}-\Delta_\phi^{(1)}+\frac{1}{2}\Delta_{0}^{(1)} \,, \nonumber\\
    B_4'&=&-2\hat{\Delta}_1^{(1)},\qquad\qquad B_5'=2\Delta_\phi^{(1)}\,. \nonumber
\end{eqnarray}
%Here, we have imposed the constraint that the coefficient of the lowest boundary conformal block ($m=0$) inferred from $H_i$ vanishes. 
Similarly, the independent inputs for the OPE and BOE coefficients at the ordinary fixed point are $\Delta_\phi^{(1)}$, $\Delta_0^{(1)}$, $\hat{\Delta}_1^{(1)}$ and $\tilde\lambda_0^{(1)}$.

% Imposing the constraint $\tilde{\mu}_{0,c}^{(1)}=0$, we obtain
% \begin{equation}
% \label{eq:constraint_mu_0_d_3_Dirichlet}
%     \tilde{\mu}_{0}^{(1)}=(-1+\log{2})(\Delta_\phi^{(1)}-\hat{\Delta}_{1}^{(1)})+\frac{\tilde{\lambda}_{0}^{(1)}}{2}.
% \end{equation}
% The next BOE coefficient is $\tilde{\mu}_{1}^{(1)}=\frac{1}{4}[-2+2\Delta_{0}^{(1)}+\log{4}(\Delta_\phi^{(1)}-\hat{\Delta}_{1}^{(1)})-\tilde{\lambda}_{0}^{(1)}]$.
%In addition, a direct resummation of all boundary conformal blocks with their BOE coefficients reproduces the same $H_i(\xi)$.

Now we apply the image symmetry constraint to the results above.
For $d=3-\epsilon$ with the Dirichlet boundary condition, image symmetry implies $\frac{H_i(\xi')}{\xi'}|_{\xi'=e^{\pm{\rm i}\pi}(\xi+1)}=-e^{\mp{\rm i}\pi/2}\frac{H_i(\xi)}{\xi}$.
Taking $\xi'=e^{{\rm i}\pi}(\xi+1)$, the constraint on $H_i(\xi)$ gives $B_1=-B_2=2B_6=-2B_7$, $B_3=-B_4$, and $B_5=0$.
Using Eq.~\eqref{eq:parameters_B_d_3_epsilon_Dirichlet}, we find
\begin{equation}
\label{eq:image_symmetry_phi6_d_3_Dirichlet}
    \Delta_\phi^{(1)}=\hat{\Delta}_{1}^{(1)},\qquad \Delta_0^{(1)}=2\hat{\Delta}_{1}^{(1)}.
\end{equation}
In contrast to the previous case, here the Dirichlet boundary condition imposes the additional requirement $\tilde{\mu}_0^{(1)}=0$.
Substituting these relations into the expressions above, we obtain the OPE and BOE coefficients
% \begin{equation}
% \label{eq:OPE_BOE_d_3_phi_6_image_symm_D}
% \begin{split}
%     &\tilde{\lambda}_{0}^{(1)}=0,\qquad
%     \tilde{\lambda}_1^{(1)}=\Delta_\phi^{(1)},\qquad
%     \tilde{\lambda}_{n\geq2}^{(1)}=\frac{(-1)^{n+1}\Gamma(n-\tfrac{1}{2})\Gamma(n+\tfrac{1}{2})}{n\sqrt{\pi}\,\Gamma(2n-\tfrac{1}{2})}\,\Delta_\phi^{(1)},\\
%     &\tilde{\mu}_0^{(1)}=0,\qquad
%     \tilde{\mu}_1^{(1)}=-\frac{1}{2}+\Delta_\phi^{(1)},\qquad
%     \tilde{\mu}_{m\geq2}^{(1)}=\frac{2^{1-2m}(2m-1)\bigl(1-(-1)^m\bigr)}{m(m-1)}\,\Delta_\phi^{(1)}.
% \end{split}
% \end{equation}
\begin{equation}
\label{eq:OPE_BOE_d_3_phi_6_image_symm_D}
\begin{split}
    &\tilde{\lambda}_n^{(1)}=\begin{cases}
        \Delta_\phi^{(1)},&\quad n=1,\\[0.7ex]
        \frac{(-1)^{n+1}\Gamma(n-\tfrac{1}{2})\Gamma(n+\tfrac{1}{2})}{n\sqrt{\pi}\,\Gamma(2n-\tfrac{1}{2})}\,\Delta_\phi^{(1)},&\quad n>1.
    \end{cases}\\[0.5ex]
    &\tilde{\mu}_m^{(1)}=\begin{cases}
        -\frac{1}{2}+\Delta_\phi^{(1)},&\quad m=1,\\[0.7ex]
        \frac{2^{1-2m}(2m-1)(1-(-1)^m)}{m(m-1)}\,\Delta_\phi^{(1)},&\quad m>1,
    \end{cases}
\end{split}
\end{equation}
and $\tilde{\lambda}_0^{(1)}=\tilde{\mu}_0^{(1)}=0$.
%The solution of $\tilde{\lambda}_0^{(1)}$ is given by the additional constraint $\tilde{\mu}_0^{(1)}=0$ with the Dirichlet boundary condition.

However, the RG analysis of the tricritical $O(N)$ theory under the Dirichlet boundary condition does not preserve the image symmetry. 
At $\mathcal{O}(\epsilon)$ order, the anomalous dimensions of bulk and boundary operators from the RG calculation are $\Delta_\phi^{(1)}=\Delta_0^{(1)}=0$ with a nontrivial $\hat{\Delta}_1^{(1)}$~\cite{speth1983tricritical}
\begin{equation}
\label{eq:d_3_phi_6_Dirichlet_ordinary_epsilon}
    \hat{\Delta}_1^{(1)}=\frac{(N+2)(N+4)}{8(3N+22)}\,.
\end{equation}
Inserting these into the bootstrap solution yields
\begin{equation}
\label{eq:OPE_BOE_d_3_phi_6_free_Dirichlet}
\begin{split}
    &\tilde{\lambda}_n^{(1)}=\begin{cases}
        \frac{1}{3}\hat{\Delta}_1^{(1)},&\qquad n=1,\\[0.7ex]
        \frac{(-1)^{n+1}\Gamma(n-\tfrac{1}{2})\Gamma(n+\tfrac{1}{2})}{n\sqrt{\pi}\,\Gamma(2n-\tfrac{1}{2})}\,\hat{\Delta}_1^{(1)},&\qquad n>1.
    \end{cases}\\[0.5ex]
    &\tilde{\mu}_m^{(1)}=\begin{cases}
        \tilde{\lambda}_0^{(1)}+(2-\log 4)\,\hat{\Delta}_1^{(1)},&\qquad m=0,\\[0.7ex]
        -\frac{1}{4}(2+\tilde{\lambda}_0^{(1)}+\log 4\ \hat{\Delta}_1^{(1)}),&\qquad m=1,\\[0.7ex]
        \frac{(-1)^{m+1}2^{1-2m}(2m-1)}{m(m-1)}\,\hat{\Delta}_1^{(1)},&\qquad m>1,
    \end{cases}
\end{split}
\end{equation}
This shows that both even and odd boundary conformal blocks appear in $H_i(\xi)$.
Here the only unknown input is $\tilde{\lambda}_0^{(1)}$. 

Finally, we examine possible $\mathcal{O}(\epsilon^{1/2})$ corrections.
Using the expansion $d=3-\epsilon$ and $\epsilon=\tilde{\epsilon}^{\,2}$, we solve the bootstrap equations in complete analogy with the previous analysis.
We find that the BOE and OPE coefficients $\tilde{\lambda}_n^{(1/2)}$ and $\tilde{\mu}_m^{(1/2)}$ retain the same functional forms as before, but the parameters $A_i$, $B_j$, and $B_j'$ in Eqs.~\eqref{eq:parameters_A_d_3_epsilon_Dirichlet}, \eqref{eq:parameters_B_d_3_epsilon_Dirichlet}, and \eqref{eq:parameters_Bp_d_3_epsilon_Dirichlet} are modified. 
Concretely, one should replace the $\mathcal{O}(\epsilon)$ entries by their $\mathcal{O}(\epsilon^{1/2})$ counterparts and remove the constant pieces in $B_{3,4}$ and $B_1'$. 

The extra constant term that appears only at $\mathcal{O}(\epsilon)$ originates from choosing $\tilde{\mu}_1^{(0)}=\frac{1}{2}$ rather than $\tilde{\mu}_1^{(0)}=\frac{3-d}{2}$.
This choice introduces an additional $-\tfrac{\epsilon}{2}$ contribution at $\mathcal{O}(\epsilon)$, and when expanding to $\mathcal{O}(\epsilon^{1/2})$, this term is absent.

Turning to concrete models, image symmetry implies that the anomalous dimensions at $\mathcal{O}(\epsilon^{1/2})$ satisfy the same properties summarized in Eq.~\eqref{eq:image_symmetry_phi6_d_3_Dirichlet}. 
Likewise, the OPE and BOE coefficients agree with Eq.~\eqref{eq:OPE_BOE_d_3_phi_6_image_symm_D} after replacing the anomalous dimensions by their $\mathcal{O}(\epsilon^{1/2})$ versions and removing the constant piece from $\tilde{\mu}_1^{(1/2)}$.
For the ordinary fixed point of the tricritical model the anomalous dimensions of the bulk and boundary primaries vanish, $\Delta_\phi^{(1/2)}=\Delta_{0}^{(1/2)}=\hat{\Delta}_{1}^{(1/2)}=0$ at $\mathcal{O}(\epsilon^{1/2})$, so the only potentially nonzero new coefficient is $\tilde{\lambda}_{0}^{(1/2)}$. 
By the same argument used above for the special point, we expect $\tilde{\lambda}_{0}^{(1/2)}$ to vanish as well, implying that there are no $\mathcal{O}(\epsilon^{1/2})$ corrections to the conformal data.

\section{Conclusion}
\label{sec:Conclusion}

In this work, we extend the analytic bootstrap framework to BCFTs with interacting boundaries. 
By exploiting analytic properties of bulk and boundary conformal blocks, we express infinitely many OPE and BOE coefficients in terms of a finite set of conformal data taken as inputs. 
 
We analyze operator expansions for general composite operators $O^{(k)}\sim\phi^k$ at zeroth order.
The result shows that, in general, the correlator of $O^{(k)}$ already decomposes into infinitely many conformal blocks. 
When $k=1$, the zeroth-order solution consists of finitely many conformal blocks, which enables the calculation for the next order in $\epsilon = d_0 - d$ expansion. 
We carry out the $\epsilon$ expansion of $\langle\phi(x)\phi(y)\rangle$ correlation for $d_0=3$ and $d_0=4$, and apply the results to broad classes of models with nontrivial boundary interactions. 
%This reproduces known conformal data and yields new results.

Besides, we study the $O(N)$ theory supplemented by cubic boundary interactions that respect $S_{N+1}$ symmetry, and identify new fixed points, the long-range Potts fixed point and the boundary Potts fixed point, together with their stability regime. 
This theory may be realized in the following lattice model,
\begin{eqnarray}
\label{eq:lattice_model}
    H = - J \sum_{\langle ij\rangle \in \rm bulk} S_i \cdot S_j - \lambda \sum_{i\in \rm bdy} \sum_{\alpha = 1}^{N+1} \left(S_i \cdot e^{\alpha}\right)^3 \,, 
\end{eqnarray}
where $S_i$ denotes an $O(N)$ vector at site $i$. 
The bulk hosts a simple ferromagnetic coupling $J$ between nearest-neighbor sites, while a cubic interaction is present at the boundary characterized by the strength $\lambda$. 
Note that the unit vector $e_i^\alpha$, defined in Eq.~\eqref{eq:condition_e}, gives precisely the Potts anisotropy. 
It would be interesting to realize the various new boundary fixed points in the lattice model.

To conclude, our analytic bootstrap provides a unified prescription and general constraints on conformal data for BCFTs, including cases with boundary interactions. 
In higher-dimensional CFTs, attention has often focused on the scaling dimensions of primary operators, while computing OPE and BOE coefficients is considerably more challenging.
Although our method does not completely fix these coefficients, it reduces them to a much smaller set of unknown inputs. 
A natural next step is to compute the remaining unknown inputs in concrete models, thereby fully determining the corresponding BCFT data~\cite{symanzik1981schrodinger,collins1984renormalization,gorishny1989construction,adzhemyan1998renormalization}. 
Moreover, as noted above, our procedure solves the conformal data starting from cases with a finite number of conformal blocks. 
At higher order, operator mixing must be carefully considered. 
As an example, bulk conformal blocks for the composite operator $O^{(k)}$ are degenerate when $d$ is an integer. 
Thus, resolving operator degeneracies is key to extending the analytic bootstrap method to next order~\cite{burrington2013operator,poghosyan2013mixing,burrington2017operator,henriksson2018critical}.
Finally, we note that the understanding from the analytic bootstrap perspective remains relatively limited for fermionic BCFTs~\cite{warner1995supersymmetry,liendo2017bootstrap,elvang2019soft,gimenez2021superconformal}. 
Moreover, conformal defects, especially lower-dimensional defects, have been widely studied~\cite{billo2016defects,liendo2018bootstrapping,gadde2020conformal,gimenez2020bootstrapping,antunes2021conformal,gimenez2022bootstrapping,trepanier2023surface,hu2024solving,meineri2025bootstrap,lanzetta2025beginning,antunes2026ising}.
We expect that our approach will motivate investigations of analytic bootstrap for fermionic BCFTs and defect CFTs, which can exhibit even richer behavior.

\begin{acknowledgments}

This work is supported in part %by MOSTC under Grant No.~2021YFA1400100 (H.Y.), 
by the NSFC under Grant Nos.~12347107 and 12334003
(X.S. and H.Y.), the New Cornerstone Science Foundation through the Xplorer Prize (H.Y.), and a start-up fund from Tulane University (S.-K.J.).
\end{acknowledgments}

\appendix

\section{Useful Technical Results}

\subsection{Expansion of conformal blocks}
\label{sec:Expansion of conformal blocks}

In this section, we calculate the series expansion of the hypergeometric function $\mathbin{_2 F_1}(a,b,c,x)$ to $\mathcal{O}(\epsilon^2)$, with $a=a_0+a_1\epsilon+a_2\epsilon^2,\ b=b_0+b_1\epsilon+b_2\epsilon^2,\ c=c_0+c_1\epsilon+c_2\epsilon^2$.

As discussed in Ref.~\cite{dey2021analytic}, we use the integral representation of hypergeometric functions:
\begin{equation}
\label{eq:integral_rep_2F1}
    \mathbin{_2 F_1}(a,b,c,z)=\frac{\Gamma(c)}{\Gamma(b)\Gamma(c-b)}\int_0^1{\rm d}t\ \frac{t^{b-1}(1-t)^{c-b-1}}{(1-tz)^a}\,.
\end{equation}
Then we expand the integrand to $\mathcal{O}(\epsilon^2)$ and do the integral order-by-order to get the expansion of the hypergeometric function.
To ensure the convergent integral, we require $c_0>b_0$. 
We can use the property of hypergeometric function
\begin{eqnarray}
\label{eq:recuresion_relation_2F1}
    && \mathbin{_2 F_1}(a,b,c,z)=-\frac{z(a+b-2c-1)+c}{c(z-1)}\mathbin{_2 F_1}(a,b,c+1,z) \nonumber \\
    && -\frac{z(a-c-1)(b-c-1)}{c(c+1)(z-1)}\mathbin{_2 F_1}(a,b,c+2,z) \,, 
\end{eqnarray}
to increase $c_0$ to achieve $c_0>b_0$. 

With $c_0>b_0$, we can safely expand the integrand in Eq.~\eqref{eq:integral_rep_2F1} to $\mathcal{O}(\epsilon^2)$.
To do the integral after the expansion, we need the function ${\rm HypExpInt}$ defined in the Mathematica package HypExp~\cite{huber2006hypexp},
\begin{equation}
\label{eq:HypExpInt}
\begin{split}
    &{\rm HypExpInt}(a_1,a_2,a_3,a_4,a_5,z)\\
    =&\int_0^1{\rm d}t\ \frac{t^{a_1}\log{(t)}^{a_2}\log{(1-t)}^{a_3}\log{(1-tz)}^{a_4}}{(tz-1)^{a_5}}\,.
\end{split}
\end{equation}
In general, we also need another function, which we call ${\rm HypExpIntS}$
\begin{equation}
\label{eq:HypExpIntS}
\begin{split}
    &{\rm HypExpIntS}(a_1,a_2,a_3,a_4,a_5,z)\\
    =&\int_0^1{\rm d}t\ \frac{(1-t)^{a_1}\log{(t)}^{a_2}\log{(1-t)}^{a_3}\log{(1-tz)}^{a_4}}{(tz-1)^{a_5}}\,,
\end{split}
\end{equation}
which can be related to the integral ${\rm HypExpInt}$ by
\begin{equation}
\label{eq:relation_HypExpIntS_HypExpInt}
\begin{split}
    &{\rm HypExpIntS}(a_1,a_2,a_3,a_4,a_5,z)
    =\frac{1}{(1-z)^{a_5}}\\
    \times&\sum_{i=0}^{a_4}C_{a_4}^i\log{(1-z)}^{a_4-i}{\rm HypExpInt}\left(a_1,a_3,a_2,i,a_5,\frac{z}{z-1}\right)\,,
\end{split}
\end{equation}
where $C_n^m=\frac{\Gamma(n)}{\Gamma(m)\Gamma(n-m)}$.
We focus on the $\mathcal{O}(\epsilon^2)$ term, because the expansion up to $\mathcal{O}(\epsilon)$ can be obtained by the Mathematica package HypExp.
After a straightforward but lengthy computation, the $\mathcal{O}(\epsilon^2)$ order of hypergeometric function is 
\begin{equation}
\label{eq:epsilon_expansion_2F1}
\begin{split}
    &\mathbin{_2 F_1}(a,b,c,x)\xrightarrow{\mathcal{O}(\epsilon^2)}\frac{\epsilon^2}{2(z-1)}\left[2(a_2+b_2-c_2)\log{(1-z)}\right.\\
    &\left.-(a_1+b_1-c_1)^2\log{(1-z)^2}-2(a_1-c_1)(b_1-c_1){\rm Li}_1(z)\right]\,.
\end{split}
\end{equation}

\begin{widetext}
\subsection{Branch cut of special functions}
\label{sec:Branch cut of special functions}
Here, we present the discontinuity of $\mathbin{_2 F_1}(a,b,c,x)$ with $x>1$~\cite{WFS-07.23.04.0021.01}
\begin{eqnarray}
\label{eq:discontinuity_2F1}
    &&\lim_{\epsilon\rightarrow0+} \mathbin{_2 F_1}(a,b,c,x+{\rm i}\epsilon)=e^{2\pi{\rm i}(a+b-c)}\mathbin{_2 F_1}(a,b,c,x)+\frac{2\pi {\rm i}e^{{\rm i}\pi(a+b-c)}\Gamma(c)}{\Gamma(c-a)\Gamma(c-b)\Gamma(a+b-c+1)}\mathbin{_2 F_1}(a,b,a+b-c+1,1-x)\,, %\quad x>1\,. 
    \nonumber\\
    &&\lim_{\epsilon\rightarrow0+} \mathbin{_2 F_1}(a,b,c,x-{\rm i}\epsilon)=\mathbin{_2 F_1}(a,b,c,x)\,,%\qquad x>1\,.
\end{eqnarray}
%https://functions.wolfram.com/HypergeometricFunctions/Hypergeometric2F1/04/ShowAll.html
\end{widetext}

\subsection{Discontinuity on the branch cut}
\label{sec:Discontinuity on the branch cut}

In this section, we list the discontinuities of some functions that are used in the main text to calculate $\underset{\xi<-1}{\rm Disc}\ (G_i(\xi)-G_b(\xi))$:
\begin{eqnarray}
\label{eq:discontinuity_functions}
    && \underset{\xi<-1}{\rm Disc}\ \log{\xi}=\underset{\xi<-1}{\rm Disc}\ \log{(1+\xi)}=2\pi {\rm i}\,,\\
    && \underset{\xi<-1}{\rm Disc}\ (\log{\xi})^2=4\pi {\rm i}\log{(-\xi)}\,, \nonumber\\
    &&\underset{\xi<-1}{\rm Disc}\ {\rm Li}_2(\xi)=-2\pi {\rm i}\log{(-\xi)}\,, \nonumber\\
    &&\underset{\xi<-1}{\rm Disc}\ \log{\xi}\log{(1+\xi)}=2\pi {\rm i}(\log{(-\xi)}+\log{(-1-\xi)})\,. \nonumber
\end{eqnarray}

\section{Normalization factor for the composite field} \label{append:normalization}

We determine the normalization factors $f_{k,N}$ in Eq.~\eqref{eq:O_k_definition}. 
For $k=2m$, consider the generating function $\langle:e^{t\phi^2}::e^{s\phi^2}:\rangle_b$, where $\langle\cdot\rangle_b$ denotes the bulk expectation value without a boundary.
Expanding them gives
\begin{equation}
\label{eq:expansion_generating_function}
    \langle:e^{t\phi^2}::e^{s\phi^2}:\rangle_b=\sum_{m,n=0}^{\infty}\frac{t^m s^n}{m! n!}\langle:(\phi^2)^m::(\phi^2)^n:\rangle_b\,.
\end{equation}
% On the other hand, the $O(N)$ symmetry requires that the correlation between different components is vanishing, hence we have 
On the other hand, the $O(N)$ symmetry enforces vanishing correlations between different components, so that
\begin{eqnarray}
\label{eq:calculation_generating_function}
    &&\langle:e^{t\phi^2}::e^{s\phi^2}:\rangle_b =\prod_{i=1}^N\sum_{m,n=0}^{\infty}\frac{t^m s^n}{m! n!}\langle:(\phi_i^2)^m::(\phi_i^2)^n:\rangle_b \nonumber\\
    &&=\prod_{i=1}^N\sum_{m,n=0}^{\infty}\frac{t^m s^n}{m! n!}(2m)!D(x-y)^{2m}\delta_{m,n}\\
    &&=\left[1-4tsD(x-y)^2\right]^{-N/2}
    =\sum_{m=0}^\infty\frac{(\frac{N}{2})_m}{m!}(4tsD(x-y)^2)^m \,, \nonumber
\end{eqnarray}
where we define $D(x-y)=\langle\phi_i(x)\phi_i(y)\rangle_b=\frac{1}{|x-y|^{2\Delta_\phi}}$ with $\Delta_\phi=\frac{d-2}{2}$, and $(a)_n=\frac{\Gamma(a+n)}{\Gamma(a)}$ is the Pochhammer symbol.
Matching Eqs.~\eqref{eq:expansion_generating_function} and \eqref{eq:calculation_generating_function} yields
\begin{equation}
\label{eq:even_phi_correlation_coefficient}
\begin{split}
    &\langle:(\phi^2)^m::(\phi^2)^n:\rangle_b\\
    =&\delta_{m,n}\frac{\Gamma(m+\frac{N}{2})\Gamma(m+1)}{\Gamma(\frac{N}{2})}(2D(x-y))^{2m},
\end{split}
\end{equation}
which gives $f_{2m,N}=(\frac{\Gamma(m+N/2)\Gamma(m+1)}{\Gamma(N/2)})^{-1/2}2^{-m}$.

For $k=2m+1$, consider the generating function $\langle:e^{t\phi^2+J\cdot\phi}::e^{s\phi^2+K\cdot\phi}:\rangle_b$, where $J\cdot\phi=\sum_i J_i \phi_i$ and similarly for $K\cdot\phi$.
On the one hand, treating $J$ and $K$ as sources, we obtain
\begin{equation}
\label{eq:expansion_generating_function_odd_k}
\begin{split}
    &\left.\frac{\partial}{\partial J_i}\frac{\partial}{\partial K_i}\right|_{J=K=0}\langle:e^{t\phi^2+J\cdot\phi}::e^{s\phi^2+K\cdot\phi}:\rangle_b\\
    =&\sum_{m,n=0}^{\infty}\frac{t^m s^n}{m! n!}\langle:(\phi^2)^m\phi_i::(\phi^2)^n\phi_i:\rangle_b \,.
\end{split}
\end{equation}
On the other hand, $\langle:e^{t\phi^2+J\cdot\phi}::e^{s\phi^2+K\cdot\phi}: \rangle_b=\prod_{i=1}^N\langle:e^{t\phi_i^2+J_i\phi_i}::e^{s\phi_i^2+K_i\phi_i}: \rangle_b$.
And for the $\mathcal{O}(J_i K_i)$ term, we have 
% $\langle:e^{t\phi_i^2}::e^{s\phi_i^2}: \rangle_b=\left[1-4tsD(x-y)^2\right]^{-1/2}$
\begin{eqnarray}
\label{eq:calculation_generating_function_odd_k}
    && \langle:e^{t\phi_i^2}\phi_i::e^{s\phi_i^2}\phi_i: \rangle_b
    = \sum_{m,n=0}^{\infty}\frac{t^m s^n}{m! n!}\langle:(\phi_i^2)^m\phi_i::(\phi_i^2)^n\phi_i: \rangle_b \nonumber\\
    &&= \sum_{m=0}^{\infty}\frac{(ts)^m}{(m!)^2}(2m+1)!D(x-y)^{2m+1} \nonumber\\
    &&=D(x-y)\left[1-4tsD(x-y)^2\right]^{-3/2}\,,
\end{eqnarray}
Therefore, it leads to
\begin{eqnarray}
\label{eq:match_both_side}
    &&\left.\frac{\partial}{\partial J_i}\frac{\partial}{\partial K_i}\right|_{J=K=0}\langle:e^{t\phi^2+J\cdot\phi}::e^{s\phi^2+K\cdot\phi}:\rangle_b \nonumber\\
    &=&D(x-y)\left[1-4tsD(x-y)^2\right]^{-(N+2)/2} \nonumber\\
    &=&D(x-y)\sum_{m=0}^\infty \frac{(\frac{N+2}{2})_m}{m!}(4tsD(x-y)^2)^m \,.
\end{eqnarray}
Matching the Eqs.~\eqref{eq:expansion_generating_function_odd_k} and~\eqref{eq:match_both_side} gives the correlation function
\begin{equation}
\label{eq:odd_phi_correlation_coefficient}
\begin{split}
    &\langle:(\phi^2)^m\phi_i::(\phi^2)^n\phi_i:\rangle_b\\
    =&\delta_{m,n}\frac{\Gamma(m+\frac{N+2}{2})\Gamma(m+1)}{2\Gamma(\frac{N+2}{2})}(2D(x-y))^{2m+1}\,.
\end{split}
\end{equation}
Thus, we have the coefficient $f_{2m+1,N}=(\frac{\Gamma(m+N/2+1)\Gamma(m+1)}{\Gamma(N/2+1)})^{-1/2}2^{-m}$.

\bibliography{reference.bib}

\begin{thebibliography}{105}%
\makeatletter
\providecommand \@ifxundefined [1]{%
 \@ifx{#1\undefined}
}%
\providecommand \@ifnum [1]{%
 \ifnum #1\expandafter \@firstoftwo
 \else \expandafter \@secondoftwo
 \fi
}%
\providecommand \@ifx [1]{%
 \ifx #1\expandafter \@firstoftwo
 \else \expandafter \@secondoftwo
 \fi
}%
\providecommand \natexlab [1]{#1}%
\providecommand \enquote  [1]{``#1''}%
\providecommand \bibnamefont  [1]{#1}%
\providecommand \bibfnamefont [1]{#1}%
\providecommand \citenamefont [1]{#1}%
\providecommand \href@noop [0]{\@secondoftwo}%
\providecommand \href [0]{\begingroup \@sanitize@url \@href}%
\providecommand \@href[1]{\@@startlink{#1}\@@href}%
\providecommand \@@href[1]{\endgroup#1\@@endlink}%
\providecommand \@sanitize@url [0]{\catcode `\\12\catcode `\$12\catcode
  `\&12\catcode `\#12\catcode `\^12\catcode `\_12\catcode `\%12\relax}%
\providecommand \@@startlink[1]{}%
\providecommand \@@endlink[0]{}%
\providecommand \url  [0]{\begingroup\@sanitize@url \@url }%
\providecommand \@url [1]{\endgroup\@href {#1}{\urlprefix }}%
\providecommand \urlprefix  [0]{URL }%
\providecommand \Eprint [0]{\href }%
\providecommand \doibase [0]{https://doi.org/}%
\providecommand \selectlanguage [0]{\@gobble}%
\providecommand \bibinfo  [0]{\@secondoftwo}%
\providecommand \bibfield  [0]{\@secondoftwo}%
\providecommand \translation [1]{[#1]}%
\providecommand \BibitemOpen [0]{}%
\providecommand \bibitemStop [0]{}%
\providecommand \bibitemNoStop [0]{.\EOS\space}%
\providecommand \EOS [0]{\spacefactor3000\relax}%
\providecommand \BibitemShut  [1]{\csname bibitem#1\endcsname}%
\let\auto@bib@innerbib\@empty
%</preamble>
\bibitem [{\citenamefont {Di~Francesco}\ \emph {et~al.}(1997)\citenamefont
  {Di~Francesco}, \citenamefont {Mathieu},\ and\ \citenamefont
  {Senechal}}]{francesco2012conformal}%
  \BibitemOpen
  \bibfield  {author} {\bibinfo {author} {\bibfnamefont {P.}~\bibnamefont
  {Di~Francesco}}, \bibinfo {author} {\bibfnamefont {P.}~\bibnamefont
  {Mathieu}},\ and\ \bibinfo {author} {\bibfnamefont {D.}~\bibnamefont
  {Senechal}},\ }\href {https://doi.org/10.1007/978-1-4612-2256-9} {\emph
  {\bibinfo {title} {{Conformal Field Theory}}}},\ Graduate Texts in
  Contemporary Physics\ (\bibinfo  {publisher} {Springer-Verlag},\ \bibinfo
  {address} {New York},\ \bibinfo {year} {1997})\BibitemShut {NoStop}%
\bibitem [{\citenamefont {Amit}\ and\ \citenamefont
  {Martin-Mayor}(2005)}]{amit2005field}%
  \BibitemOpen
  \bibfield  {author} {\bibinfo {author} {\bibfnamefont {D.~J.}\ \bibnamefont
  {Amit}}\ and\ \bibinfo {author} {\bibfnamefont {V.}~\bibnamefont
  {Martin-Mayor}},\ }\href@noop {} {\emph {\bibinfo {title} {Field theory, the
  renormalization group, and critical phenomena: graphs to computers}}}\
  (\bibinfo  {publisher} {World Scientific Publishing Company},\ \bibinfo
  {year} {2005})\BibitemShut {NoStop}%
\bibitem [{\citenamefont {Cardy}(1984)}]{cardy1984conformal}%
  \BibitemOpen
  \bibfield  {author} {\bibinfo {author} {\bibfnamefont {J.~L.}\ \bibnamefont
  {Cardy}},\ }\href {https://doi.org/10.1016/0550-3213(84)90241-4} {\bibfield
  {journal} {\bibinfo  {journal} {Nucl. Phys. B}\ }\textbf {\bibinfo {volume}
  {240}},\ \bibinfo {pages} {514} (\bibinfo {year} {1984})}\BibitemShut
  {NoStop}%
\bibitem [{\citenamefont {Diehl}(1986)}]{Domb1986PHASE}%
  \BibitemOpen
  \bibfield  {author} {\bibinfo {author} {\bibfnamefont {H.}~\bibnamefont
  {Diehl}},\ }\href@noop {} {\bibfield  {journal} {\bibinfo  {journal} {Phase
  transitions and and Critical Phenomena, edited by C. Domb and JL Lebowitz}\
  }\textbf {\bibinfo {volume} {10}},\ \bibinfo {pages} {75} (\bibinfo {year}
  {1986})}\BibitemShut {NoStop}%
\bibitem [{\citenamefont {Affleck}\ and\ \citenamefont
  {Ludwig}(1991)}]{affleck1991universal}%
  \BibitemOpen
  \bibfield  {author} {\bibinfo {author} {\bibfnamefont {I.}~\bibnamefont
  {Affleck}}\ and\ \bibinfo {author} {\bibfnamefont {A.~W.~W.}\ \bibnamefont
  {Ludwig}},\ }\href {https://doi.org/10.1103/PhysRevLett.67.161} {\bibfield
  {journal} {\bibinfo  {journal} {Phys. Rev. Lett.}\ }\textbf {\bibinfo
  {volume} {67}},\ \bibinfo {pages} {161} (\bibinfo {year} {1991})}\BibitemShut
  {NoStop}%
\bibitem [{\citenamefont {Bariev}\ and\ \citenamefont
  {Turban}(1992)}]{bariev1992surface}%
  \BibitemOpen
  \bibfield  {author} {\bibinfo {author} {\bibfnamefont {R.~Z.}\ \bibnamefont
  {Bariev}}\ and\ \bibinfo {author} {\bibfnamefont {L.}~\bibnamefont
  {Turban}},\ }\href {https://doi.org/10.1103/PhysRevB.45.10761} {\bibfield
  {journal} {\bibinfo  {journal} {Phys. Rev. B}\ }\textbf {\bibinfo {volume}
  {45}},\ \bibinfo {pages} {10761} (\bibinfo {year} {1992})}\BibitemShut
  {NoStop}%
\bibitem [{\citenamefont {Diehl}(1997)}]{diehl1997theory}%
  \BibitemOpen
  \bibfield  {author} {\bibinfo {author} {\bibfnamefont {H.~W.}\ \bibnamefont
  {Diehl}},\ }\href {https://doi.org/10.1142/S0217979297001751} {\bibfield
  {journal} {\bibinfo  {journal} {Int. J. Mod. Phys. B}\ }\textbf {\bibinfo
  {volume} {11}},\ \bibinfo {pages} {3503} (\bibinfo {year} {1997})},\ \Eprint
  {https://arxiv.org/abs/cond-mat/9610143} {arXiv:cond-mat/9610143}
  \BibitemShut {NoStop}%
\bibitem [{\citenamefont {Cardy}(2004)}]{cardy2004boundary}%
  \BibitemOpen
  \bibfield  {author} {\bibinfo {author} {\bibfnamefont {J.}~\bibnamefont
  {Cardy}},\ }\href {https://arxiv.org/abs/hep-th/0411189} {\bibinfo {title}
  {Boundary conformal field theory}} (\bibinfo {year} {2004}),\ \Eprint
  {https://arxiv.org/abs/hep-th/0411189} {arXiv:hep-th/0411189 [hep-th]}
  \BibitemShut {NoStop}%
\bibitem [{\citenamefont {Herzog}\ \emph {et~al.}(2016)\citenamefont {Herzog},
  \citenamefont {Huang},\ and\ \citenamefont {Jensen}}]{herzog2016universal}%
  \BibitemOpen
  \bibfield  {author} {\bibinfo {author} {\bibfnamefont {C.~P.}\ \bibnamefont
  {Herzog}}, \bibinfo {author} {\bibfnamefont {K.-W.}\ \bibnamefont {Huang}},\
  and\ \bibinfo {author} {\bibfnamefont {K.}~\bibnamefont {Jensen}},\ }\href
  {https://doi.org/10.1007/JHEP01(2016)162} {\bibfield  {journal} {\bibinfo
  {journal} {JHEP}\ }\textbf {\bibinfo {volume} {01}},\ \bibinfo {pages}
  {162}},\ \Eprint {https://arxiv.org/abs/1510.00021} {arXiv:1510.00021
  [hep-th]} \BibitemShut {NoStop}%
\bibitem [{\citenamefont {Hogervorst}(2017)}]{hogervorst2017crossing}%
  \BibitemOpen
  \bibfield  {author} {\bibinfo {author} {\bibfnamefont {M.}~\bibnamefont
  {Hogervorst}},\ }\href {https://arxiv.org/abs/1703.08159} {\bibinfo {title}
  {Crossing kernels for boundary and crosscap cfts}} (\bibinfo {year} {2017}),\
  \Eprint {https://arxiv.org/abs/1703.08159} {arXiv:1703.08159 [hep-th]}
  \BibitemShut {NoStop}%
\bibitem [{\citenamefont {Andrei}\ \emph {et~al.}(2020)\citenamefont {Andrei},
  \citenamefont {Bissi}, \citenamefont {Buican}, \citenamefont {Cardy},
  \citenamefont {Dorey}, \citenamefont {Drukker}, \citenamefont {Erdmenger},
  \citenamefont {Friedan}, \citenamefont {Fursaev}, \citenamefont {Konechny},
  \citenamefont {Kristjansen}, \citenamefont {Makabe}, \citenamefont
  {Nakayama}, \citenamefont {O’Bannon}, \citenamefont {Parini}, \citenamefont
  {Robinson}, \citenamefont {Ryu}, \citenamefont {Schmidt-Colinet},
  \citenamefont {Schomerus}, \citenamefont {Schweigert},\ and\ \citenamefont
  {Watts}}]{andrei2020boundary}%
  \BibitemOpen
  \bibfield  {author} {\bibinfo {author} {\bibfnamefont {N.}~\bibnamefont
  {Andrei}}, \bibinfo {author} {\bibfnamefont {A.}~\bibnamefont {Bissi}},
  \bibinfo {author} {\bibfnamefont {M.}~\bibnamefont {Buican}}, \bibinfo
  {author} {\bibfnamefont {J.}~\bibnamefont {Cardy}}, \bibinfo {author}
  {\bibfnamefont {P.}~\bibnamefont {Dorey}}, \bibinfo {author} {\bibfnamefont
  {N.}~\bibnamefont {Drukker}}, \bibinfo {author} {\bibfnamefont
  {J.}~\bibnamefont {Erdmenger}}, \bibinfo {author} {\bibfnamefont
  {D.}~\bibnamefont {Friedan}}, \bibinfo {author} {\bibfnamefont
  {D.}~\bibnamefont {Fursaev}}, \bibinfo {author} {\bibfnamefont
  {A.}~\bibnamefont {Konechny}}, \bibinfo {author} {\bibfnamefont
  {C.}~\bibnamefont {Kristjansen}}, \bibinfo {author} {\bibfnamefont
  {I.}~\bibnamefont {Makabe}}, \bibinfo {author} {\bibfnamefont
  {Y.}~\bibnamefont {Nakayama}}, \bibinfo {author} {\bibfnamefont
  {A.}~\bibnamefont {O’Bannon}}, \bibinfo {author} {\bibfnamefont
  {R.}~\bibnamefont {Parini}}, \bibinfo {author} {\bibfnamefont
  {B.}~\bibnamefont {Robinson}}, \bibinfo {author} {\bibfnamefont
  {S.}~\bibnamefont {Ryu}}, \bibinfo {author} {\bibfnamefont {C.}~\bibnamefont
  {Schmidt-Colinet}}, \bibinfo {author} {\bibfnamefont {V.}~\bibnamefont
  {Schomerus}}, \bibinfo {author} {\bibfnamefont {C.}~\bibnamefont
  {Schweigert}},\ and\ \bibinfo {author} {\bibfnamefont {G.~M.~T.}\
  \bibnamefont {Watts}},\ }\href {https://doi.org/10.1088/1751-8121/abb0fe}
  {\bibfield  {journal} {\bibinfo  {journal} {Journal of Physics A:
  Mathematical and Theoretical}\ }\textbf {\bibinfo {volume} {53}},\ \bibinfo
  {pages} {453002} (\bibinfo {year} {2020})}\BibitemShut {NoStop}%
\bibitem [{\citenamefont {Lubensky}\ and\ \citenamefont
  {Rubin}(1975)}]{lubensky1975critical}%
  \BibitemOpen
  \bibfield  {author} {\bibinfo {author} {\bibfnamefont {T.~C.}\ \bibnamefont
  {Lubensky}}\ and\ \bibinfo {author} {\bibfnamefont {M.~H.}\ \bibnamefont
  {Rubin}},\ }\href {https://doi.org/10.1103/PhysRevB.12.3885} {\bibfield
  {journal} {\bibinfo  {journal} {Phys. Rev. B}\ }\textbf {\bibinfo {volume}
  {12}},\ \bibinfo {pages} {3885} (\bibinfo {year} {1975})}\BibitemShut
  {NoStop}%
\bibitem [{\citenamefont {Ohno}\ and\ \citenamefont
  {Okabe}(1984)}]{ohno1984The}%
  \BibitemOpen
  \bibfield  {author} {\bibinfo {author} {\bibfnamefont {K.}~\bibnamefont
  {Ohno}}\ and\ \bibinfo {author} {\bibfnamefont {Y.}~\bibnamefont {Okabe}},\
  }\href {https://doi.org/10.1143/PTP.72.736} {\bibfield  {journal} {\bibinfo
  {journal} {Progress of Theoretical Physics}\ }\textbf {\bibinfo {volume}
  {72}},\ \bibinfo {pages} {736} (\bibinfo {year} {1984})}\BibitemShut
  {NoStop}%
\bibitem [{\citenamefont {Diehi}\ and\ \citenamefont
  {Eisenriegler}(1987)}]{diehi1987walks}%
  \BibitemOpen
  \bibfield  {author} {\bibinfo {author} {\bibfnamefont {H.~W.}\ \bibnamefont
  {Diehi}}\ and\ \bibinfo {author} {\bibfnamefont {E.}~\bibnamefont
  {Eisenriegler}},\ }\href {https://doi.org/10.1209/0295-5075/4/6/012}
  {\bibfield  {journal} {\bibinfo  {journal} {Europhysics Letters}\ }\textbf
  {\bibinfo {volume} {4}},\ \bibinfo {pages} {709} (\bibinfo {year}
  {1987})}\BibitemShut {NoStop}%
\bibitem [{\citenamefont {Lauria}\ \emph {et~al.}(2018)\citenamefont {Lauria},
  \citenamefont {Meineri},\ and\ \citenamefont {Trevisani}}]{lauria2018radial}%
  \BibitemOpen
  \bibfield  {author} {\bibinfo {author} {\bibfnamefont {E.}~\bibnamefont
  {Lauria}}, \bibinfo {author} {\bibfnamefont {M.}~\bibnamefont {Meineri}},\
  and\ \bibinfo {author} {\bibfnamefont {E.}~\bibnamefont {Trevisani}},\ }\href
  {https://doi.org/10.1007/JHEP11(2018)148} {\bibfield  {journal} {\bibinfo
  {journal} {JHEP}\ }\textbf {\bibinfo {volume} {11}},\ \bibinfo {pages}
  {148}},\ \Eprint {https://arxiv.org/abs/1712.07668} {arXiv:1712.07668
  [hep-th]} \BibitemShut {NoStop}%
\bibitem [{\citenamefont {Dey}\ \emph {et~al.}(2020)\citenamefont {Dey},
  \citenamefont {Hansen},\ and\ \citenamefont {Shpot}}]{dey2020operator}%
  \BibitemOpen
  \bibfield  {author} {\bibinfo {author} {\bibfnamefont {P.}~\bibnamefont
  {Dey}}, \bibinfo {author} {\bibfnamefont {T.}~\bibnamefont {Hansen}},\ and\
  \bibinfo {author} {\bibfnamefont {M.}~\bibnamefont {Shpot}},\ }\href
  {https://doi.org/10.1007/JHEP12(2020)051} {\bibfield  {journal} {\bibinfo
  {journal} {JHEP}\ }\textbf {\bibinfo {volume} {12}},\ \bibinfo {pages}
  {051}},\ \Eprint {https://arxiv.org/abs/2006.11253} {arXiv:2006.11253
  [hep-th]} \BibitemShut {NoStop}%
\bibitem [{\citenamefont {Krishnan}\ and\ \citenamefont
  {Metlitski}(2023)}]{krishnan2023plane}%
  \BibitemOpen
  \bibfield  {author} {\bibinfo {author} {\bibfnamefont {A.}~\bibnamefont
  {Krishnan}}\ and\ \bibinfo {author} {\bibfnamefont {M.~A.}\ \bibnamefont
  {Metlitski}},\ }\href {https://doi.org/10.21468/SciPostPhys.15.3.090}
  {\bibfield  {journal} {\bibinfo  {journal} {SciPost Phys.}\ }\textbf
  {\bibinfo {volume} {15}},\ \bibinfo {pages} {090} (\bibinfo {year}
  {2023})}\BibitemShut {NoStop}%
\bibitem [{\citenamefont {Ishibashi}(1989)}]{ishibashi1989boundary}%
  \BibitemOpen
  \bibfield  {author} {\bibinfo {author} {\bibfnamefont {N.}~\bibnamefont
  {Ishibashi}},\ }\href {https://doi.org/10.1142/S0217732389000320} {\bibfield
  {journal} {\bibinfo  {journal} {Mod. Phys. Lett. A}\ }\textbf {\bibinfo
  {volume} {4}},\ \bibinfo {pages} {251} (\bibinfo {year} {1989})}\BibitemShut
  {NoStop}%
\bibitem [{\citenamefont {Cardy}(1989)}]{cardy1989boundary}%
  \BibitemOpen
  \bibfield  {author} {\bibinfo {author} {\bibfnamefont {J.~L.}\ \bibnamefont
  {Cardy}},\ }\href {https://doi.org/10.1016/0550-3213(89)90521-X} {\bibfield
  {journal} {\bibinfo  {journal} {Nucl. Phys. B}\ }\textbf {\bibinfo {volume}
  {324}},\ \bibinfo {pages} {581} (\bibinfo {year} {1989})}\BibitemShut
  {NoStop}%
\bibitem [{\citenamefont {Polyakov}(1973)}]{polyakov1974nonhamiltonian}%
  \BibitemOpen
  \bibfield  {author} {\bibinfo {author} {\bibfnamefont {A.~M.}\ \bibnamefont
  {Polyakov}},\ }\href {https://www.osti.gov/biblio/4331386} {\bibfield
  {journal} {\bibinfo  {journal} {Zh. Eksp. Teor. Fiz., v. 66, no. 1, pp.
  23-42}\ } (\bibinfo {year} {1973})}\BibitemShut {NoStop}%
\bibitem [{\citenamefont {Belavin}\ \emph {et~al.}(1984)\citenamefont
  {Belavin}, \citenamefont {Polyakov},\ and\ \citenamefont
  {Zamolodchikov}}]{belavin1984infinite}%
  \BibitemOpen
  \bibfield  {author} {\bibinfo {author} {\bibfnamefont {A.~A.}\ \bibnamefont
  {Belavin}}, \bibinfo {author} {\bibfnamefont {A.~M.}\ \bibnamefont
  {Polyakov}},\ and\ \bibinfo {author} {\bibfnamefont {A.~B.}\ \bibnamefont
  {Zamolodchikov}},\ }\href {https://doi.org/10.1016/0550-3213(84)90052-X}
  {\bibfield  {journal} {\bibinfo  {journal} {Nucl. Phys. B}\ }\textbf
  {\bibinfo {volume} {241}},\ \bibinfo {pages} {333} (\bibinfo {year}
  {1984})}\BibitemShut {NoStop}%
\bibitem [{\citenamefont {Dolan}\ and\ \citenamefont
  {Osborn}(2004)}]{dolan2004conformal}%
  \BibitemOpen
  \bibfield  {author} {\bibinfo {author} {\bibfnamefont {F.~A.}\ \bibnamefont
  {Dolan}}\ and\ \bibinfo {author} {\bibfnamefont {H.}~\bibnamefont {Osborn}},\
  }\href {https://doi.org/10.1016/j.nuclphysb.2003.11.016} {\bibfield
  {journal} {\bibinfo  {journal} {Nucl. Phys. B}\ }\textbf {\bibinfo {volume}
  {678}},\ \bibinfo {pages} {491} (\bibinfo {year} {2004})},\ \Eprint
  {https://arxiv.org/abs/hep-th/0309180} {arXiv:hep-th/0309180} \BibitemShut
  {NoStop}%
\bibitem [{\citenamefont {Poland}\ \emph {et~al.}(2019)\citenamefont {Poland},
  \citenamefont {Rychkov},\ and\ \citenamefont {Vichi}}]{poland2019conformal}%
  \BibitemOpen
  \bibfield  {author} {\bibinfo {author} {\bibfnamefont {D.}~\bibnamefont
  {Poland}}, \bibinfo {author} {\bibfnamefont {S.}~\bibnamefont {Rychkov}},\
  and\ \bibinfo {author} {\bibfnamefont {A.}~\bibnamefont {Vichi}},\ }\href
  {https://doi.org/10.1103/RevModPhys.91.015002} {\bibfield  {journal}
  {\bibinfo  {journal} {Rev. Mod. Phys.}\ }\textbf {\bibinfo {volume} {91}},\
  \bibinfo {pages} {015002} (\bibinfo {year} {2019})}\BibitemShut {NoStop}%
\bibitem [{\citenamefont {Cardy}\ and\ \citenamefont
  {Lewellen}(1991)}]{cardy1991bulk}%
  \BibitemOpen
  \bibfield  {author} {\bibinfo {author} {\bibfnamefont {J.~L.}\ \bibnamefont
  {Cardy}}\ and\ \bibinfo {author} {\bibfnamefont {D.~C.}\ \bibnamefont
  {Lewellen}},\ }\href {https://doi.org/10.1016/0370-2693(91)90828-E}
  {\bibfield  {journal} {\bibinfo  {journal} {Phys. Lett. B}\ }\textbf
  {\bibinfo {volume} {259}},\ \bibinfo {pages} {274} (\bibinfo {year}
  {1991})}\BibitemShut {NoStop}%
\bibitem [{\citenamefont {Rastelli}\ and\ \citenamefont
  {Zhou}(2017)}]{rastelli2017mellin}%
  \BibitemOpen
  \bibfield  {author} {\bibinfo {author} {\bibfnamefont {L.}~\bibnamefont
  {Rastelli}}\ and\ \bibinfo {author} {\bibfnamefont {X.}~\bibnamefont
  {Zhou}},\ }\href {https://doi.org/10.1007/JHEP10(2017)146} {\bibfield
  {journal} {\bibinfo  {journal} {JHEP}\ }\textbf {\bibinfo {volume} {10}},\
  \bibinfo {pages} {146}},\ \Eprint {https://arxiv.org/abs/1705.05362}
  {arXiv:1705.05362 [hep-th]} \BibitemShut {NoStop}%
\bibitem [{\citenamefont {Maz{\'a}{\v{c}}}\ \emph {et~al.}(2019)\citenamefont
  {Maz{\'a}{\v{c}}}, \citenamefont {Rastelli},\ and\ \citenamefont
  {Zhou}}]{mazavc2019analytic}%
  \BibitemOpen
  \bibfield  {author} {\bibinfo {author} {\bibfnamefont {D.}~\bibnamefont
  {Maz{\'a}{\v{c}}}}, \bibinfo {author} {\bibfnamefont {L.}~\bibnamefont
  {Rastelli}},\ and\ \bibinfo {author} {\bibfnamefont {X.}~\bibnamefont
  {Zhou}},\ }\href {https://doi.org/10.1007/JHEP12(2019)004} {\bibfield
  {journal} {\bibinfo  {journal} {JHEP}\ }\textbf {\bibinfo {volume} {12}},\
  \bibinfo {pages} {004}},\ \Eprint {https://arxiv.org/abs/1812.09314}
  {arXiv:1812.09314 [hep-th]} \BibitemShut {NoStop}%
\bibitem [{\citenamefont {Pagani}\ and\ \citenamefont
  {Sonoda}(2020)}]{pagani2020operator}%
  \BibitemOpen
  \bibfield  {author} {\bibinfo {author} {\bibfnamefont {C.}~\bibnamefont
  {Pagani}}\ and\ \bibinfo {author} {\bibfnamefont {H.}~\bibnamefont
  {Sonoda}},\ }\href {https://doi.org/10.1103/PhysRevD.101.105007} {\bibfield
  {journal} {\bibinfo  {journal} {Phys. Rev. D}\ }\textbf {\bibinfo {volume}
  {101}},\ \bibinfo {pages} {105007} (\bibinfo {year} {2020})},\ \Eprint
  {https://arxiv.org/abs/2001.07015} {arXiv:2001.07015 [hep-th]} \BibitemShut
  {NoStop}%
\bibitem [{\citenamefont {Kaviraj}\ and\ \citenamefont
  {Paulos}(2020)}]{kaviraj2020functional}%
  \BibitemOpen
  \bibfield  {author} {\bibinfo {author} {\bibfnamefont {A.}~\bibnamefont
  {Kaviraj}}\ and\ \bibinfo {author} {\bibfnamefont {M.~F.}\ \bibnamefont
  {Paulos}},\ }\href {https://doi.org/10.1007/JHEP04(2020)135} {\bibfield
  {journal} {\bibinfo  {journal} {JHEP}\ }\textbf {\bibinfo {volume} {04}},\
  \bibinfo {pages} {135}},\ \Eprint {https://arxiv.org/abs/1812.04034}
  {arXiv:1812.04034 [hep-th]} \BibitemShut {NoStop}%
\bibitem [{\citenamefont {Nishioka}\ \emph {et~al.}(2023)\citenamefont
  {Nishioka}, \citenamefont {Okuyama},\ and\ \citenamefont
  {Shimamori}}]{nishioka2023comments}%
  \BibitemOpen
  \bibfield  {author} {\bibinfo {author} {\bibfnamefont {T.}~\bibnamefont
  {Nishioka}}, \bibinfo {author} {\bibfnamefont {Y.}~\bibnamefont {Okuyama}},\
  and\ \bibinfo {author} {\bibfnamefont {S.}~\bibnamefont {Shimamori}},\ }\href
  {https://doi.org/10.1007/JHEP03(2023)051} {\bibfield  {journal} {\bibinfo
  {journal} {JHEP}\ }\textbf {\bibinfo {volume} {03}},\ \bibinfo {pages}
  {051}},\ \Eprint {https://arxiv.org/abs/2212.04078} {arXiv:2212.04078
  [hep-th]} \BibitemShut {NoStop}%
\bibitem [{\citenamefont {Liendo}\ \emph {et~al.}(2013)\citenamefont {Liendo},
  \citenamefont {Rastelli},\ and\ \citenamefont {van
  Rees}}]{liendo2013bootstrap}%
  \BibitemOpen
  \bibfield  {author} {\bibinfo {author} {\bibfnamefont {P.}~\bibnamefont
  {Liendo}}, \bibinfo {author} {\bibfnamefont {L.}~\bibnamefont {Rastelli}},\
  and\ \bibinfo {author} {\bibfnamefont {B.~C.}\ \bibnamefont {van Rees}},\
  }\href {https://doi.org/10.1007/JHEP07(2013)113} {\bibfield  {journal}
  {\bibinfo  {journal} {JHEP}\ }\textbf {\bibinfo {volume} {07}},\ \bibinfo
  {pages} {113}},\ \Eprint {https://arxiv.org/abs/1210.4258} {arXiv:1210.4258
  [hep-th]} \BibitemShut {NoStop}%
\bibitem [{\citenamefont {Gliozzi}\ \emph {et~al.}(2015)\citenamefont
  {Gliozzi}, \citenamefont {Liendo}, \citenamefont {Meineri},\ and\
  \citenamefont {Rago}}]{gliozzi2015boundary}%
  \BibitemOpen
  \bibfield  {author} {\bibinfo {author} {\bibfnamefont {F.}~\bibnamefont
  {Gliozzi}}, \bibinfo {author} {\bibfnamefont {P.}~\bibnamefont {Liendo}},
  \bibinfo {author} {\bibfnamefont {M.}~\bibnamefont {Meineri}},\ and\ \bibinfo
  {author} {\bibfnamefont {A.}~\bibnamefont {Rago}},\ }\href
  {https://doi.org/10.1007/JHEP05(2015)036} {\bibfield  {journal} {\bibinfo
  {journal} {JHEP}\ }\textbf {\bibinfo {volume} {05}},\ \bibinfo {pages}
  {036}},\ \bibinfo {note} {[Erratum: JHEP 12, 093 (2021)]},\ \Eprint
  {https://arxiv.org/abs/1502.07217} {arXiv:1502.07217 [hep-th]} \BibitemShut
  {NoStop}%
\bibitem [{\citenamefont {Behan}\ \emph {et~al.}(2020)\citenamefont {Behan},
  \citenamefont {Di~Pietro}, \citenamefont {Lauria},\ and\ \citenamefont
  {Van~Rees}}]{behan2020bootstrapping}%
  \BibitemOpen
  \bibfield  {author} {\bibinfo {author} {\bibfnamefont {C.}~\bibnamefont
  {Behan}}, \bibinfo {author} {\bibfnamefont {L.}~\bibnamefont {Di~Pietro}},
  \bibinfo {author} {\bibfnamefont {E.}~\bibnamefont {Lauria}},\ and\ \bibinfo
  {author} {\bibfnamefont {B.~C.}\ \bibnamefont {Van~Rees}},\ }\href
  {https://doi.org/10.1007/JHEP12(2020)182} {\bibfield  {journal} {\bibinfo
  {journal} {JHEP}\ }\textbf {\bibinfo {volume} {12}},\ \bibinfo {pages}
  {182}},\ \Eprint {https://arxiv.org/abs/2009.03336} {arXiv:2009.03336
  [hep-th]} \BibitemShut {NoStop}%
\bibitem [{\citenamefont {Bissi}\ \emph {et~al.}(2019)\citenamefont {Bissi},
  \citenamefont {Hansen},\ and\ \citenamefont
  {S{\"o}derberg}}]{bissi2019analytic}%
  \BibitemOpen
  \bibfield  {author} {\bibinfo {author} {\bibfnamefont {A.}~\bibnamefont
  {Bissi}}, \bibinfo {author} {\bibfnamefont {T.}~\bibnamefont {Hansen}},\ and\
  \bibinfo {author} {\bibfnamefont {A.}~\bibnamefont {S{\"o}derberg}},\ }\href
  {https://doi.org/10.1007/JHEP01(2019)010} {\bibfield  {journal} {\bibinfo
  {journal} {JHEP}\ }\textbf {\bibinfo {volume} {01}},\ \bibinfo {pages}
  {010}},\ \Eprint {https://arxiv.org/abs/1808.08155} {arXiv:1808.08155
  [hep-th]} \BibitemShut {NoStop}%
\bibitem [{\citenamefont {Dey}\ and\ \citenamefont
  {S{\"o}derberg}(2021)}]{dey2021analytic}%
  \BibitemOpen
  \bibfield  {author} {\bibinfo {author} {\bibfnamefont {P.}~\bibnamefont
  {Dey}}\ and\ \bibinfo {author} {\bibfnamefont {A.}~\bibnamefont
  {S{\"o}derberg}},\ }\href {https://doi.org/10.1007/JHEP07(2021)013}
  {\bibfield  {journal} {\bibinfo  {journal} {JHEP}\ }\textbf {\bibinfo
  {volume} {07}},\ \bibinfo {pages} {013}},\ \Eprint
  {https://arxiv.org/abs/2012.11344} {arXiv:2012.11344 [hep-th]} \BibitemShut
  {NoStop}%
\bibitem [{\citenamefont {Behan}\ \emph {et~al.}(2022)\citenamefont {Behan},
  \citenamefont {Di~Pietro}, \citenamefont {Lauria},\ and\ \citenamefont {van
  Rees}}]{behan2022bootstrapping}%
  \BibitemOpen
  \bibfield  {author} {\bibinfo {author} {\bibfnamefont {C.}~\bibnamefont
  {Behan}}, \bibinfo {author} {\bibfnamefont {L.}~\bibnamefont {Di~Pietro}},
  \bibinfo {author} {\bibfnamefont {E.}~\bibnamefont {Lauria}},\ and\ \bibinfo
  {author} {\bibfnamefont {B.~C.}\ \bibnamefont {van Rees}},\ }\href
  {https://doi.org/10.1007/JHEP03(2022)146} {\bibfield  {journal} {\bibinfo
  {journal} {JHEP}\ }\textbf {\bibinfo {volume} {03}},\ \bibinfo {pages}
  {146}},\ \Eprint {https://arxiv.org/abs/2111.04747} {arXiv:2111.04747
  [hep-th]} \BibitemShut {NoStop}%
\bibitem [{\citenamefont {Diehl}\ and\ \citenamefont
  {Dietrich}(1981)}]{diehl1981field}%
  \BibitemOpen
  \bibfield  {author} {\bibinfo {author} {\bibfnamefont {H.~W.}\ \bibnamefont
  {Diehl}}\ and\ \bibinfo {author} {\bibfnamefont {S.}~\bibnamefont
  {Dietrich}},\ }\href {https://doi.org/10.1007/BF01298293} {\bibfield
  {journal} {\bibinfo  {journal} {Z. Phys. B}\ }\textbf {\bibinfo {volume}
  {42}},\ \bibinfo {pages} {65} (\bibinfo {year} {1981})}\BibitemShut {NoStop}%
\bibitem [{\citenamefont {Diehl}\ and\ \citenamefont
  {Dietrich}(1983)}]{diehl1983multicritical}%
  \BibitemOpen
  \bibfield  {author} {\bibinfo {author} {\bibfnamefont {H.}~\bibnamefont
  {Diehl}}\ and\ \bibinfo {author} {\bibfnamefont {S.}~\bibnamefont
  {Dietrich}},\ }\href {https://doi.org/10.1007/BF01304094} {\bibfield
  {journal} {\bibinfo  {journal} {Zeitschrift f{\"u}r Physik B Condensed
  Matter}\ }\textbf {\bibinfo {volume} {50}},\ \bibinfo {pages} {117} (\bibinfo
  {year} {1983})}\BibitemShut {NoStop}%
\bibitem [{\citenamefont {Burkhardt}\ and\ \citenamefont
  {Cardy}(1987)}]{burkhardt1987surface}%
  \BibitemOpen
  \bibfield  {author} {\bibinfo {author} {\bibfnamefont {T.~W.}\ \bibnamefont
  {Burkhardt}}\ and\ \bibinfo {author} {\bibfnamefont {J.~L.}\ \bibnamefont
  {Cardy}},\ }\href {https://doi.org/10.1088/0305-4470/20/4/010} {\bibfield
  {journal} {\bibinfo  {journal} {Journal of Physics A: Mathematical and
  General}\ }\textbf {\bibinfo {volume} {20}},\ \bibinfo {pages} {L233}
  (\bibinfo {year} {1987})}\BibitemShut {NoStop}%
\bibitem [{\citenamefont {Batchelor}\ and\ \citenamefont
  {Yung}(1995)}]{batchelor1995surface}%
  \BibitemOpen
  \bibfield  {author} {\bibinfo {author} {\bibfnamefont {M.~T.}\ \bibnamefont
  {Batchelor}}\ and\ \bibinfo {author} {\bibfnamefont {C.~M.}\ \bibnamefont
  {Yung}},\ }\href {https://doi.org/10.1088/0305-4470/28/16/001} {\bibfield
  {journal} {\bibinfo  {journal} {J. Phys. A}\ }\textbf {\bibinfo {volume}
  {28}},\ \bibinfo {pages} {L421} (\bibinfo {year} {1995})},\ \Eprint
  {https://arxiv.org/abs/cond-mat/9507010} {arXiv:cond-mat/9507010}
  \BibitemShut {NoStop}%
\bibitem [{\citenamefont {Diehl}\ and\ \citenamefont
  {Shpot}(1998)}]{diehl1998massive}%
  \BibitemOpen
  \bibfield  {author} {\bibinfo {author} {\bibfnamefont {H.~W.}\ \bibnamefont
  {Diehl}}\ and\ \bibinfo {author} {\bibfnamefont {M.}~\bibnamefont {Shpot}},\
  }\href {https://doi.org/10.1016/S0550-3213(98)00489-1} {\bibfield  {journal}
  {\bibinfo  {journal} {Nucl. Phys. B}\ }\textbf {\bibinfo {volume} {528}},\
  \bibinfo {pages} {595} (\bibinfo {year} {1998})},\ \Eprint
  {https://arxiv.org/abs/cond-mat/9804083} {arXiv:cond-mat/9804083}
  \BibitemShut {NoStop}%
\bibitem [{\citenamefont {Toldin}\ and\ \citenamefont
  {Metlitski}(2022)}]{parisen2022boundary}%
  \BibitemOpen
  \bibfield  {author} {\bibinfo {author} {\bibfnamefont {F.~P.}\ \bibnamefont
  {Toldin}}\ and\ \bibinfo {author} {\bibfnamefont {M.~A.}\ \bibnamefont
  {Metlitski}},\ }\href {https://doi.org/10.1103/PhysRevLett.128.215701}
  {\bibfield  {journal} {\bibinfo  {journal} {Phys. Rev. Lett.}\ }\textbf
  {\bibinfo {volume} {128}},\ \bibinfo {pages} {215701} (\bibinfo {year}
  {2022})},\ \Eprint {https://arxiv.org/abs/2111.03613} {arXiv:2111.03613
  [cond-mat.stat-mech]} \BibitemShut {NoStop}%
\bibitem [{\citenamefont {Metlitski}(2022)}]{metlitski2022boundary}%
  \BibitemOpen
  \bibfield  {author} {\bibinfo {author} {\bibfnamefont {M.~A.}\ \bibnamefont
  {Metlitski}},\ }\href {https://doi.org/10.21468/SciPostPhys.12.4.131}
  {\bibfield  {journal} {\bibinfo  {journal} {SciPost Phys.}\ }\textbf
  {\bibinfo {volume} {12}},\ \bibinfo {pages} {131} (\bibinfo {year} {2022})},\
  \Eprint {https://arxiv.org/abs/2009.05119} {arXiv:2009.05119
  [cond-mat.str-el]} \BibitemShut {NoStop}%
\bibitem [{\citenamefont {Hu}\ \emph {et~al.}(2021)\citenamefont {Hu},
  \citenamefont {Deng},\ and\ \citenamefont {Lv}}]{hu2021extraordinary}%
  \BibitemOpen
  \bibfield  {author} {\bibinfo {author} {\bibfnamefont {M.}~\bibnamefont
  {Hu}}, \bibinfo {author} {\bibfnamefont {Y.}~\bibnamefont {Deng}},\ and\
  \bibinfo {author} {\bibfnamefont {J.-P.}\ \bibnamefont {Lv}},\ }\href
  {https://doi.org/10.1103/PhysRevLett.127.120603} {\bibfield  {journal}
  {\bibinfo  {journal} {Phys. Rev. Lett.}\ }\textbf {\bibinfo {volume} {127}},\
  \bibinfo {pages} {120603} (\bibinfo {year} {2021})},\ \Eprint
  {https://arxiv.org/abs/2104.05152} {arXiv:2104.05152 [cond-mat.stat-mech]}
  \BibitemShut {NoStop}%
\bibitem [{\citenamefont {Padayasi}\ \emph {et~al.}(2022)\citenamefont
  {Padayasi}, \citenamefont {Krishnan}, \citenamefont {Metlitski},
  \citenamefont {Gruzberg},\ and\ \citenamefont
  {Meineri}}]{padayasi2022extraordinary}%
  \BibitemOpen
  \bibfield  {author} {\bibinfo {author} {\bibfnamefont {J.}~\bibnamefont
  {Padayasi}}, \bibinfo {author} {\bibfnamefont {A.}~\bibnamefont {Krishnan}},
  \bibinfo {author} {\bibfnamefont {M.~A.}\ \bibnamefont {Metlitski}}, \bibinfo
  {author} {\bibfnamefont {I.~A.}\ \bibnamefont {Gruzberg}},\ and\ \bibinfo
  {author} {\bibfnamefont {M.}~\bibnamefont {Meineri}},\ }\href
  {https://doi.org/10.21468/SciPostPhys.12.6.190} {\bibfield  {journal}
  {\bibinfo  {journal} {SciPost Phys.}\ }\textbf {\bibinfo {volume} {12}},\
  \bibinfo {pages} {190} (\bibinfo {year} {2022})},\ \Eprint
  {https://arxiv.org/abs/2111.03071} {arXiv:2111.03071 [cond-mat.stat-mech]}
  \BibitemShut {NoStop}%
\bibitem [{\citenamefont {Sun}\ \emph {et~al.}(2023)\citenamefont {Sun},
  \citenamefont {Hu}, \citenamefont {Deng},\ and\ \citenamefont
  {Lv}}]{sun2023extraordinary}%
  \BibitemOpen
  \bibfield  {author} {\bibinfo {author} {\bibfnamefont {Y.}~\bibnamefont
  {Sun}}, \bibinfo {author} {\bibfnamefont {M.}~\bibnamefont {Hu}}, \bibinfo
  {author} {\bibfnamefont {Y.}~\bibnamefont {Deng}},\ and\ \bibinfo {author}
  {\bibfnamefont {J.-P.}\ \bibnamefont {Lv}},\ }\href
  {https://doi.org/10.1103/PhysRevLett.131.207101} {\bibfield  {journal}
  {\bibinfo  {journal} {Phys. Rev. Lett.}\ }\textbf {\bibinfo {volume} {131}},\
  \bibinfo {pages} {207101} (\bibinfo {year} {2023})},\ \Eprint
  {https://arxiv.org/abs/2301.11720} {arXiv:2301.11720 [cond-mat.stat-mech]}
  \BibitemShut {NoStop}%
\bibitem [{\citenamefont {Toldin}\ \emph {et~al.}(2025)\citenamefont {Toldin},
  \citenamefont {Krishnan},\ and\ \citenamefont
  {Metlitski}}]{parisen2024universal}%
  \BibitemOpen
  \bibfield  {author} {\bibinfo {author} {\bibfnamefont {F.~P.}\ \bibnamefont
  {Toldin}}, \bibinfo {author} {\bibfnamefont {A.}~\bibnamefont {Krishnan}},\
  and\ \bibinfo {author} {\bibfnamefont {M.~A.}\ \bibnamefont {Metlitski}},\
  }\href {https://doi.org/10.1103/PhysRevResearch.7.023052} {\bibfield
  {journal} {\bibinfo  {journal} {Phys. Rev. Res.}\ }\textbf {\bibinfo {volume}
  {7}},\ \bibinfo {pages} {023052} (\bibinfo {year} {2025})},\ \Eprint
  {https://arxiv.org/abs/2411.05089} {arXiv:2411.05089 [cond-mat.stat-mech]}
  \BibitemShut {NoStop}%
\bibitem [{\citenamefont {Cuomo}\ and\ \citenamefont
  {Zhang}(2024)}]{cuomo2024spontaneous}%
  \BibitemOpen
  \bibfield  {author} {\bibinfo {author} {\bibfnamefont {G.}~\bibnamefont
  {Cuomo}}\ and\ \bibinfo {author} {\bibfnamefont {S.}~\bibnamefont {Zhang}},\
  }\href {https://doi.org/10.1007/JHEP03(2024)022} {\bibfield  {journal}
  {\bibinfo  {journal} {JHEP}\ }\textbf {\bibinfo {volume} {03}},\ \bibinfo
  {pages} {022}},\ \Eprint {https://arxiv.org/abs/2306.00085} {arXiv:2306.00085
  [hep-th]} \BibitemShut {NoStop}%
\bibitem [{\citenamefont {Sun}\ and\ \citenamefont
  {Jian}(2025)}]{sun2025boundary}%
  \BibitemOpen
  \bibfield  {author} {\bibinfo {author} {\bibfnamefont {X.}~\bibnamefont
  {Sun}}\ and\ \bibinfo {author} {\bibfnamefont {S.-K.}\ \bibnamefont {Jian}},\
  }\href {https://doi.org/10.21468/SciPostPhys.18.6.210} {\bibfield  {journal}
  {\bibinfo  {journal} {SciPost Phys.}\ }\textbf {\bibinfo {volume} {18}},\
  \bibinfo {pages} {210} (\bibinfo {year} {2025})},\ \Eprint
  {https://arxiv.org/abs/2501.06287} {arXiv:2501.06287 [cond-mat.str-el]}
  \BibitemShut {NoStop}%
\bibitem [{\citenamefont {van Loon}(2018)}]{van2018analytic}%
  \BibitemOpen
  \bibfield  {author} {\bibinfo {author} {\bibfnamefont {M.}~\bibnamefont {van
  Loon}},\ }\href {https://doi.org/10.1007/JHEP01(2018)104} {\bibfield
  {journal} {\bibinfo  {journal} {JHEP}\ }\textbf {\bibinfo {volume} {01}},\
  \bibinfo {pages} {104}},\ \Eprint {https://arxiv.org/abs/1711.02099}
  {arXiv:1711.02099 [hep-th]} \BibitemShut {NoStop}%
\bibitem [{\citenamefont {Giombi}\ \emph {et~al.}(2022)\citenamefont {Giombi},
  \citenamefont {Helfenberger},\ and\ \citenamefont
  {Khanchandani}}]{giombi2022fermions}%
  \BibitemOpen
  \bibfield  {author} {\bibinfo {author} {\bibfnamefont {S.}~\bibnamefont
  {Giombi}}, \bibinfo {author} {\bibfnamefont {E.}~\bibnamefont
  {Helfenberger}},\ and\ \bibinfo {author} {\bibfnamefont {H.}~\bibnamefont
  {Khanchandani}},\ }\href {https://doi.org/10.1007/JHEP07(2022)018} {\bibfield
   {journal} {\bibinfo  {journal} {JHEP}\ }\textbf {\bibinfo {volume} {07}},\
  \bibinfo {pages} {018}},\ \Eprint {https://arxiv.org/abs/2110.04268}
  {arXiv:2110.04268 [hep-th]} \BibitemShut {NoStop}%
\bibitem [{\citenamefont {Herzog}\ and\ \citenamefont
  {Schaub}(2023)}]{herzog2023fermions}%
  \BibitemOpen
  \bibfield  {author} {\bibinfo {author} {\bibfnamefont {C.~P.}\ \bibnamefont
  {Herzog}}\ and\ \bibinfo {author} {\bibfnamefont {V.}~\bibnamefont
  {Schaub}},\ }\href {https://doi.org/10.1007/JHEP02(2023)129} {\bibfield
  {journal} {\bibinfo  {journal} {JHEP}\ }\textbf {\bibinfo {volume} {02}},\
  \bibinfo {pages} {129}},\ \Eprint {https://arxiv.org/abs/2209.05511}
  {arXiv:2209.05511 [hep-th]} \BibitemShut {NoStop}%
\bibitem [{\citenamefont {Giombi}\ \emph {et~al.}(2023)\citenamefont {Giombi},
  \citenamefont {Helfenberger},\ and\ \citenamefont
  {Khanchandani}}]{giombi2023line}%
  \BibitemOpen
  \bibfield  {author} {\bibinfo {author} {\bibfnamefont {S.}~\bibnamefont
  {Giombi}}, \bibinfo {author} {\bibfnamefont {E.}~\bibnamefont
  {Helfenberger}},\ and\ \bibinfo {author} {\bibfnamefont {H.}~\bibnamefont
  {Khanchandani}},\ }\href {https://doi.org/10.1007/JHEP08(2023)224} {\bibfield
   {journal} {\bibinfo  {journal} {JHEP}\ }\textbf {\bibinfo {volume} {08}},\
  \bibinfo {pages} {224}},\ \Eprint {https://arxiv.org/abs/2211.11073}
  {arXiv:2211.11073 [hep-th]} \BibitemShut {NoStop}%
\bibitem [{\citenamefont {Shen}\ \emph {et~al.}(2025)\citenamefont {Shen},
  \citenamefont {Wu},\ and\ \citenamefont {Jian}}]{shen2024new}%
  \BibitemOpen
  \bibfield  {author} {\bibinfo {author} {\bibfnamefont {X.}~\bibnamefont
  {Shen}}, \bibinfo {author} {\bibfnamefont {Z.}~\bibnamefont {Wu}},\ and\
  \bibinfo {author} {\bibfnamefont {S.-K.}\ \bibnamefont {Jian}},\ }\href
  {https://doi.org/10.1103/4lv4-mc81} {\bibfield  {journal} {\bibinfo
  {journal} {Phys. Rev. B}\ }\textbf {\bibinfo {volume} {112}},\ \bibinfo
  {pages} {L041118} (\bibinfo {year} {2025})}\BibitemShut {NoStop}%
\bibitem [{\citenamefont {Jiang}\ \emph {et~al.}(2025)\citenamefont {Jiang},
  \citenamefont {Ge},\ and\ \citenamefont {Jian}}]{jiang2025boundary}%
  \BibitemOpen
  \bibfield  {author} {\bibinfo {author} {\bibfnamefont {H.}~\bibnamefont
  {Jiang}}, \bibinfo {author} {\bibfnamefont {Y.}~\bibnamefont {Ge}},\ and\
  \bibinfo {author} {\bibfnamefont {S.-K.}\ \bibnamefont {Jian}},\ }\href
  {https://doi.org/10.1103/tjfk-84f8} {\bibfield  {journal} {\bibinfo
  {journal} {Phys. Rev. Lett.}\ }\textbf {\bibinfo {volume} {135}},\ \bibinfo
  {pages} {141602} (\bibinfo {year} {2025})}\BibitemShut {NoStop}%
\bibitem [{\citenamefont {Barrat}\ \emph {et~al.}(2025)\citenamefont {Barrat},
  \citenamefont {Liendo},\ and\ \citenamefont {van Vliet}}]{barrat2025line}%
  \BibitemOpen
  \bibfield  {author} {\bibinfo {author} {\bibfnamefont {J.}~\bibnamefont
  {Barrat}}, \bibinfo {author} {\bibfnamefont {P.}~\bibnamefont {Liendo}},\
  and\ \bibinfo {author} {\bibfnamefont {P.}~\bibnamefont {van Vliet}},\ }\href
  {https://doi.org/10.1007/JHEP05(2025)146} {\bibfield  {journal} {\bibinfo
  {journal} {JHEP}\ }\textbf {\bibinfo {volume} {05}},\ \bibinfo {pages}
  {146}},\ \Eprint {https://arxiv.org/abs/2304.13588} {arXiv:2304.13588
  [hep-th]} \BibitemShut {NoStop}%
\bibitem [{\citenamefont {Diehl}\ and\ \citenamefont
  {Ciach}(1991)}]{diehl1991surface}%
  \BibitemOpen
  \bibfield  {author} {\bibinfo {author} {\bibfnamefont {H.~W.}\ \bibnamefont
  {Diehl}}\ and\ \bibinfo {author} {\bibfnamefont {A.}~\bibnamefont {Ciach}},\
  }\href {https://doi.org/10.1103/PhysRevB.44.6642} {\bibfield  {journal}
  {\bibinfo  {journal} {Phys. Rev. B}\ }\textbf {\bibinfo {volume} {44}},\
  \bibinfo {pages} {6642} (\bibinfo {year} {1991})}\BibitemShut {NoStop}%
\bibitem [{\citenamefont {Eisenriegler}\ and\ \citenamefont
  {Diehl}(1988)}]{eisenriegler1988surface}%
  \BibitemOpen
  \bibfield  {author} {\bibinfo {author} {\bibfnamefont {E.}~\bibnamefont
  {Eisenriegler}}\ and\ \bibinfo {author} {\bibfnamefont {H.~W.}\ \bibnamefont
  {Diehl}},\ }\href {https://doi.org/10.1103/physrevb.37.5257} {\bibfield
  {journal} {\bibinfo  {journal} {Phys. Rev. B}\ }\textbf {\bibinfo {volume}
  {37}},\ \bibinfo {pages} {5257} (\bibinfo {year} {1988})}\BibitemShut
  {NoStop}%
\bibitem [{\citenamefont {Herzog}\ and\ \citenamefont
  {Shamir}(2019)}]{herzog2019marginal}%
  \BibitemOpen
  \bibfield  {author} {\bibinfo {author} {\bibfnamefont {C.~P.}\ \bibnamefont
  {Herzog}}\ and\ \bibinfo {author} {\bibfnamefont {I.}~\bibnamefont
  {Shamir}},\ }\href {https://doi.org/10.1007/JHEP10(2019)088} {\bibfield
  {journal} {\bibinfo  {journal} {JHEP}\ }\textbf {\bibinfo {volume} {10}},\
  \bibinfo {pages} {088}},\ \Eprint {https://arxiv.org/abs/1906.11281}
  {arXiv:1906.11281 [hep-th]} \BibitemShut {NoStop}%
\bibitem [{\citenamefont {Harribey}\ \emph {et~al.}(2023)\citenamefont
  {Harribey}, \citenamefont {Klebanov},\ and\ \citenamefont
  {Sun}}]{harribey2023boundaries}%
  \BibitemOpen
  \bibfield  {author} {\bibinfo {author} {\bibfnamefont {S.}~\bibnamefont
  {Harribey}}, \bibinfo {author} {\bibfnamefont {I.~R.}\ \bibnamefont
  {Klebanov}},\ and\ \bibinfo {author} {\bibfnamefont {Z.}~\bibnamefont
  {Sun}},\ }\href {https://doi.org/10.1007/JHEP10(2023)017} {\bibfield
  {journal} {\bibinfo  {journal} {JHEP}\ }\textbf {\bibinfo {volume} {10}},\
  \bibinfo {pages} {017}},\ \Eprint {https://arxiv.org/abs/2307.00072}
  {arXiv:2307.00072 [hep-th]} \BibitemShut {NoStop}%
\bibitem [{\citenamefont {Speth}(1983)}]{speth1983tricritical}%
  \BibitemOpen
  \bibfield  {author} {\bibinfo {author} {\bibfnamefont {W.}~\bibnamefont
  {Speth}},\ }\href {https://doi.org/10.1007/BF01319219} {\bibfield  {journal}
  {\bibinfo  {journal} {Zeitschrift f{\"u}r Physik B Condensed Matter}\
  }\textbf {\bibinfo {volume} {51}},\ \bibinfo {pages} {361} (\bibinfo {year}
  {1983})}\BibitemShut {NoStop}%
\bibitem [{\citenamefont {Guo}\ and\ \citenamefont
  {Li}(2026)}]{guo2026boundary}%
  \BibitemOpen
  \bibfield  {author} {\bibinfo {author} {\bibfnamefont {Y.}~\bibnamefont
  {Guo}}\ and\ \bibinfo {author} {\bibfnamefont {W.}~\bibnamefont {Li}},\
  }\href {https://arxiv.org/abs/2605.16119} {\bibinfo {title} {Boundary
  anomalous dimensions from bcft: $\phi^{3}$ theories with a boundary and
  higher-derivative generalizations}} (\bibinfo {year} {2026}),\ \Eprint
  {https://arxiv.org/abs/2605.16119} {arXiv:2605.16119 [hep-th]} \BibitemShut
  {NoStop}%
\bibitem [{\citenamefont {Proch{\'a}zka}\ and\ \citenamefont
  {S{\"o}derberg}(2020)}]{prochazka2020composite}%
  \BibitemOpen
  \bibfield  {author} {\bibinfo {author} {\bibfnamefont {V.}~\bibnamefont
  {Proch{\'a}zka}}\ and\ \bibinfo {author} {\bibfnamefont {A.}~\bibnamefont
  {S{\"o}derberg}},\ }\href {https://doi.org/10.1007/JHEP03(2020)114}
  {\bibfield  {journal} {\bibinfo  {journal} {JHEP}\ }\textbf {\bibinfo
  {volume} {03}},\ \bibinfo {pages} {114}},\ \Eprint
  {https://arxiv.org/abs/1912.07505} {arXiv:1912.07505 [hep-th]} \BibitemShut
  {NoStop}%
\bibitem [{\citenamefont {McAvity}\ and\ \citenamefont
  {Osborn}(1995)}]{mcavity1995conformal}%
  \BibitemOpen
  \bibfield  {author} {\bibinfo {author} {\bibfnamefont {D.~M.}\ \bibnamefont
  {McAvity}}\ and\ \bibinfo {author} {\bibfnamefont {H.}~\bibnamefont
  {Osborn}},\ }\href {https://doi.org/10.1016/0550-3213(95)00476-9} {\bibfield
  {journal} {\bibinfo  {journal} {Nucl. Phys. B}\ }\textbf {\bibinfo {volume}
  {455}},\ \bibinfo {pages} {522} (\bibinfo {year} {1995})},\ \Eprint
  {https://arxiv.org/abs/cond-mat/9505127} {arXiv:cond-mat/9505127}
  \BibitemShut {NoStop}%
\bibitem [{Note1()}]{Note1}%
  \BibitemOpen
  \bibinfo {note} {We consider the dimension $d_0$ to be either $d_0 = 3$ or
  $d_0=4$ in this paper}\BibitemShut {NoStop}%
\bibitem [{Note2()}]{Note2}%
  \BibitemOpen
  \bibinfo {note} {Note that for noninteger $\Delta _n/2$ and $\protect \hat
  {\Delta }_m$, both bulk and boundary conformal blocks have a branch cut for
  $\xi <0$ due to the prefactors multiplying the hypergeometric function in
  Eqs.~\protect \textup {\hbox {\mathsurround \z@ \protect \normalfont
  (\ignorespaces \ref {eq:bulk_conformal_block}\unskip \@@italiccorr )}}
  and~\protect \textup {\hbox {\mathsurround \z@ \protect \normalfont
  (\ignorespaces \ref {eq:boundary_conformal_block}\unskip \@@italiccorr
  )}}.}\BibitemShut {Stop}%
\bibitem [{Note3()}]{Note3}%
  \BibitemOpen
  \bibinfo {note} {Note that for $d=4$ with a bulk-$\phi ^4$ interaction, the
  equation of motion implies $\DOTSB \sum@ \slimits@ _i\phi _i\partial ^2\phi
  _i\sim \phi ^4$.}\BibitemShut {Stop}%
\bibitem [{{\relax DLMF}()}]{NIST:DLMF}%
  \BibitemOpen
  {\relax DLMF},\ \href {https://dlmf.nist.gov/} {\bibinfo {title} {{\it NIST
  Digital Library of Mathematical Functions}}},\ \bibinfo {howpublished}
  {\url{https://dlmf.nist.gov/}, Release 1.2.4 of 2025-03-15},\ \bibinfo {note}
  {f.~W.~J. Olver, A.~B. {Olde Daalhuis}, D.~W. Lozier, B.~I. Schneider, R.~F.
  Boisvert, C.~W. Clark, B.~R. Miller, B.~V. Saunders, H.~S. Cohl, and M.~A.
  McClain, eds.}\BibitemShut {Stop}%
\bibitem [{\citenamefont {Heemskerk}\ \emph {et~al.}(2009)\citenamefont
  {Heemskerk}, \citenamefont {Penedones}, \citenamefont {Polchinski},\ and\
  \citenamefont {Sully}}]{heemskerk2009holography}%
  \BibitemOpen
  \bibfield  {author} {\bibinfo {author} {\bibfnamefont {I.}~\bibnamefont
  {Heemskerk}}, \bibinfo {author} {\bibfnamefont {J.}~\bibnamefont
  {Penedones}}, \bibinfo {author} {\bibfnamefont {J.}~\bibnamefont
  {Polchinski}},\ and\ \bibinfo {author} {\bibfnamefont {J.}~\bibnamefont
  {Sully}},\ }\href {https://doi.org/10.1088/1126-6708/2009/10/079} {\bibfield
  {journal} {\bibinfo  {journal} {JHEP}\ }\textbf {\bibinfo {volume} {10}},\
  \bibinfo {pages} {079}},\ \Eprint {https://arxiv.org/abs/0907.0151}
  {arXiv:0907.0151 [hep-th]} \BibitemShut {NoStop}%
\bibitem [{\citenamefont {Osborn}(1991)}]{osborn1991weyl}%
  \BibitemOpen
  \bibfield  {author} {\bibinfo {author} {\bibfnamefont {H.}~\bibnamefont
  {Osborn}},\ }\href
  {https://doi.org/https://doi.org/10.1016/0550-3213(91)80030-P} {\bibfield
  {journal} {\bibinfo  {journal} {Nuclear Physics B}\ }\textbf {\bibinfo
  {volume} {363}},\ \bibinfo {pages} {486} (\bibinfo {year}
  {1991})}\BibitemShut {NoStop}%
\bibitem [{Note4()}]{Note4}%
  \BibitemOpen
  \bibinfo {note} {Here we do not present a full proof. Instead, we verify the
  identity numerically. An analytic proof may be obtained by expanding both
  hypergeometric functions in series and isolating the \(w^{-1}\) term in the
  integrand, and the coefficient of this term reproduces the \(\delta
  \)-function.}\BibitemShut {Stop}%
\bibitem [{\citenamefont {Wilson}(1973)}]{wilson1973quantum}%
  \BibitemOpen
  \bibfield  {author} {\bibinfo {author} {\bibfnamefont {K.~G.}\ \bibnamefont
  {Wilson}},\ }\href {https://doi.org/10.1103/PhysRevD.7.2911} {\bibfield
  {journal} {\bibinfo  {journal} {Phys. Rev. D}\ }\textbf {\bibinfo {volume}
  {7}},\ \bibinfo {pages} {2911} (\bibinfo {year} {1973})}\BibitemShut
  {NoStop}%
\bibitem [{Note5()}]{Note5}%
  \BibitemOpen
  \bibinfo {note} {If we include composite operators on the boundary, they may
  acquire nontrivial anomalous dimensions~\cite
  {prochazka2020composite}.}\BibitemShut {Stop}%
\bibitem [{\citenamefont {de~Alcantara~Bonfim}\ \emph
  {et~al.}(1981)\citenamefont {de~Alcantara~Bonfim}, \citenamefont {Kirkham},\
  and\ \citenamefont {McKane}}]{de1981critical}%
  \BibitemOpen
  \bibfield  {author} {\bibinfo {author} {\bibfnamefont {O.~F.}\ \bibnamefont
  {de~Alcantara~Bonfim}}, \bibinfo {author} {\bibfnamefont {J.~E.}\
  \bibnamefont {Kirkham}},\ and\ \bibinfo {author} {\bibfnamefont {A.~J.}\
  \bibnamefont {McKane}},\ }\href {https://doi.org/10.1088/0305-4470/14/9/034}
  {\bibfield  {journal} {\bibinfo  {journal} {J. Phys. A}\ }\textbf {\bibinfo
  {volume} {14}},\ \bibinfo {pages} {2391} (\bibinfo {year}
  {1981})}\BibitemShut {NoStop}%
\bibitem [{Note6()}]{Note6}%
  \BibitemOpen
  \bibinfo {note} {For $N=1$, the cubic tensor $d_{ijk}$ vanishes, and the
  Potts model is identical to the Ising model. Consequently, the Yang-Lee edge
  singularity is not obtained by simply setting $N=1$ in the Potts $\phi ^3$
  theory.}\BibitemShut {Stop}%
\bibitem [{Note7()}]{Note7}%
  \BibitemOpen
  \bibinfo {note} {The special fixed point is stable; the Yang-Lee fixed point
  is attractive along the boundary coupling but unstable against the bulk
  coupling; and the new fixed point is a saddle with one attractive and one
  repulsive direction.}\BibitemShut {Stop}%
\bibitem [{\citenamefont {Prosen}(2011)}]{Prosen2011Exact}%
  \BibitemOpen
  \bibfield  {author} {\bibinfo {author} {\bibfnamefont {T.}~\bibnamefont
  {Prosen}},\ }\href {https://doi.org/10.1103/PhysRevLett.107.137201}
  {\bibfield  {journal} {\bibinfo  {journal} {Phys. Rev. Lett.}\ }\textbf
  {\bibinfo {volume} {107}},\ \bibinfo {pages} {137201} (\bibinfo {year}
  {2011})},\ \Eprint {https://arxiv.org/abs/1106.2978} {arXiv:1106.2978
  [quant-ph]} \BibitemShut {NoStop}%
\bibitem [{\citenamefont {Wang}\ \emph {et~al.}(2023)\citenamefont {Wang},
  \citenamefont {Li}, \citenamefont {Song},\ and\ \citenamefont
  {Wang}}]{Wang2023Scale}%
  \BibitemOpen
  \bibfield  {author} {\bibinfo {author} {\bibfnamefont {H.-R.}\ \bibnamefont
  {Wang}}, \bibinfo {author} {\bibfnamefont {B.}~\bibnamefont {Li}}, \bibinfo
  {author} {\bibfnamefont {F.}~\bibnamefont {Song}},\ and\ \bibinfo {author}
  {\bibfnamefont {Z.}~\bibnamefont {Wang}},\ }\href
  {https://doi.org/10.21468/SciPostPhys.15.5.191} {\bibfield  {journal}
  {\bibinfo  {journal} {SciPost Phys.}\ }\textbf {\bibinfo {volume} {15}},\
  \bibinfo {pages} {191} (\bibinfo {year} {2023})},\ \Eprint
  {https://arxiv.org/abs/2301.11896} {arXiv:2301.11896 [quant-ph]} \BibitemShut
  {NoStop}%
\bibitem [{Note8()}]{Note8}%
  \BibitemOpen
  \bibinfo {note} {Setting $C_5'=1$ for the Yang-Lee type interaction, this
  reproduces the results in Ref.~\cite {diehl1991surface}.}\BibitemShut {Stop}%
\bibitem [{Note9()}]{Note9}%
  \BibitemOpen
  \bibinfo {note} {Note that $C_5'=1$ for the boundary Yang-Lee type
  interaction.}\BibitemShut {Stop}%
\bibitem [{Note10()}]{Note10}%
  \BibitemOpen
  \bibinfo {note} {Note, however, that composite boundary operators such as
  $\protect \hat {\phi }^2$ can acquire nontrivial $\protect \mathcal
  {O}(\epsilon ^{1/2})$ anomalous dimensions~\cite
  {eisenriegler1988surface}.}\BibitemShut {Stop}%
\bibitem [{\citenamefont {Symanzik}(1981)}]{symanzik1981schrodinger}%
  \BibitemOpen
  \bibfield  {author} {\bibinfo {author} {\bibfnamefont {K.}~\bibnamefont
  {Symanzik}},\ }\href {https://doi.org/10.1016/0550-3213(81)90482-X}
  {\bibfield  {journal} {\bibinfo  {journal} {Nucl. Phys. B}\ }\textbf
  {\bibinfo {volume} {190}},\ \bibinfo {pages} {1} (\bibinfo {year}
  {1981})}\BibitemShut {NoStop}%
\bibitem [{\citenamefont {Collins}(1984)}]{collins1984renormalization}%
  \BibitemOpen
  \bibfield  {author} {\bibinfo {author} {\bibfnamefont {J.~C.}\ \bibnamefont
  {Collins}},\ }\href {https://doi.org/10.1017/9781009401807} {\emph {\bibinfo
  {title} {{Renormalization : An Introduction to Renormalization, the
  Renormalization Group and the Operator-Product Expansion}}}},\ \bibinfo
  {series} {Cambridge Monographs on Mathematical Physics}, Vol.~\bibinfo
  {volume} {26}\ (\bibinfo  {publisher} {Cambridge University Press},\ \bibinfo
  {address} {Cambridge},\ \bibinfo {year} {1984})\BibitemShut {NoStop}%
\bibitem [{\citenamefont {Gorishnii}(1989)}]{gorishny1989construction}%
  \BibitemOpen
  \bibfield  {author} {\bibinfo {author} {\bibfnamefont {S.~G.}\ \bibnamefont
  {Gorishnii}},\ }\href {https://doi.org/10.1016/0550-3213(89)90622-6}
  {\bibfield  {journal} {\bibinfo  {journal} {Nucl. Phys. B}\ }\textbf
  {\bibinfo {volume} {319}},\ \bibinfo {pages} {633} (\bibinfo {year}
  {1989})}\BibitemShut {NoStop}%
\bibitem [{\citenamefont {Adzhemyan}\ \emph {et~al.}(1998)\citenamefont
  {Adzhemyan}, \citenamefont {Antonov},\ and\ \citenamefont
  {Vasil'ev}}]{adzhemyan1998renormalization}%
  \BibitemOpen
  \bibfield  {author} {\bibinfo {author} {\bibfnamefont {L.~T.}\ \bibnamefont
  {Adzhemyan}}, \bibinfo {author} {\bibfnamefont {N.~V.}\ \bibnamefont
  {Antonov}},\ and\ \bibinfo {author} {\bibfnamefont {A.~N.}\ \bibnamefont
  {Vasil'ev}},\ }\href {https://doi.org/10.1103/PhysRevE.58.1823} {\bibfield
  {journal} {\bibinfo  {journal} {Phys. Rev. E}\ }\textbf {\bibinfo {volume}
  {58}},\ \bibinfo {pages} {1823} (\bibinfo {year} {1998})}\BibitemShut
  {NoStop}%
\bibitem [{\citenamefont {Burrington}\ \emph {et~al.}(2013)\citenamefont
  {Burrington}, \citenamefont {Peet},\ and\ \citenamefont
  {Zadeh}}]{burrington2013operator}%
  \BibitemOpen
  \bibfield  {author} {\bibinfo {author} {\bibfnamefont {B.~A.}\ \bibnamefont
  {Burrington}}, \bibinfo {author} {\bibfnamefont {A.~W.}\ \bibnamefont
  {Peet}},\ and\ \bibinfo {author} {\bibfnamefont {I.~G.}\ \bibnamefont
  {Zadeh}},\ }\href {https://doi.org/10.1103/PhysRevD.87.106001} {\bibfield
  {journal} {\bibinfo  {journal} {Phys. Rev. D}\ }\textbf {\bibinfo {volume}
  {87}},\ \bibinfo {pages} {106001} (\bibinfo {year} {2013})},\ \Eprint
  {https://arxiv.org/abs/1211.6699} {arXiv:1211.6699 [hep-th]} \BibitemShut
  {NoStop}%
\bibitem [{\citenamefont {Poghosyan}\ and\ \citenamefont
  {Poghosyan}(2013)}]{poghosyan2013mixing}%
  \BibitemOpen
  \bibfield  {author} {\bibinfo {author} {\bibfnamefont {A.}~\bibnamefont
  {Poghosyan}}\ and\ \bibinfo {author} {\bibfnamefont {H.}~\bibnamefont
  {Poghosyan}},\ }\href {https://doi.org/10.1007/JHEP10(2013)131} {\bibfield
  {journal} {\bibinfo  {journal} {JHEP}\ }\textbf {\bibinfo {volume} {10}},\
  \bibinfo {pages} {131}},\ \Eprint {https://arxiv.org/abs/1305.6066}
  {arXiv:1305.6066 [hep-th]} \BibitemShut {NoStop}%
\bibitem [{\citenamefont {Burrington}\ \emph {et~al.}(2017)\citenamefont
  {Burrington}, \citenamefont {Jardine},\ and\ \citenamefont
  {Peet}}]{burrington2017operator}%
  \BibitemOpen
  \bibfield  {author} {\bibinfo {author} {\bibfnamefont {B.~A.}\ \bibnamefont
  {Burrington}}, \bibinfo {author} {\bibfnamefont {I.~T.}\ \bibnamefont
  {Jardine}},\ and\ \bibinfo {author} {\bibfnamefont {A.~W.}\ \bibnamefont
  {Peet}},\ }\href {https://doi.org/10.1007/JHEP06(2017)149} {\bibfield
  {journal} {\bibinfo  {journal} {JHEP}\ }\textbf {\bibinfo {volume} {06}},\
  \bibinfo {pages} {149}},\ \Eprint {https://arxiv.org/abs/1703.04744}
  {arXiv:1703.04744 [hep-th]} \BibitemShut {NoStop}%
\bibitem [{\citenamefont {Henriksson}\ and\ \citenamefont
  {Van~Loon}(2019)}]{henriksson2018critical}%
  \BibitemOpen
  \bibfield  {author} {\bibinfo {author} {\bibfnamefont {J.}~\bibnamefont
  {Henriksson}}\ and\ \bibinfo {author} {\bibfnamefont {M.}~\bibnamefont
  {Van~Loon}},\ }\href {https://doi.org/10.1088/1751-8121/aaf1e2} {\bibfield
  {journal} {\bibinfo  {journal} {J. Phys. A}\ }\textbf {\bibinfo {volume}
  {52}},\ \bibinfo {pages} {025401} (\bibinfo {year} {2019})},\ \Eprint
  {https://arxiv.org/abs/1801.03512} {arXiv:1801.03512 [hep-th]} \BibitemShut
  {NoStop}%
\bibitem [{\citenamefont {Warner}(1995)}]{warner1995supersymmetry}%
  \BibitemOpen
  \bibfield  {author} {\bibinfo {author} {\bibfnamefont {N.~P.}\ \bibnamefont
  {Warner}},\ }\href {https://doi.org/10.1016/0550-3213(95)00402-E} {\bibfield
  {journal} {\bibinfo  {journal} {Nucl. Phys. B}\ }\textbf {\bibinfo {volume}
  {450}},\ \bibinfo {pages} {663} (\bibinfo {year} {1995})},\ \Eprint
  {https://arxiv.org/abs/hep-th/9506064} {arXiv:hep-th/9506064} \BibitemShut
  {NoStop}%
\bibitem [{\citenamefont {Liendo}\ and\ \citenamefont
  {Meneghelli}(2017)}]{liendo2017bootstrap}%
  \BibitemOpen
  \bibfield  {author} {\bibinfo {author} {\bibfnamefont {P.}~\bibnamefont
  {Liendo}}\ and\ \bibinfo {author} {\bibfnamefont {C.}~\bibnamefont
  {Meneghelli}},\ }\href {https://doi.org/10.1007/JHEP01(2017)122} {\bibfield
  {journal} {\bibinfo  {journal} {JHEP}\ }\textbf {\bibinfo {volume} {01}},\
  \bibinfo {pages} {122}},\ \Eprint {https://arxiv.org/abs/1608.05126}
  {arXiv:1608.05126 [hep-th]} \BibitemShut {NoStop}%
\bibitem [{\citenamefont {Elvang}\ \emph {et~al.}(2019)\citenamefont {Elvang},
  \citenamefont {Hadjiantonis}, \citenamefont {Jones},\ and\ \citenamefont
  {Paranjape}}]{elvang2019soft}%
  \BibitemOpen
  \bibfield  {author} {\bibinfo {author} {\bibfnamefont {H.}~\bibnamefont
  {Elvang}}, \bibinfo {author} {\bibfnamefont {M.}~\bibnamefont
  {Hadjiantonis}}, \bibinfo {author} {\bibfnamefont {C.~R.~T.}\ \bibnamefont
  {Jones}},\ and\ \bibinfo {author} {\bibfnamefont {S.}~\bibnamefont
  {Paranjape}},\ }\href {https://doi.org/10.1007/JHEP01(2019)195} {\bibfield
  {journal} {\bibinfo  {journal} {JHEP}\ }\textbf {\bibinfo {volume} {01}},\
  \bibinfo {pages} {195}},\ \Eprint {https://arxiv.org/abs/1806.06079}
  {arXiv:1806.06079 [hep-th]} \BibitemShut {NoStop}%
\bibitem [{\citenamefont {Gimenez-Grau}\ \emph {et~al.}(2021)\citenamefont
  {Gimenez-Grau}, \citenamefont {Liendo},\ and\ \citenamefont {van
  Vliet}}]{gimenez2021superconformal}%
  \BibitemOpen
  \bibfield  {author} {\bibinfo {author} {\bibfnamefont {A.}~\bibnamefont
  {Gimenez-Grau}}, \bibinfo {author} {\bibfnamefont {P.}~\bibnamefont
  {Liendo}},\ and\ \bibinfo {author} {\bibfnamefont {P.}~\bibnamefont {van
  Vliet}},\ }\href {https://doi.org/10.1007/JHEP04(2021)167} {\bibfield
  {journal} {\bibinfo  {journal} {JHEP}\ }\textbf {\bibinfo {volume} {04}},\
  \bibinfo {pages} {167}},\ \Eprint {https://arxiv.org/abs/2012.00018}
  {arXiv:2012.00018 [hep-th]} \BibitemShut {NoStop}%
\bibitem [{\citenamefont {Bill{\`o}}\ \emph {et~al.}(2016)\citenamefont
  {Bill{\`o}}, \citenamefont {Gon{\c{c}}alves}, \citenamefont {Lauria},\ and\
  \citenamefont {Meineri}}]{billo2016defects}%
  \BibitemOpen
  \bibfield  {author} {\bibinfo {author} {\bibfnamefont {M.}~\bibnamefont
  {Bill{\`o}}}, \bibinfo {author} {\bibfnamefont {V.}~\bibnamefont
  {Gon{\c{c}}alves}}, \bibinfo {author} {\bibfnamefont {E.}~\bibnamefont
  {Lauria}},\ and\ \bibinfo {author} {\bibfnamefont {M.}~\bibnamefont
  {Meineri}},\ }\href {https://doi.org/10.1007/JHEP04(2016)091} {\bibfield
  {journal} {\bibinfo  {journal} {JHEP}\ }\textbf {\bibinfo {volume} {04}},\
  \bibinfo {pages} {091}},\ \Eprint {https://arxiv.org/abs/1601.02883}
  {arXiv:1601.02883 [hep-th]} \BibitemShut {NoStop}%
\bibitem [{\citenamefont {Liendo}\ \emph {et~al.}(2018)\citenamefont {Liendo},
  \citenamefont {Meneghelli},\ and\ \citenamefont
  {Mitev}}]{liendo2018bootstrapping}%
  \BibitemOpen
  \bibfield  {author} {\bibinfo {author} {\bibfnamefont {P.}~\bibnamefont
  {Liendo}}, \bibinfo {author} {\bibfnamefont {C.}~\bibnamefont {Meneghelli}},\
  and\ \bibinfo {author} {\bibfnamefont {V.}~\bibnamefont {Mitev}},\ }\href
  {https://doi.org/10.1007/JHEP10(2018)077} {\bibfield  {journal} {\bibinfo
  {journal} {JHEP}\ }\textbf {\bibinfo {volume} {10}},\ \bibinfo {pages}
  {077}},\ \Eprint {https://arxiv.org/abs/1806.01862} {arXiv:1806.01862
  [hep-th]} \BibitemShut {NoStop}%
\bibitem [{\citenamefont {Gadde}(2020)}]{gadde2020conformal}%
  \BibitemOpen
  \bibfield  {author} {\bibinfo {author} {\bibfnamefont {A.}~\bibnamefont
  {Gadde}},\ }\href {https://doi.org/10.1007/JHEP01(2020)038} {\bibfield
  {journal} {\bibinfo  {journal} {JHEP}\ }\textbf {\bibinfo {volume} {01}},\
  \bibinfo {pages} {038}},\ \Eprint {https://arxiv.org/abs/1602.06354}
  {arXiv:1602.06354 [hep-th]} \BibitemShut {NoStop}%
\bibitem [{\citenamefont {Gimenez-Grau}\ and\ \citenamefont
  {Liendo}(2020)}]{gimenez2020bootstrapping}%
  \BibitemOpen
  \bibfield  {author} {\bibinfo {author} {\bibfnamefont {A.}~\bibnamefont
  {Gimenez-Grau}}\ and\ \bibinfo {author} {\bibfnamefont {P.}~\bibnamefont
  {Liendo}},\ }\href {https://doi.org/10.1007/JHEP03(2020)121} {\bibfield
  {journal} {\bibinfo  {journal} {JHEP}\ }\textbf {\bibinfo {volume} {03}},\
  \bibinfo {pages} {121}},\ \Eprint {https://arxiv.org/abs/1907.04345}
  {arXiv:1907.04345 [hep-th]} \BibitemShut {NoStop}%
\bibitem [{\citenamefont {Antunes}(2021)}]{antunes2021conformal}%
  \BibitemOpen
  \bibfield  {author} {\bibinfo {author} {\bibfnamefont {A.}~\bibnamefont
  {Antunes}},\ }\href {https://doi.org/10.1007/JHEP10(2021)057} {\bibfield
  {journal} {\bibinfo  {journal} {JHEP}\ }\textbf {\bibinfo {volume} {10}},\
  \bibinfo {pages} {057}},\ \Eprint {https://arxiv.org/abs/2103.03132}
  {arXiv:2103.03132 [hep-th]} \BibitemShut {NoStop}%
\bibitem [{\citenamefont {Gimenez-Grau}\ \emph {et~al.}(2022)\citenamefont
  {Gimenez-Grau}, \citenamefont {Lauria}, \citenamefont {Liendo},\ and\
  \citenamefont {van Vliet}}]{gimenez2022bootstrapping}%
  \BibitemOpen
  \bibfield  {author} {\bibinfo {author} {\bibfnamefont {A.}~\bibnamefont
  {Gimenez-Grau}}, \bibinfo {author} {\bibfnamefont {E.}~\bibnamefont
  {Lauria}}, \bibinfo {author} {\bibfnamefont {P.}~\bibnamefont {Liendo}},\
  and\ \bibinfo {author} {\bibfnamefont {P.}~\bibnamefont {van Vliet}},\ }\href
  {https://doi.org/10.1007/JHEP11(2022)018} {\bibfield  {journal} {\bibinfo
  {journal} {JHEP}\ }\textbf {\bibinfo {volume} {11}},\ \bibinfo {pages}
  {018}},\ \Eprint {https://arxiv.org/abs/2208.11715} {arXiv:2208.11715
  [hep-th]} \BibitemShut {NoStop}%
\bibitem [{\citenamefont {Tr{\'e}panier}(2023)}]{trepanier2023surface}%
  \BibitemOpen
  \bibfield  {author} {\bibinfo {author} {\bibfnamefont {M.}~\bibnamefont
  {Tr{\'e}panier}},\ }\href {https://doi.org/10.1007/JHEP09(2023)074}
  {\bibfield  {journal} {\bibinfo  {journal} {JHEP}\ }\textbf {\bibinfo
  {volume} {09}},\ \bibinfo {pages} {074}},\ \Eprint
  {https://arxiv.org/abs/2305.10486} {arXiv:2305.10486 [hep-th]} \BibitemShut
  {NoStop}%
\bibitem [{\citenamefont {Hu}\ \emph {et~al.}(2024)\citenamefont {Hu},
  \citenamefont {He},\ and\ \citenamefont {Zhu}}]{hu2024solving}%
  \BibitemOpen
  \bibfield  {author} {\bibinfo {author} {\bibfnamefont {L.}~\bibnamefont
  {Hu}}, \bibinfo {author} {\bibfnamefont {Y.-C.}\ \bibnamefont {He}},\ and\
  \bibinfo {author} {\bibfnamefont {W.}~\bibnamefont {Zhu}},\ }\href
  {https://doi.org/10.1038/s41467-024-52959-2} {\bibfield  {journal} {\bibinfo
  {journal} {Nature Commun.}\ }\textbf {\bibinfo {volume} {15}},\ \bibinfo
  {pages} {9013} (\bibinfo {year} {2024})},\ \Eprint
  {https://arxiv.org/abs/2308.01903} {arXiv:2308.01903 [cond-mat.stat-mech]}
  \BibitemShut {NoStop}%
\bibitem [{\citenamefont {Meineri}\ and\ \citenamefont
  {Radhakrishnan}(2025)}]{meineri2025bootstrap}%
  \BibitemOpen
  \bibfield  {author} {\bibinfo {author} {\bibfnamefont {M.}~\bibnamefont
  {Meineri}}\ and\ \bibinfo {author} {\bibfnamefont {B.}~\bibnamefont
  {Radhakrishnan}},\ }\href {https://arxiv.org/abs/2506.17382} {\bibinfo
  {title} {The bootstrap of points and lines}} (\bibinfo {year} {2025}),\
  \Eprint {https://arxiv.org/abs/2506.17382} {arXiv:2506.17382 [hep-th]}
  \BibitemShut {NoStop}%
\bibitem [{\citenamefont {Lanzetta}\ \emph {et~al.}(2025)\citenamefont
  {Lanzetta}, \citenamefont {Liu},\ and\ \citenamefont
  {Metlitski}}]{lanzetta2025beginning}%
  \BibitemOpen
  \bibfield  {author} {\bibinfo {author} {\bibfnamefont {R.~A.}\ \bibnamefont
  {Lanzetta}}, \bibinfo {author} {\bibfnamefont {S.}~\bibnamefont {Liu}},\ and\
  \bibinfo {author} {\bibfnamefont {M.~A.}\ \bibnamefont {Metlitski}},\ }\href
  {https://arxiv.org/abs/2508.14964} {\bibinfo {title} {The beginning of the
  endpoint bootstrap for conformal line defects}} (\bibinfo {year} {2025}),\
  \Eprint {https://arxiv.org/abs/2508.14964} {arXiv:2508.14964
  [cond-mat.str-el]} \BibitemShut {NoStop}%
\bibitem [{\citenamefont {Antunes}\ \emph {et~al.}(2026)\citenamefont
  {Antunes}, \citenamefont {Kaviraj},\ and\ \citenamefont
  {Roy}}]{antunes2026ising}%
  \BibitemOpen
  \bibfield  {author} {\bibinfo {author} {\bibfnamefont {A.}~\bibnamefont
  {Antunes}}, \bibinfo {author} {\bibfnamefont {A.}~\bibnamefont {Kaviraj}},\
  and\ \bibinfo {author} {\bibfnamefont {B.}~\bibnamefont {Roy}},\ }\href
  {https://arxiv.org/abs/2605.22628} {\bibinfo {title} {Ising surface defects
  can get dirty}} (\bibinfo {year} {2026}),\ \Eprint
  {https://arxiv.org/abs/2605.22628} {arXiv:2605.22628 [hep-th]} \BibitemShut
  {NoStop}%
\bibitem [{\citenamefont {Huber}\ and\ \citenamefont
  {Ma{\^\i}tre}(2006)}]{huber2006hypexp}%
  \BibitemOpen
  \bibfield  {author} {\bibinfo {author} {\bibfnamefont {T.}~\bibnamefont
  {Huber}}\ and\ \bibinfo {author} {\bibfnamefont {D.}~\bibnamefont
  {Ma{\^\i}tre}},\ }\href {https://doi.org/10.1016/j.cpc.2006.01.007}
  {\bibfield  {journal} {\bibinfo  {journal} {Comput. Phys. Commun.}\ }\textbf
  {\bibinfo {volume} {175}},\ \bibinfo {pages} {122} (\bibinfo {year}
  {2006})},\ \Eprint {https://arxiv.org/abs/hep-ph/0507094}
  {arXiv:hep-ph/0507094} \BibitemShut {NoStop}%
\bibitem [{\citenamefont {{Wolfram Research}}()}]{WFS-07.23.04.0021.01}%
  \BibitemOpen
  \bibfield  {author} {\bibinfo {author} {\bibnamefont {{Wolfram Research}}},\
  }\href {https://functions.wolfram.com/07.23.04.0021.01} {\bibinfo {title}
  {Gauss hypergeometric function — specific values: Values at \(z=1\)
  (formula 07.23.04.0021.01)}}\BibitemShut {NoStop}%
\end{thebibliography}%

\end{document}